\documentclass[prd,aps,twocolumn,superscriptaddress,nofootinbib]{revtex4-2}

\usepackage{amsmath,graphicx,epsfig}
\usepackage{epstopdf}
\usepackage{euscript}
\usepackage{amsfonts}
\usepackage{amssymb}
\usepackage{float}

\usepackage{dcolumn}

\usepackage[
    colorlinks=true,
    linkcolor=blue,
    urlcolor=blue]{hyperref}

\def\prn#1{{\left(#1\right)}}

\def\sbrk#1{{\left[#1\right]}}
\def\abrk#1{{\langle#1\rangle}}

\def\bra#1{{\langle#1|}}

\def\cg(#1,#2)(#3,#4)(#5,#6){\bra{#1,#2,#3,#4}#5,#6\rangle}

\def\ts#1{{_{\mbox{\scriptsize #1}}}}

\def\threej(#1,#2)(#3,#4)(#5,#6){\begin{pmatrix}#1&#3&#5\\#2&#4&#6\end{pmatrix}}
\def\sixj(#1,#2,#3)(#4,#5,#6){\begin{Bmatrix}#1&#2&#3\\#4&#5&#6\end{Bmatrix}}
\def\ninej(#1,#2,#3)(#4,#5,#6)(#7,#8,#9){\begin{Bmatrix}#1&#2&#3\\#4&#5&#6\\#7&#8&#9\end{Bmatrix}}

\def\bs{\boldsymbol}
\def\mc{\mathcal}

\begin{document}

\title{Search for a solar-bound axion halo using the\\ Global Network of Optical Magnetometers for Exotic physics searches} 

\author{Tatum Z. Wilson}
\thanks{present address: Vector Atomic Inc., Pleasanton, California 94566, USA}
\affiliation{Department of Physics, California State University -- East Bay, Hayward, California 94542, USA}
\affiliation{Department of Physics, University of Illinois at Urbana-Champaign, Urbana, Illinois 61801, USA}

\author{Derek F. Jackson Kimball}
\email{derek.jacksonkimball@csueastbay.edu}
\affiliation{Department of Physics, California State University -- East Bay, Hayward, California 94542, USA}

\author{Samer Afach}
\affiliation{Helmholtz-Institut Mainz, 55099 Mainz, Germany}
\affiliation{Johannes Gutenberg-Universit{\"a}t Mainz, 55128 Mainz, Germany}
\affiliation{GSI Helmholtzzentrum f{\"u}r Schwerionenforschung GmbH, 64291 Darmstadt, Germany}

\author{Jiexiao	Bi}
\affiliation{State Key Laboratory of Photonics and Communications, School of Electronics, and Center for Quantum Information Technology, Peking University, Beijing 100871, China}

\author{B. C. Buchler}
\affiliation{Research School of Physics, Australian National University, Canberra, Australian Capital Territory 2601, Australia}

\author{Dmitry Budker} 
\affiliation{Helmholtz-Institut Mainz, 55099 Mainz, Germany}
\affiliation{Johannes Gutenberg-Universit{\"a}t Mainz, 55128 Mainz, Germany}
\affiliation{GSI Helmholtzzentrum f{\"u}r Schwerionenforschung GmbH, 64291 Darmstadt, Germany}
\affiliation{Department of Physics, University of California, Berkeley, California 94720, USA}

\author{Kaleb Cervantes}
\affiliation{Department of Physics, California State University -- East Bay, Hayward, California 94542, USA}

\author{Joshua Eby}
\affiliation{Kavli Institute for the Physics and Mathematics of the Universe (WPI), The University of Tokyo Institutes for Advanced Study, The University of Tokyo, Kashiwa, Chiba 277-8583, Japan}

\author{Nataniel L.	Figueroa}
\affiliation{Helmholtz-Institut Mainz, 55099 Mainz, Germany}
\affiliation{Johannes Gutenberg-Universit{\"a}t Mainz, 55128 Mainz, Germany}
\affiliation{GSI Helmholtzzentrum f{\"u}r Schwerionenforschung GmbH, 64291 Darmstadt, Germany}

\author{Ron	Folman}
\affiliation{Department of Physics, Ben-Gurion University of the Negev, Be’er Sheva 84105, Israel}

\author{Jiawei Gao}
\affiliation{Department of Physics, University of California, Santa Barbara, CA 93106, USA}

\author{Daniel Gavilán-Martín}
\affiliation{Helmholtz-Institut Mainz, 55099 Mainz, Germany}
\affiliation{Johannes Gutenberg-Universit{\"a}t Mainz, 55128 Mainz, Germany}
\affiliation{GSI Helmholtzzentrum f{\"u}r Schwerionenforschung GmbH, 64291 Darmstadt, Germany}

\author{Menachem Givon}
\affiliation{Department of Physics, Ben-Gurion University of the Negev, Be’er Sheva 84105, Israel}

\author{Zoran D. Grujić}
\affiliation{Institute of Physic Belgrade, University of Belgrade, 11080 Belgrade, Serbia}

\author{Hong Guo}
\affiliation{State Key Laboratory of Photonics and Communications, School of Electronics, and Center for Quantum Information Technology, Peking University, Beijing 100871, China}

\author{Paul Hamilton}
\affiliation{Department of Physics and Astronomy, University of California, Los Angeles, California 90095, USA}

\author{M. P. Hedges}
\affiliation{Centre for Quantum Computation and Communication Technology, Research School of Physics, Australian National University, Canberra, Australian Capital Territory 2601, Australia}

\author{Zhejun Huang}
\affiliation{State Key Laboratory of Photonics and Communications, School of Electronics, and Center for Quantum Information Technology, Peking University, Beijing 100871, China}

\author{Dongok Kim}
\thanks{present address: Mechatronics Research, Samsung Electronics, Hwaseong 18448, South Korea}
\affiliation{Department of Physics, Korea Advanced Institute of Science and Technology, 34141, Republic of Korea}
\affiliation{Center for Axion and Precision Physics Research, Institute for Basic Science, 34051, Republic of Korea}

\author{Younggeun Kim}
\affiliation{Department of Physics, Korea Advanced Institute of Science and Technology, 34141, Republic of Korea}
\affiliation{Center for Axion and Precision Physics Research, Institute for Basic Science, 34051, Republic of Korea}
\affiliation{Helmholtz-Institut Mainz, 55099 Mainz, Germany}
\affiliation{Johannes Gutenberg-Universit{\"a}t Mainz, 55128 Mainz, Germany}
\affiliation{GSI Helmholtzzentrum f{\"u}r Schwerionenforschung GmbH, 64291 Darmstadt, Germany}

\author{Sami S. Khamis}
\affiliation{Department of Physics and Astronomy, University of California, Los Angeles, California 90095, USA}

\author{Emmanuel Klinger}
\affiliation{Université Marie et Louis Pasteur, CNRS, SUPMICROTECH, Institut FEMTO-ST, Besançon, France}

\author{Abaz Kryemadhi}
\affiliation{Department of Computing, Math, and Physics, Messiah University, Mechanicsburg, Pennsylvania 17055, USA}

\author{Nina Kukowski}
\affiliation{Friedrich Schiller University Jena, Institute of Geosciences, Burgweg 11, 07749 Jena, Germany}

\author{Jianjun	Li}
\affiliation{State Key Laboratory of Photonics and Communications, School of Electronics, and Center for Quantum Information Technology, Peking University, Beijing 100871, China}

\author{Grzegorz Łukasiewicz}
\affiliation{Institute of Physics, Jagiellonian Universty in Kraków, Łojasiewicza 11, 30-368, Kraków, Poland}
\affiliation{Doctoral School of Exact and Natural Sciences, Jagiellonian University in Kraków, Kraków, Poland}

\author{Hector Masia-Roig}
\affiliation{Helmholtz-Institut Mainz, 55099 Mainz, Germany}
\affiliation{Johannes Gutenberg-Universit{\"a}t Mainz, 55128 Mainz, Germany}
\affiliation{GSI Helmholtzzentrum f{\"u}r Schwerionenforschung GmbH, 64291 Darmstadt, Germany}

\author{Tafai Muck}
\affiliation{Department of Physics, California State University -- East Bay, Hayward, California 94542, USA}

\author{Michał Padniuk}
\affiliation{Institute of Physics, Jagiellonian Universty in Kraków, Łojasiewicza 11, 30-368, Kraków, Poland}

\author{Christopher A. Palm}
\affiliation{Department of Physics, California State University -- East Bay, Hayward, California 94542, USA}

\author{Chaitanya Paranjape}
\affiliation{Department of Particle Physics and Astrophysics, Weizmann Institute of Science, Rehovot 7610001, Israel}

\author{Sun Yool Park}
\thanks{present address: Department of Physics, University of Colorado, Boulder, Colorado 80309, USA}
\affiliation{Department of Physics and Astronomy, Oberlin College, Oberlin, Ohio 44074, USA}

\author{Xiang Peng}
\affiliation{State Key Laboratory of Photonics and Communications, School of Electronics, and Center for Quantum Information Technology, Peking University, Beijing 100871, China}

\author{Gilad Perez}
\affiliation{Department of Particle Physics and Astrophysics, Weizmann Institute of Science, Rehovot 7610001, Israel}

\author{Szymon Pustelny}
\affiliation{Institute of Physics, Jagiellonian Universty in Kraków, Łojasiewicza 11, 30-368, Kraków, Poland}
\affiliation{Department of Physics, Harvard University, 17 Oxford Street, Cambridge, 02138 USA}

\author{Wolfram	Ratzinger}
\affiliation{Department of Particle Physics and Astrophysics, Weizmann Institute of Science, Rehovot 7610001, Israel}

\author{Yossi Rosenzweig}
\affiliation{Department of Physics, Ben-Gurion University of the Negev, Be’er Sheva 84105, Israel}

\author{Ophir M. Ruimi}
\affiliation{Racah Institute of Physics, Hebrew University of Jerusalem, Jerusalem 9190401, Israel}
\affiliation{Helmholtz-Institut Mainz, 55099 Mainz, Germany}
\affiliation{Johannes Gutenberg-Universit{\"a}t Mainz, 55128 Mainz, Germany}
\affiliation{GSI Helmholtzzentrum f{\"u}r Schwerionenforschung GmbH, 64291 Darmstadt, Germany}

\author{Amy Saputo}
\affiliation{Department of Physics, California State University -- East Bay, Hayward, California 94542, USA}

\author{Theo Scholtes}
\affiliation{Leibniz Institute of Photonic Technology, Albert-Einstein-Straße 9, 07745 Jena, Germany}

\author{P. C. Segura}
\thanks{present address: Department of Physics \& Astrononomy, Harvard University, Cambridge Massachusetts  02138, USA}
\affiliation{Department of Physics and Astronomy, Oberlin College, Oberlin, Ohio 44074, USA}

\author{Yannis K. Semertzidis}
\affiliation{Department of Physics, Korea Advanced Institute of Science and Technology, 34141, Republic of Korea}
\affiliation{Center for Axion and Precision Physics Research, Institute for Basic Science, 34051, Republic of Korea}

\author{Yun Chang Shin}
\affiliation{Center for Axion and Precision Physics Research, Institute for Basic Science, 34051, Republic of Korea}

\author{Jason E. Stalnaker}
\affiliation{Department of Physics and Astronomy, Oberlin College, Oberlin, Ohio 44074, USA}

\author{Ibrahim Sulai}
\affiliation{Department of Physics and Astronomy, Bucknell University. 1 Dent Drive, Lewisburg PA 17837, USA}

\author{Dhruv Tandon}
\thanks{present address: Department of Physics, Stanford University, Palo Alto, California 94305, USA}
\affiliation{Department of Physics and Astronomy, Oberlin College, Oberlin, Ohio 44074, USA}

\author{Rayshaun Thompson}
\affiliation{Department of Physics, California State University -- East Bay, Hayward, California 94542, USA}

\author{Ken Vu}
\affiliation{Department of Physics, California State University -- East Bay, Hayward, California 94542, USA}

\author{Arne Wickenbrock}
\affiliation{Helmholtz-Institut Mainz, 55099 Mainz, Germany}
\affiliation{Johannes Gutenberg-Universit{\"a}t Mainz, 55128 Mainz, Germany}
\affiliation{GSI Helmholtzzentrum f{\"u}r Schwerionenforschung GmbH, 64291 Darmstadt, Germany}

\author{Teng Wu}
\affiliation{State Key Laboratory of Photonics and Communications, School of Electronics, and Center for Quantum Information Technology, Peking University, Beijing 100871, China}

\author{Yucheng Yang}
\affiliation{State Key Laboratory of Photonics and Communications, School of Electronics, and Center for Quantum Information Technology, Peking University, Beijing 100871, China}

\author{Yixin Zhao}
\affiliation{State Key Laboratory of Photonics and Communications, School of Electronics, and Center for Quantum Information Technology, Peking University, Beijing 100871, China}

\date{\today}



\begin{abstract}

We report on a search for a gravitationally bound solar axion halo using data from the Global Network of Optical Magnetometers for Exotic physics searches (GNOME), a worldwide array of magnetically shielded atomic magnetometers with sensitivity to exotic spin couplings. Motivated by recent theoretical work suggesting that self-interacting ultralight axions can be captured by the Sun’s gravitational field and thermalize into the ground state, we develop a signal model for the pseudo-magnetic fields generated by axion–proton gradient couplings in such a halo. The analysis focuses on the fifth GNOME Science Run (69 days, 12 stations), employing a cross-correlation pipeline with time-shifted daily modulation templates to search for the global, direction-dependent, monochromatic signal expected from a solar axion halo. No statistically significant candidate signals are observed. We set 95\% confidence-level upper limits on the amplitude of the axion-induced pseudo-magnetic field over the frequency range $\approx 0.05-20$~Hz, translating to constraints on the linear and quadratic axion–proton couplings for halo densities predicted by gravitational capture models and for the maximum overdensities allowed by planetary ephemerides. In the quadratic coupling case, our limits surpass existing astrophysical bounds by over two orders of magnitude across much of the accessible parameter space.

\end{abstract}



\maketitle


\section{Introduction}

Understanding the nature of dark matter is of utmost significance to the fields of astrophysics, cosmology, and particle physics.
A well-motivated possibility is that dark matter consists of ultralight bosons such as axions or axion-like particles (ALPs) with  masses $m_a \ll 1\,{\rm{eV}}$ \cite{graham2015experimental,kimball2022search}, hereafter generically referred to as axions.
The phenomenology of axion dark matter is well described by a classical field oscillating near the Compton frequency,
\begin{align}
    \omega_a = \frac{m_a c^2}{\hbar}~,
\label{eq:ComptonFrequency}
\end{align}
where $c$ is the speed of light and $\hbar$ is Planck's constant.

\begin{figure}
\center
\includegraphics[width=\columnwidth]{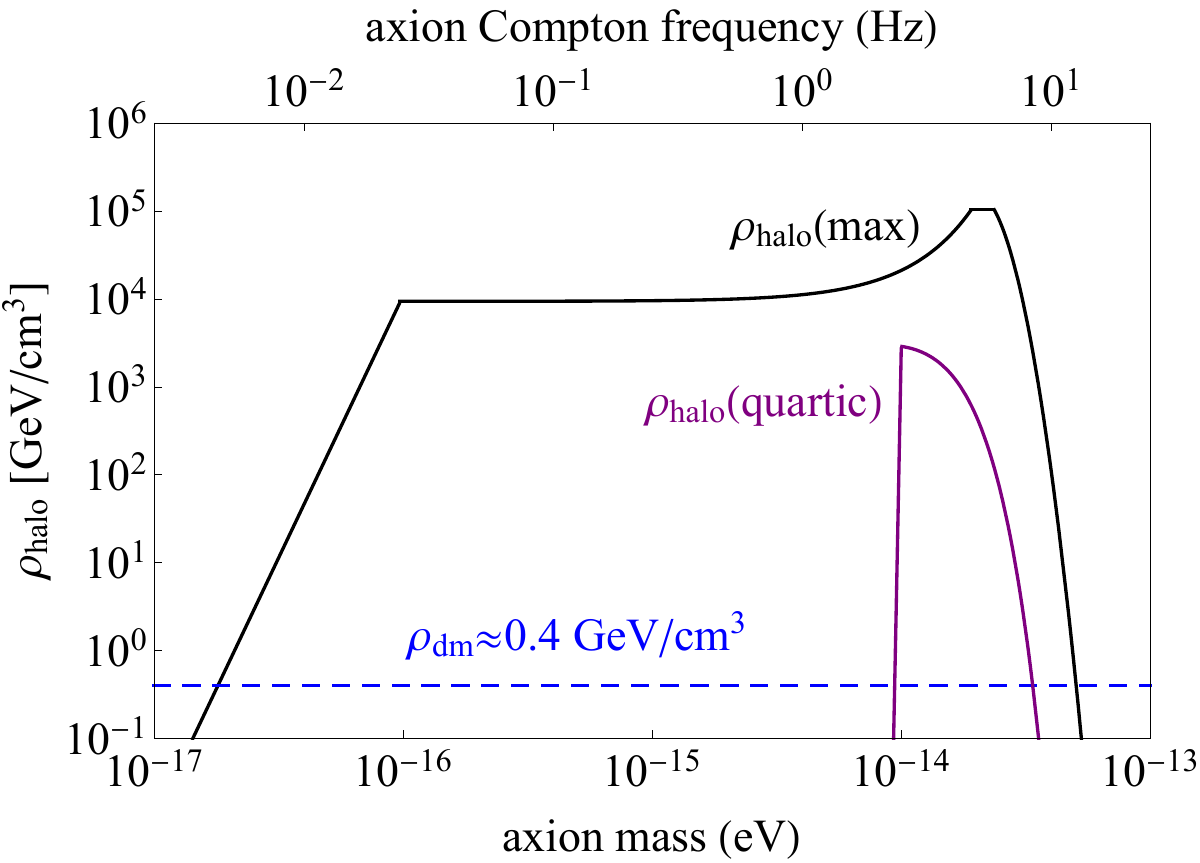}
\caption{The potential enhanced density at the position of the Earth for a solar axion halo as compared to the average local dark matter density $\rho\ts{dm}$ (dashed blue line). The black line represents the maximum solar axion halo mass density, $\rho\ts{halo}({\rm{max}})$, allowed by gravitational constraints as evaluated in Refs.\,\cite{banerjee2020relaxion,banerjee2020searching}. The purple line represents the solar axion halo mass density, $\rho\ts{halo}({\rm{quartic}})$, predicted from gravitational capture of axions by the Sun where the dissipation mechanism is provided by a quartic self-interaction of axions as described in detail in Ref.\,\cite{budker2023generic}.}
\label{Fig:max-solar-ALP-halo-density}
\end{figure}

It has recently been suggested that bodies such as the Earth and the Sun may accumulate a halo of axions \cite{banerjee2020relaxion,banerjee2020searching,tretiak2022improved}, leading to a substantial overdensity of the axion field near these bodies compared to the average dark matter density (Fig.\,\ref{Fig:max-solar-ALP-halo-density}).
Possible mechanisms for the capture of dark matter by dense astrophysical bodies have been studied, for example, in Refs.\,\cite{xu2008dark,khriplovich2009capture,khriplovich2011capture,khriplovich2011capture2,brito2015accretion,brito2016interaction}, and they continue to be actively investigated.
Recent calculations have revealed that the quartic self-interaction of axions typically allows axion dark matter to be captured by external gravitational forces \cite{budker2023generic}, such as those of stars like the Sun.
This results in the formation of axion bound states, which can, in a sense, be thought of as a sort of ``gravitational atom.''
Through this generic mechanism, a solar axion halo can be created around the Sun.\footnote{We note that the term ``gravitational atom'' has appeared previously in the literature to denote two-body gravitational bound states of very heavy dark-sector particles \cite{nielsen2019gravitational}, as well as related ``atom'' analogies involving a superradiant black-hole ``nucleus'' surrounded by an axion cloud (see, for example, Refs.\,\cite{arvanitaki2011exploring,baryakhtar2021black}). In contrast, the present work uses ``gravitational atom'' as an \emph{analogy} for an ultralight axion(-like) field gravitationally bound to an astrophysical body like the Sun and predominantly occupying the ground-state configuration. To avoid confusion we otherwise refer to this system as a \emph{ground-state solar axion halo}, or, simply, a \emph{solar axion halo}.}

Assuming that axions constitute the majority of dark matter, the crucial missing ingredient for the formation of a solar axion halo is some form of dissipation.
Unbound axions have positive total energy, then accelerate as they fall into the solar gravitational potential, and without dissipation the kinetic energy is reconverted into gravitational potential energy and the axions subsequently depart the solar system.
Consequently, an efficient energy dissipation mechanism is required for axion dark matter to be captured by the Sun’s gravitational field.
The energy dissipation mechanism described in Ref.\,\cite{budker2023generic} involves gravitational focusing of axion dark matter by the Sun \cite{kryemadhi2023gravitational,kim2022gravitational,patla2013flux}, leading to an enhancement of the axion density.
There is efficient gravitational focusing if the de Broglie wavelength $\lambdabar\ts{dB} = \hbar/\prn{ m_a v }$ of the axions (where $v$ is the relative speed of the axions with respect to the Sun) is greater than the ``gravitational Bohr radius'' $R_\star$ of the system,
\begin{align}
    R_\star \approx \frac{\hbar^2}{G_N M_\odot m_a^2}~, \label{eq:Rstar}
\end{align}
where $G_N$ is Newton's gravitational constant, and $M_\odot$ is the mass of the Sun.
Thus, efficient gravitational focusing is achieved when
\begin{align}
m_a \gtrsim \frac{\hbar v}{G_N M_\odot}\,.
\end{align}
In this regime, the Sun's $1/r$ potential coherently amplifies and distorts the incoming axion wave within $r \lesssim R_\star$.
If the axion field $a(\bs{r},t)$ has sufficiently strong quartic self-interactions, $\propto a^4$, the enhanced density from gravitational focusing can lead to significant axion-axion scattering, resulting in velocity-changing collisions in which one axion is scattered into lower-energy gravitational bound states while the other is ejected to infinity with increased energy.
Eventually, the bound-state axion density becomes sufficiently large so that Bose-enhanced stimulated transitions from the ensemble of unbound dark matter axions into the ground state of the solar axion halo becomes significant and a relatively large axion density accumulates in the halo ground state \cite{budker2023generic}.
In the model of Ref.\,\cite{budker2023generic}, the strength of the quartic axion self-interactions sets the gravitational capture timescale, the relaxation rate of axions to the ground state, and the necessary critical density created by gravitational focusing for the axions to efficiently accumulate in the halo.
Note that the halo creation process can be efficient on timescales of billions of years, even in the case where the axion quartic self-interactions are relatively small, namely when $m_a/f_a \ll 1$, where $f_a$ is the axion decay constant that sets the scale of axion interactions.
Therefore, on the time scale of the formation of the solar system, an axion halo can plausibly form in the gravitational potential well of the Sun.

The term \emph{solar axion halo} used in this work explicitly refers to a classical stable configuration of a scalar (axion) field in the ground state centered around the Sun as found in~\cite{banerjee2020relaxion,banerjee2020searching}, where the axion density is proportional to $e^{-2r/R_\star}$.
Fundamentally, the axion field configuration resulting from the aforementioned formation process \cite{budker2023generic} can be understood as a non-relativistic field that corresponds to an ensemble of ultralight scalar dark matter particles with macroscopic occupation in the ground state (principal quantum number $n=1$, orbital angular momentum quantum number $\ell = 0$).
This basically matches the configuration described in Ref.\,\cite{foster2018revealing}, but only the $1s$ ground state of a gravitational ``hydrogen'' atom is populated and in a coherent state.
See also Ref.\,\cite{cheong2025quantum} for related discussion.


Figure 1 shows the possible enhancement of the axion density $\rho\ts{halo}$ at the location of the Earth in comparison to the commonly assumed dark matter density $\rho\ts{dm}$ (dashed blue line) in the solar system \cite{de2021dark,particle2022review}.
The maximum possible axion density in the solar halo, $\rho\ts{halo}({\rm{max}})$, based on gravitational constraints derived from planetary ephemerides \cite{pitjev2013constraints}, is shown by the black line (see discussion in Refs.\,\cite{banerjee2020relaxion,banerjee2020searching}).
We take $\rho\ts{halo}({\rm{max}})$ to be the upper limit on signal enhancement from an axion solar halo, being agnostic as to the possible capture mechanism.
The predicted axion density $\rho\ts{halo}({\rm{quartic}})$ based on the model discussed above and described in detail in Ref.\,\cite{budker2023generic} is shown by the purple line.
There is a sharp decrease of $\rho\ts{halo}({\rm{quartic}})$ for $m_a \lesssim 10^{-14}$\,eV due to the fact that gravitational focusing becomes inefficient at such low masses and so the Sun cannot capture the axions, and a more gradual decrease of $\rho\ts{halo}({\rm{quartic}})$ at larger masses as the effective radius of the solar axion halo, $R_\star$, shrinks below the radius of Earth's orbit.

The axion field $a(\bs{r},t)$ may couple to atomic spins through the linear or quadratic gradient interactions,\footnote{Note that here we refer to the interaction of the axion field with fermion spins \cite{graham2013new,stadnik2014axion,cong2025spin}, which is distinct from the axion self-interaction that provides the dissipation mechanism for the gravitational capture driving the formation of the axion solar halo \cite{budker2023generic}.} which in the nonrelativistic limit are described by the Hamiltonians \cite{Pos13,safronova2018search,dailey2021quantum,afach2023WhatCanGNOMEdo}
\begin{align}
    \mc{H}_l = -\frac{2 (\hbar c)^{3/2}}{f_l} \bs{S} \cdot \bs{\nabla} a(\bs{r},t)~, \label{eq:linear-Hamiltonian} \\
    \mc{H}_q = -\frac{2\hbar^2c^2}{f_q^2} \bs{S} \cdot \bs{\nabla} a^2(\bs{r},t)~, \label{eq:quadratic-Hamiltonian}
\end{align}
where $\bs{S}$ is the atomic spin in units of $\hbar$, and $f_l$ and $f_q$ are the respective coupling constants (which in turn are related to the axion symmetry breaking scale).
The physically observable effects of an axion field coupling through the gradient interactions can be parametrized in terms of a ``pseudo-magnetic'' field $\bs{\mc{B}}_p$; this is possible because of the close similarity between the form of Eqs.~\eqref{eq:linear-Hamiltonian} and \eqref{eq:quadratic-Hamiltonian} and the form of the Zeeman Hamiltonian,
\begin{align}
    \mc{H}_Z = - \gamma \bs{S} \cdot \bs{B}~, \label{eq:Zeeman-Hamiltonian}
\end{align}
where $\gamma$ is the gyromagnetic ratio for the atom and $\bs{B}$ is a real magnetic field.
Crucially, the pseudo-magnetic field $\bs{\mc{B}}_p$ \emph{does not couple to magnetic moments} (since it is \emph{not} a real magnetic field), but rather $\bs{\mc{B}}_p$ couples to electron and/or nuclear spin, and thus generally the magnitude of the relative atomic response to $\bs{\mc{B}}_p$ can be quite different compared to the atomic response to a magnetic field $\bs{B}$ \cite{Kim15,kimball2016magnetic}.

\begin{figure}
\center
\includegraphics[width=\columnwidth]{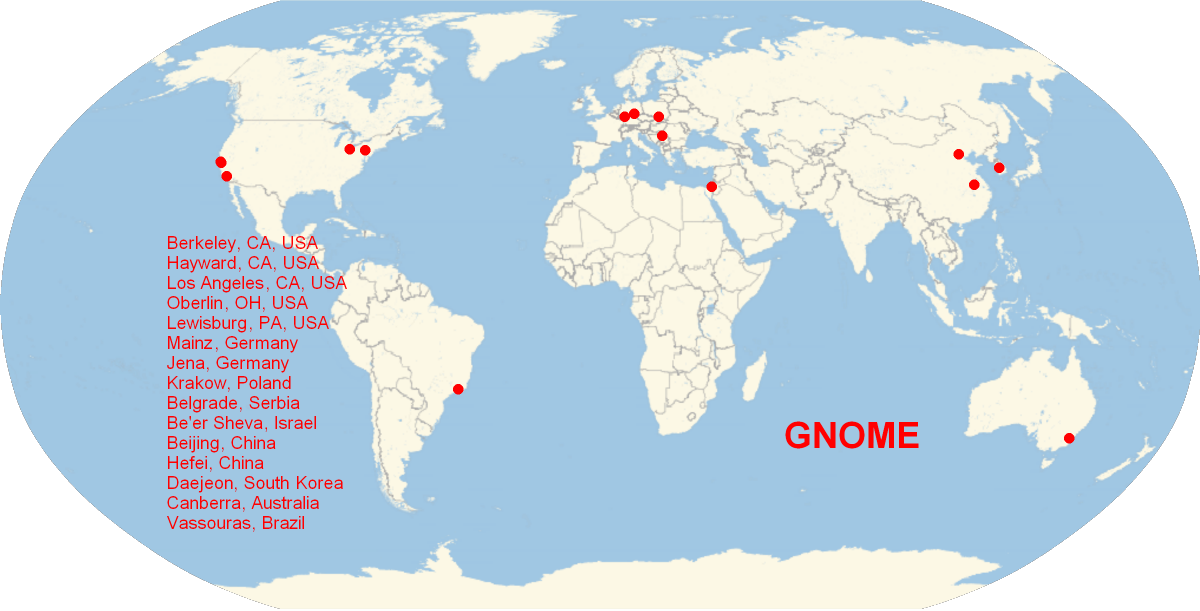}
\caption{Map and list of locations of presently active GNOME stations (summer 2025).}
\label{Fig:GNOME-map}
\end{figure}

In this work we search for a solar axion halo by analyzing data from the Global Network of Optical Magnetometers for Exotic physics searches (GNOME) \cite{Pos13,pustelny2013global,kimball2018searching,afach2018characterization,masia2020analysis,afach2021search,afach2023WhatCanGNOMEdo}, a worldwide network of magnetically shielded optical atomic magnetometers \cite{budker2007optical,Bud13} designed to search for beyond-the-Standard-Model physics (see Fig.~\ref{Fig:GNOME-map} for a map of active GNOME stations).
The GNOME magnetometers have sensitivities of roughly between $0.1 - 1.0$~${\rm pT/\sqrt{Hz}}$ over bandwidths $\approx 100~{\rm Hz}$ \cite{afach2018characterization,afach2021search,afach2023WhatCanGNOMEdo}.
Because all GNOME magnetometers are placed within multi-layer mu-metal or ferrite magnetic shields, sensitivity to predominantly electron-spin-coupled interactions (such as the Zeeman interaction) is greatly reduced \cite{kimball2016magnetic}.
Thus GNOME is primarily sensitive to the interaction of exotic fields with nuclear spins.
Since, at present, all GNOME magnetometers are based on measurement of spin precession in alkali atoms, whose nuclei have valence protons, our experiment searches in particular for the coupling of the axion field to proton spins \cite{Kim15}.

Individual GNOME magnetometers record changes in the local magnetic field $\bs{B}$ within the shields by measuring magneto-optical properties of spin-polarized atomic vapors \cite{budker2002resonant,budker2007optical,Bud13}.
The measured signals are translated into magnetic field units via Eq.\,\eqref{eq:Zeeman-Hamiltonian}.
In order to interpret data in terms of a global pseudo-magnetic field $\bs{\mc{B}}_p$ coupling specifically to proton spins, the measured field at each station $j$ is re-scaled according to
\begin{align}
    \bs{\mc{B}}_j = \frac{\sigma_j}{g_{F,j}} \bs{\mc{B}}_p~, \label{eq:pseudo-mag-field-in-each-sensor}
\end{align}
where $\sigma_j$ is the effective proton spin polarization \cite{Kim15}, $g_{F,j}$ is the Land\'e g-factor for the atoms used in magnetometer $j$, and $\bs{\mc{B}}_p$ is the pseudo-magnetic field related to the axion gradient couplings to proton spins described by Eqs.~\eqref{eq:linear-Hamiltonian} and \eqref{eq:quadratic-Hamiltonian}:
\begin{align}
    \bs{\mc{B}}_{p,l}(\bs{r},t) & = -\frac{2 (\hbar c)^{3/2}}{\mu_B f_l}  \bs{\nabla} a(\bs{r},t) ~, \label{eq:pseudo-mag-field-linear} \\
    \bs{\mc{B}}_{p,q}(\bs{r},t) & = -\frac{2 (\hbar c)^2}{\mu_B f_q^2}  \bs{\nabla} a^2(\bs{r},t)~, \label{eq:pseudo-mag-field-quadratic}
\end{align}
where $\mu_B$ is the Bohr magneton.
Tables listing $\sigma_j/g_{F,j}$ for various GNOME magnetometers are given in Appendix~\ref{appendix:GNOME-mag-characteristics}.

In the present work, the spatiotemporal characteristics of the field $a(\bs{r},t)$ are based on the model of a gravitationally bound solar axion halo described in Refs.\,\cite{banerjee2020relaxion,banerjee2020searching,budker2023generic}.
In particular, our analysis assumes that the axions in the solar halo are predominantly in the ground state, described by the $n=1$, $\ell=0$, $m_\ell = 0$ wave function as discussed in Ref.\,\cite{budker2023generic}, where $n$ is the radial quantum number, $\ell$ is the quantum number associated with the orbital angular momentum, and $m_\ell$ is the quantum number associated with the projection of the orbital angular momentum along a particular quantization axis.
This will be the case if the gravitational capture mechanism involves, as it likely must, an efficient dissipation mechanism coupled with the Bose enhancement of scattering into the ground state.
For the remainder of this work we explicitly assume that we are measuring a solar halo where essentially all the axions are in the ground $n=1$, $\ell=0$, $m_\ell = 0$ state, and ignore any excited state axions.
Therefore we can assume that there is no transverse momentum of the axions in the rest frame of the Sun (and, also, no dispersion of the measurable transverse momentum).\footnote{One may wonder if the gravitational pull from planets might tidally disrupt the solar axion halo and excite a non-neglible population of $l \neq 0$ states of the solar axion halo.  The largest contribution to such an effect would come from Jupiter. The quadrupolar tidal potential due to Jupiter acting on the solar axion halo is given by $\Phi\ts{tid} = - G_N M_J R_\star^2 P_2(\cos\vartheta) / r_J^3 $, where $M_J$ is the mass of Jupiter, $r_J$ is the radius of Jupiter's orbit, and $P_2(\cos\vartheta) = \prn{ 3\cos^2\vartheta - 1 }/2$ is the second degree Legendre polynomial with $\vartheta$ being the angle between Jupiter and a given location in the solar axion halo. For the purposes of this estimate, we take $P_2(\cos\vartheta) \sim 1$. The significance of tidal effects due to Jupiter can be gleaned from estimating the fractional tidal bulge of the solar axion halo, $\left| \delta R_\star/R_\star \right| \approx \Phi\ts{tid} R_\star^2 / \prn{G_N M_\odot R_\star } \sim M_J R_\star^3/\prn{ M_\odot r_J^3 } $. For $m_a \sim 10^{-14}$\,eV we find that $\left| \delta R_\star/R_\star \right| \lesssim 10^{-6}$, in which case excitation of $l \neq 0$ states of the solar axion halo by tidal forces can be neglected. However, already for $m_a \lesssim 10^{-15}$\,eV the estimated tidal forces due to Jupiter become non-negligible. Thus we conclude that while there is no effect on our analysis pertaining to the quartic model, there may be a non-negligible effect for smaller $m_a$. Detailed considerations of these effects are beyond the scope of the present work and we assume the dominant fraction of the axions remain in the ground state. Finally, our constraints pertain to a present-day solar axion halo. We implicitly assume either (i) late-time formation after the gas disk dispersed and planetary migration largely ceased, or (ii) survival through early Solar-System epochs. The latter is model-dependent: drag in the protoplanetary disk and dynamical heating during migration could deplete a bound population. Our limits should therefore be read as conditional on one of these assumptions holding.}
This is in contrast to a halo of virialized axions (occupying a range of excited states) that has a random (stochastic) component of momentum comparable to the relative velocity of the Earth with respect to the Sun.\footnote{If a significant excited-state population were present, azimuthal flow could modestly increase the transverse component of the signal, but the resulting angular/phase dispersion would likely degrade the daily-modulation coherence that our analysis exploits, so the net effect would typically weaken our sensitivity. Thus our result should be interpreted as specifically constraining the existence of ground-state solar axion halos.}

For a solar axion halo in the ground state, the axion field amplitude at the position of the Earth is exponentially decaying over a characteristic length scale given by $R_\star$ [Eq.\,\eqref{eq:Rstar}].
For the solar axion halo to extend to the position of the Earth, $R_\star \gtrsim R\ts{ES}$ where $R\ts{ES}$ is the distance between the Earth and Sun; this imposes the requirement that $m_a c^2 \lesssim 10^{-14}~{\rm eV}$ (although we note, as can be seen from Fig.~\ref{Fig:max-solar-ALP-halo-density}, that the particular gravitational constraints on the possible solar axion halo density allow significant overdensity for axion masses up to $m_a c^2 \sim 5 \times 10^{-14}~{\rm eV}$).
The axion field oscillates at $\omega_a$ with a coherence time of at least $\tau\ts{coh} \gtrsim m_a R_\star^2 / \hbar$ \cite{banerjee2020relaxion,banerjee2020searching} (for $m_a c^2 \lesssim 10^{-14}~{\rm eV}$, $\tau\ts{coh}$ is longer than $\sim 10^7$~s), and much longer if the axions are in the ground state as we assume.
Therefore, for our analysis we can treat the axion field as single frequency.
In the rest frame of the halo, which is also the rest frame of the Sun, the axion field can be described as \cite{banerjee2020relaxion,banerjee2020searching}:
\begin{align}
    a(\bs{r},t) \approx a_0 \cos\prn{ \omega_a t + \theta } e^{ - r / R_\star }~, \label{eq:solar-halo-axion-field-Sun-frame}
\end{align}
where $a_0$ is a constant determined by the overall energy density in the solar axion halo and $\theta$ is a random phase, constant over the coherence time and coherence length.
The motion of a sensor through the solar axion halo generates an additional phase shift (obtained by Lorentz boosting the axion field into the observer frame), and so the axion field probed in our experiment is:
\begin{align}
    a(\bs{r},t) \approx a_0 \cos\prn{ \omega_a t - \bs{k}\cdot\bs{r} + \theta } e^{ - r / R_\star }~, \label{eq:solar-halo-ALP-field}
\end{align}
where $\bs{k}$ is the axion field wave vector due to the relative motion between the sensor and the halo's rest frame.
The average energy density of the axion solar halo at a distance $r$ from the Sun is given by
\begin{align}
    \rho\ts{halo} (r) = \frac{m_a^2 c^2}{2\hbar^2} \abrk{ a^2(\bs{r},t) } = \frac{m_a^2 c^2}{4\hbar^2} a_0^2 e^{ - 2r / R_\star }~, \label{eq:avg-halo-density}
\end{align}
which allows estimation of the field amplitude assuming a particular axion overdensity, where $\abrk{ \cdots }$ indicates the time-averaged quantity.

\begin{figure}
\center
\includegraphics[width=0.8\columnwidth]{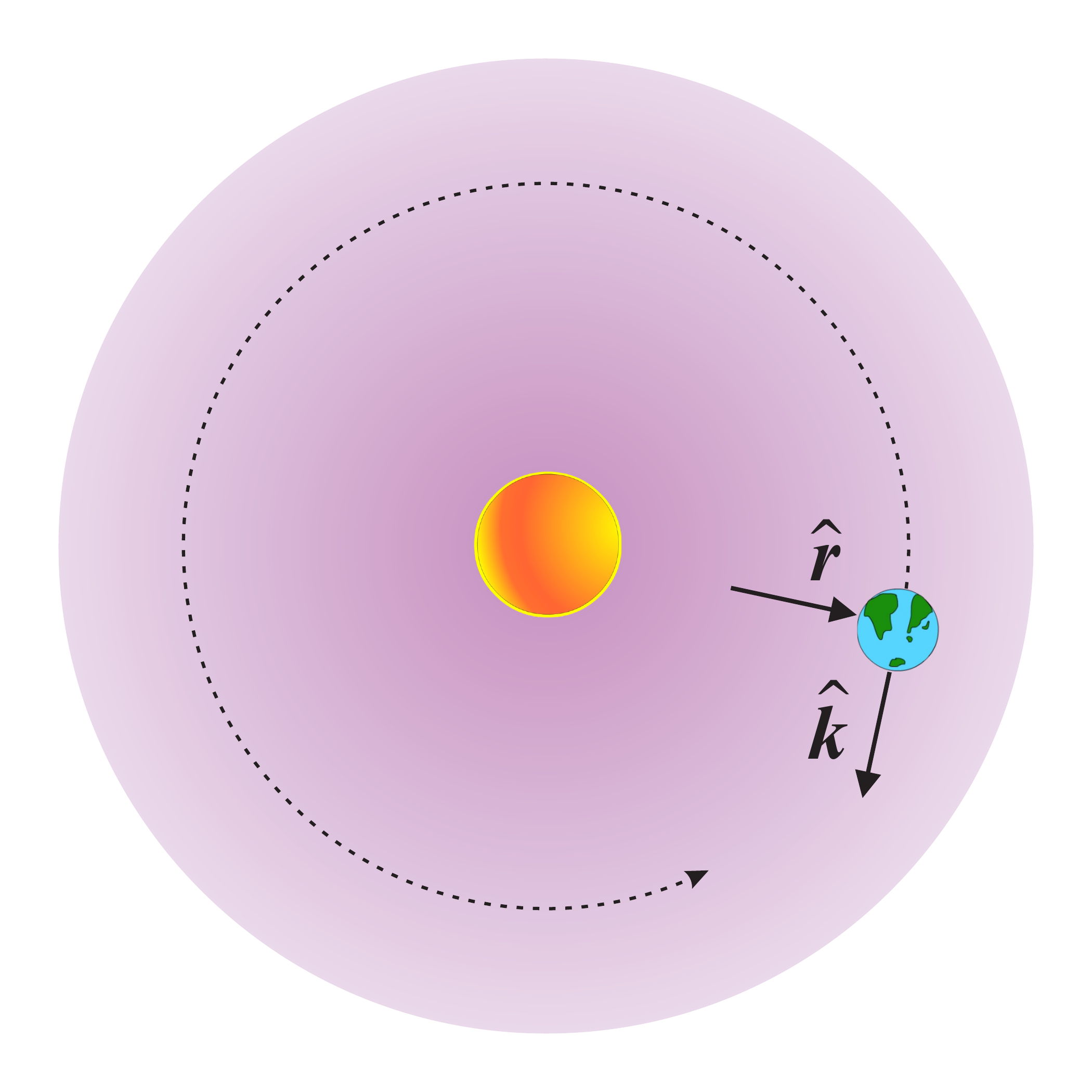}
\caption{Schematic diagram of the Earth moving through a solar axion halo (purple shaded region). The pseudo-magnetic field associated with a solar axion halo coupling to atomic spins has both a radial component in the $-\hat{\bs{r}}$ direction due to the spatial gradient and a transverse component due to the ``axion  wind'' directed along the axion halo's relative velocity with respect to the lab frame, $\hat{\bs{k}}$.}
\label{Fig:earth-around-Sun-in-halo}
\end{figure}

The global signal pattern due to a solar axion halo that we search for using GNOME can be derived by using the form of $a(\bs{r},t)$ given by Eq.\,\eqref{eq:solar-halo-ALP-field} in Eqs.~\eqref{eq:pseudo-mag-field-linear} and \eqref{eq:pseudo-mag-field-quadratic}.
We note that there are two components of the gradient interaction in the lab frame \cite{banerjee2020searching} (Fig.~\ref{Fig:earth-around-Sun-in-halo}): (1) a radial component from the spatial axion gradient directed toward the Sun's position,
\begin{align}
    \nabla_r a(\bs{r},t) = -\frac{a(\bs{r},t)}{R_\star} \hat{\bs{r}}~,
    \label{eq:radial-gradient}
\end{align}
and (2) a tangential component directed along the axion halo's relative velocity with respect to the lab frame $\bs{v}\ts{rel} = \hbar\bs{k}/m_a$ (an effect referred to as the axion ``wind'' interaction; see, for example, Refs.\,\cite{graham2018spin,stadnik2014axion}),
\begin{align}
    \nabla_\perp a(\bs{r},t) = a_0 \frac{m_a \bs{v}\ts{rel}}{\hbar} \sin\prn{ \omega_a t - \bs{k}\cdot\bs{r} + \theta } e^{ - r / R_\star }~.
    \label{eq:tangential-gradient}
\end{align}

Recent work \cite{del2025enhanced,banerjee2025momentum} has discovered that in the case of QCD axions, quadratic interactions of axions with matter can lead to significant modifications of the axion field's amplitude and gradient near the Earth's surface. Quadratic couplings between axions and terrestrial matter result in wave scattering phenomena related to the modified effective axion mass \cite{hees2018violation,banerjee2023phenomenology}, which generically leads to enhancement of the axion gradient, thereby improving the sensitivity of experiments like GNOME \cite{del2025enhanced,banerjee2025momentum}. In the present work, we assume no such enhancement from a quadratic axion-matter interaction, and leave consideration of this effect for future work.

Since we assume that the axions are in the ground state with $\ell=0$, in the following we assume that the solar axion halo is non-rotating and so is at rest with respect to the Sun.
Consequently, because the relative velocity of the Earth with respect to the Sun is dominated by its orbital motion ($\sim 100$ times faster than the velocity component due to the Earth's rotation about its axis), we can assume in our analysis that all GNOME magnetometers have approximately the same relative velocity with respect to the solar axion halo.
Furthermore, as the relative speed of the Earth with respect to the Sun varies only by $\approx 3\%$ over the year (due to the ellipticity of Earth's orbit), for our purposes we can assume a constant value of $v\ts{rel} \approx 3 \times 10^4~{\rm m/s}$.

\begin{figure}
\center
\includegraphics[width=7.5cm]{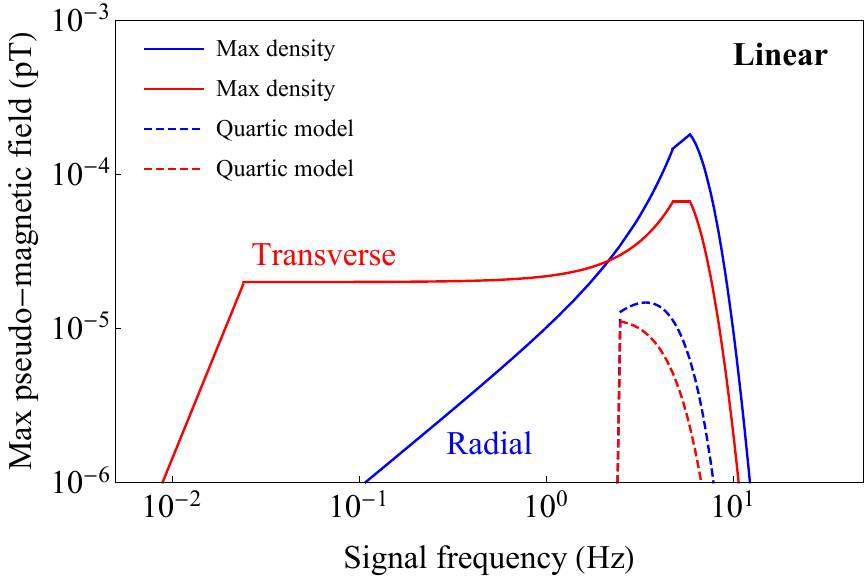}
\includegraphics[width=7.5cm]{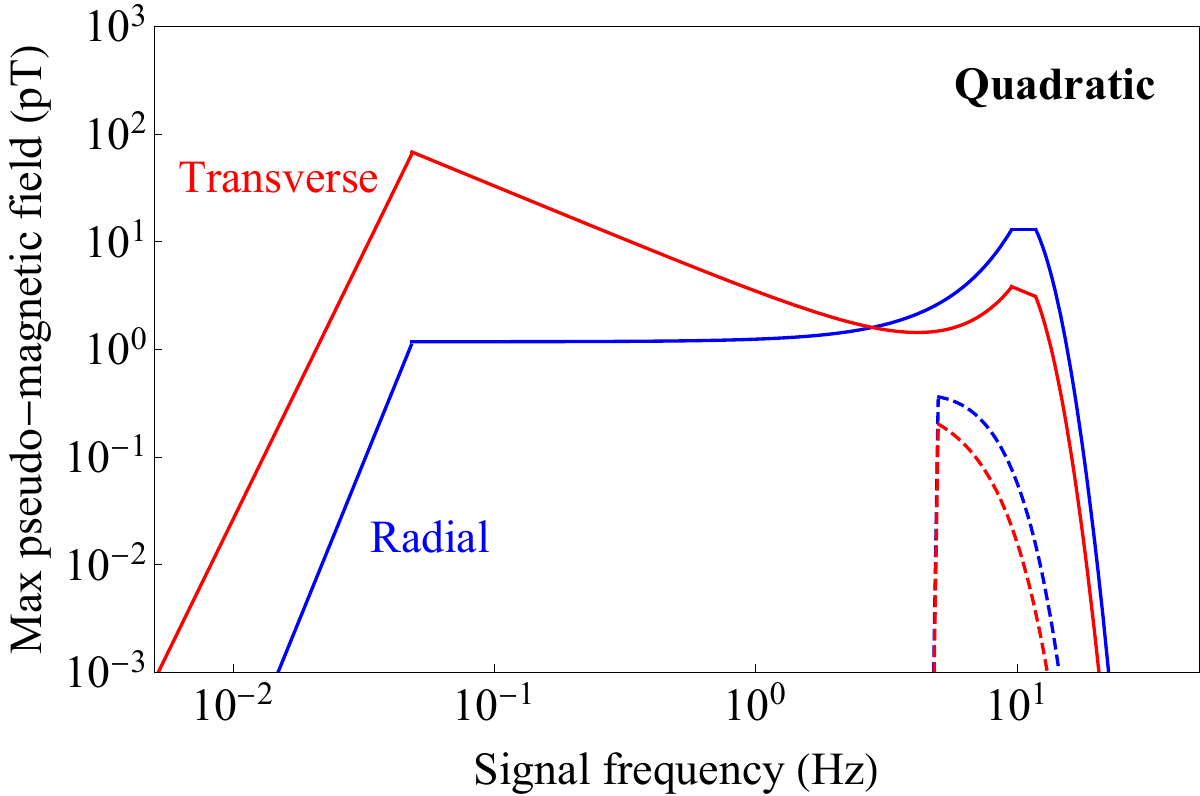}
\caption{The solid lines show the maximum possible pseudo-magnetic field amplitudes from a solar axion halo as a function of signal frequency based on Eqs.~\eqref{eq:Bpl-rad} -- \eqref{eq:Bpq-tan}, the dashed lines show the maximum possible pseudo-magnetic field based on the quartic self-interaction capture model of Ref.\,\cite{budker2023generic}. For the solid lines the solar axion halo density is set to its maximum value $\rho\ts{halo}({\rm{max}})$ and for the dashed lines the solar axion halo density is set to the model value $\rho\ts{halo}({\rm{quartic}})$ as shown in Fig.~\ref{Fig:max-solar-ALP-halo-density}. The upper plot shows the pseudo-magnetic field amplitudes for the radial (blue lines) and transverse (red lines) components of the linear gradient interaction, $\mc{B}_{p,l}^{(r)}$ and $\mc{B}_{p,l}^{(\perp)}$, respectively. The coupling is set to the strongest value consistent with astrophysical limits, corresponding to $f_l \approx 10^8~{\rm{GeV}}$. The lower plot shows the pseudo-magnetic field amplitudes for the radial (blue lines) and transverse (red lines) components of the quadratic gradient interaction, $\mc{B}_{p,q}^{(r)}$ and $\mc{B}_{p,q}^{(\perp)}$, respectively. The coupling is set to the strongest value consistent with astrophysical limits, corresponding to $f_q \approx 10^4~{\rm{GeV}}$. Note the different vertical scales on the two plots: the maximum possible pseudo-magnetic field amplitude for the quadratic interaction is orders of magnitude larger than that for the linear interaction.}
\label{Fig:max-pseudo-magnetic-fields}
\end{figure}

For $m_ac^2 \lesssim 10^{-14}~{\rm eV}$, the tangential component of the axion gradient is equal to or larger than the radial component.
The pseudo-magnetic field components for the linear gradient interactions, derived by combining Eqs.\,\eqref{eq:pseudo-mag-field-linear}, \eqref{eq:avg-halo-density}, \eqref{eq:radial-gradient}, and \eqref{eq:tangential-gradient}, are given by
\begin{widetext}
\begin{align}
    \mc{B}_{p,l}^{(r)} &\approx \prn{ 1.5 \times 10^{-7}~{\rm{pT}} } \cos\prn{ \omega_a t + \theta' } \prn{ \frac{m_a c^2}{ 10^{-14}~{\rm{ eV }} } } \prn{ \frac{10^8~{\rm{GeV}}}{f_l} } \sqrt{ \frac{\rho\ts{halo}(R\ts{ES})}{ 0.4~{\rm{GeV/cm^3}} } }~, \label{eq:Bpl-rad} \\
    \mc{B}_{p,l}^{(\perp)} &\approx \prn{ 1.3 \times 10^{-7}~{\rm{pT}} } \sin\prn{ \omega_a t + \theta' } \prn{ \frac{10^8~{\rm{GeV}}}{f_l} } \sqrt{ \frac{\rho\ts{halo}(R\ts{ES})}{ 0.4~{\rm{GeV/cm^3}} } }~, \label{eq:Bpl-tan}
\end{align}
where $\theta' = \theta - \bs{k}\cdot\bs{r}$ is a slowly varying phase, and the parameters $f_l$, $m_a$, and $\rho\ts{halo}(R\ts{ES})$ are referenced to benchmark values. The pseudo-magnetic field components for the quadratic gradient interactions, derived similarly, are given by
\begin{align}
    \mc{B}_{p,q}^{(r)} &\approx \prn{ 5.0 \times 10^{-5}~{\rm{pT}} } \sbrk{1+\cos\prn{ 2\omega_a t + 2\theta' }} \prn{ \frac{10^4~{\rm{GeV}}}{f_q} }^2 \prn{ \frac{\rho\ts{halo}(R\ts{ES})}{ 0.4~{\rm{GeV/cm^3}} } }~, \label{eq:Bpq-rad} \\
    \mc{B}_{p,q}^{(\perp)} &\approx \prn{ 4.3 \times 10^{-5}~{\rm{pT}} } \sin\prn{ 2\omega_a t + 2\theta' } \prn{ \frac{ 10^{-14}~{\rm{ eV }} }{m_a c^2} } \prn{ \frac{10^4~{\rm{GeV}}}{f_q} }^2 \prn{ \frac{\rho\ts{halo}(R\ts{ES})}{ 0.4~{\rm{GeV/cm^3}} } }~, \label{eq:Bpq-tan}
\end{align}
\end{widetext}
where $f_q$ is referenced to a benchmark value.
As can be seen from Eqs.~\eqref{eq:Bpl-rad} -- \eqref{eq:Bpq-tan} and in Fig.~\ref{Fig:max-pseudo-magnetic-fields}, for different axion masses, corresponding to different axion field frequencies, the radial or tangential components are relatively weaker or stronger. We account for this behavior in our analysis method as discussed in Sec.~\ref{sec:methods}.

In summary, if axions of sufficiently small mass ($m_a \lesssim 10^{-14}~{\rm{eV}}/c^2$) coalesce into a halo around the Sun and the axion field interacts with proton spins through either the Hamiltonian described by Eq.\,\eqref{eq:linear-Hamiltonian} or Eq.\,\eqref{eq:quadratic-Hamiltonian}, then GNOME magnetometers could detect a common pseudo-magnetic field oscillating at $\omega_a$ [Eqs.~\eqref{eq:Bpl-rad} and \eqref{eq:Bpl-tan}] or $2\omega_a$ [Eqs.~\eqref{eq:Bpq-rad} and \eqref{eq:Bpq-tan}] pointing in a definite direction relative to the Sun's position.\footnote{Note that the quartic axion self-interaction invoked for halo formation does not enter the axion-proton gradient coupling directly; it affects our signal only through its role in setting the existence and density profile of the solar-bound configuration. In the mean-field description of Ref.\,\cite{budker2023generic} (using the Gross-Pitaevskii equation with a nonlinear term $\propto |\psi|^2\psi$), self-interactions are perturbative for ``dilute'' halos with density below the critical value where self-interaction and gravitational energies compete, so the bound state remains well approximated by the linear hydrogenic ground state with $\rho(r)\propto e^{-2r/R_\star}$ and the standard relation between field amplitude and energy density used in Eq.\,\eqref{eq:avg-halo-density} applies, and Eqs.\,\eqref{eq:Bpl-rad} -- \eqref{eq:Bpq-tan} are applicable to our considered case. Possible self-interaction-induced anharmonicities (e.g., frequency shifts or higher harmonics) are suppressed in this regime; if instead the present-day halo were in a dense/near-critical phase (e.g., approaching the collapse/Bosenova behavior discussed in Ref.\,\cite{budker2023generic}), a dedicated non-linear signal model would be required and the daily-modulated monochromatic template used here would generally not be optimal.}
Based on this model of a solar axion halo \cite{banerjee2020relaxion,banerjee2020searching,budker2023generic}, we have developed an analysis algorithm based on cross-correlation between data from different GNOME magnetometers in order to search for the corresponding solar axion halo signal pattern.
The data analysis is described in Sec.~\ref{sec:methods} and the results of the search are presented and interpreted in terms of constraints on solar axion halo properties in Sec.~\ref{sec:interpretation}.

\section{Data and Analysis}
\label{sec:methods}

\subsection{Data Overview}
\label{sec:methods:overview}

The data used to search for evidence of a solar axion halo are from the 5th Science Run of GNOME, comprising 69 days of data from 12 different stations.
Science Run 5 started on the 23rd of August 2021 and continued until the 31st of October 2021.
Although none of the 12 GNOME stations were able to acquire continuous data for all 69 days, there were at least 5 stations active at all times throughout Science Run 5.
Science Run 5 achieved a longer total time of continuous data acquisition than prior GNOME Science Runs, better overall sensitivity to exotic physics, and, as described below, incorporated hourly calibration pulses to enhance data reliability.\footnote{Calibration pulses were also employed in Science Run 4, but Science Run 4, like Science Runs 1-3, did not achieve the same level of continuous operation or sensitivity \cite{afach2023WhatCanGNOMEdo}. The data from Science Run 2 were used to search for axion domain walls \cite{afach2021search} and the data from Science Run 4 were used to search for exotic field emission from a black hole merger \cite{khamis2024multi}.}
For these reasons, we chose to focus our analysis solely on Science Run 5 data.

The technical details and characterization data for many of the GNOME magnetometers are described in Ref.\,\cite{afach2018characterization}, and further information about GNOME magnetometers is available in Refs.\,\cite{afach2021search,afach2023WhatCanGNOMEdo}.
Key information concerning the translation of the magnetic fields to the pseudo-magnetic fields generated by axion-spin couplings for the GNOME magnetometers active during Science Run 5 is given in Appendix~\ref{appendix:GNOME-mag-characteristics} and summarized in Table~\ref{table:SR5-mags}.
In brief, the GNOME magnetometers exhibit a combination of (i) broadband noise that is approximately white over a substantial portion of the analysis band, (ii) enhanced low-frequency noise and drift, and (iii) narrow spectral features from technical/environmental interference.
As discussed in Ref.\,\cite{afach2018characterization}, the square-root PSDs of representative GNOME stations are typically ``reasonably flat'' down to $\sim 0.1\,\mathrm{Hz}$, while a rise in noise at lower frequencies primarily reflects slow drifts (e.g., in residual magnetic fields within the shields, temperature-dependent offsets, and other long-timescale technical variations). Discrete monochromatic lines are also commonly observed, most notably at the mains frequencies (50/60\,Hz) and their harmonics; importantly, in several cases such features appear even outside a station's magnetic bandwidth, indicating that they are dominated by electronic pickup/interference rather than true magnetic field signals coupled through the sensor response\,\cite{afach2018characterization}.
At frequencies above the effective magnetometer bandwidth, the recorded spectrum is dominated by non-magnetic electronic contributions.

To enhance the accuracy and reliability of the data, various measures are employed.
A custom automated system with auxiliary sensors (including accelerometers, gyroscopes and unshielded magnetometers) is used at each GNOME station to continuously monitor for any environmental disturbances that could lead to transient excess noise in the data, such as mechanical shocks or magnetic pulses from nearby devices.
The system also keeps track of each station's operational status, such as the magnetometer feedback and/or laser lock error signal, temperature sensor readings, and the photodiode signal that monitors the laser or ambient light power.
Data flagged by the automated system are excluded from the analysis.

Throughout the 69-day Science Run 5, there were occasional changes in experimental parameters that influenced the calibration and bandwidth of the GNOME magnetometers.
These changes arose due to drifts of laser power or frequency, vapor cell temperature, magnetic field gradients within the shields, and so on.
To monitor these drifts and understand how they affected the acquired data, a series of oscillating magnetic fields were periodically applied to each magnetometer station through coils inside the magnetic shields.
The frequency of the applied magnetic field was stepped from 1\,Hz to 180\,Hz over the course of 9 s using a programmable function generator.
During Science Run 5, the pulse sequence was applied every hour.
The response of the magnetometers at the different frequencies provided a convenient way to check the operation of the magnetometers as well as a method to measure the frequency response and bandwidth of the detectors.

The calibration pulses of some of the magnetometers in the network revealed variations in the calibration and/or bandwidth of the detector.
These variations were usually less than 15\% for the GNOME magnetometers with the most significant drift.
To adjust for these changes, the data is rescaled hourly based on the known amplitude of the calibration pulse at 1~Hz.
This method of re-calibration does not take into account the bandwidth changes, but it does account for the calibration variations.
The search for solar axion halos is mainly focused on signals oscillating at frequencies below 10~Hz, and the effect of the bandwidth changes was found to be insignificant in this region.

Our analysis procedure is based on the premise that, if a solar axion halo exists and couples to proton spins, all GNOME magnetometers will measure an oscillating, single-frequency pseudo-magnetic field of the form given by either Eqs.~\eqref{eq:Bpl-rad} and \eqref{eq:Bpl-tan} or Eqs.~\eqref{eq:Bpq-rad} and \eqref{eq:Bpq-tan}, which we represent as:
\begin{align}
    \bs{\mc{B}}_p(t) = \mc{B}_p(t) \hat{\bs{\beta}} = \mc{B}_p^{(r)}(t) \hat{\bs{r}} + \mc{B}_p^{(\perp)}(t) \hat{\bs{k}}~,
    \label{eq:pseudo-magnetic-field-global}
\end{align}
where the unit vector $\hat{\bs{\beta}}$ gives the field direction.
Each GNOME magnetometer $j$ measures the field component along a particular sensitive axis $\hat{\bs{m}}_j$.
Also, as previously noted, the effective field measured by a particular magnetometer must be re-scaled according to Eq.\,\eqref{eq:pseudo-mag-field-in-each-sensor}.
Thus, the signal measured by a particular GNOME magnetometer is described by
\begin{align}
    B_j(t) & = \frac{\sigma_j}{g_{F,j}} \mc{B}_p(t) \hat{\bs{\beta}} \cdot \hat{\bs{m}}_j + B_{n,j}(t)~, \nonumber \\
    & = \frac{\sigma_j}{g_{F,j}}\sbrk{ \mc{B}_p^{(r)}(t) \hat{\bs{r}}\cdot\hat{\bs{m}}_j + \mc{B}_p^{(\perp)}(t) \hat{\bs{k}}\cdot\hat{\bs{m}}_j } + B_{n,j}(t)~, \nonumber \\
    & = B_{p,j}(t) + B_{n,j}(t)\,,
    \label{eq:pseudo-magnetic-field-local}
\end{align}
where $B_{p,j}(t)$ is the apparent magnetic field measured by magnetometer $j$ due to the existence of a pseudo-magnetic field $\bs{\mc{B}}_p(t)$ and $B_{n,j}(t)$ is the background noise measured by magnetometer $j$ in the absence of any signal from a solar axion halo.
The amplitudes of $\mc{B}_p^{(r)}(t)$ and $\mc{B}_p^{(\perp)}(t)$ are relatively constant, so the sought-after signal has a predictable daily modulation due to the time dependence of $\hat{\bs{r}}\cdot\hat{\bs{m}}_j$ and $\hat{\bs{k}}\cdot\hat{\bs{m}}_j$ caused by the rotation of the Earth.

\subsection{Data pre-processing}
\label{sec:methods:pre-process}

\begin{figure}
\center
\includegraphics[width=\columnwidth]{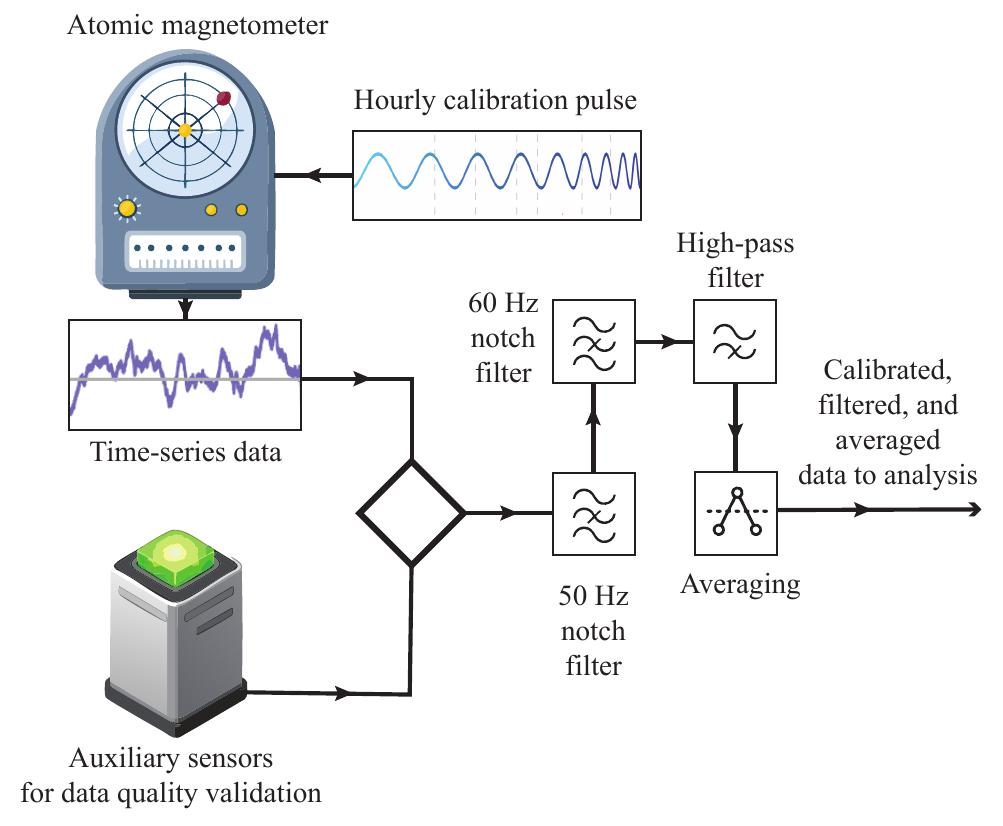}
\caption{Schematic flow chart diagramming key elements of data collection and pre-processing as described in the text.}
\label{Fig:pre-processing-flowchart}
\end{figure}

Before the analysis can begin, the data are pre-processed as shown in the schematic flow chart in Fig.\,\ref{Fig:pre-processing-flowchart}.
The raw time-series data from each GNOME magnetometer are sent through multiple filters: a notch filter at the power line frequency corresponding to the region of the GNOME station [a digital second-order infinite impulse response (IIR) filter at $50$~Hz or $60$~Hz] and a high-pass filter [a linear digital IIR filter] set at $0.0003$~Hz to eliminate slow drifts and reduce low-frequency noise outside the range of frequencies searched.
While the data were originally collected with a sampling rate of 512~Hz, GNOME magnetometers have bandwidths of $\lesssim 100~{\rm Hz}$ and our analysis only studies frequencies up to 32~Hz (well above the axion field frequency for which an enhanced signal is predicted \cite{banerjee2020relaxion,banerjee2020searching,budker2023generic}, see Fig.~\ref{Fig:max-solar-ALP-halo-density}).
Therefore, we bin the data into groups of eight samples and take the average value of the samples in each bin, which lowers the effective sample rate to 64~Hz, giving a corresponding Nyquist frequency for the data of 32~Hz.

\subsection{Solar axion halo signal simulation}
\label{sec:methods:simulation}

As mentioned in Sec.~\ref{sec:methods:overview}, the amplitude of the solar axion halo signal is modulated throughout the day (see also Refs.\,\cite{banerjee2020relaxion,banerjee2020searching}).
The time dependence of the solar axion halo signal depends on the magnetometer location and orientation of the sensitive axis $\hat{\bs{m}}_j$, as well as on the axion Compton frequency $\omega_a$, which affects the relative strength of the radial and tangential components of $\bs{\mc{B}}_p(t)$ (as seen in Fig.~\ref{Fig:max-pseudo-magnetic-fields}).
Our analysis method is based on searching for cross-correlation between signals in different magnetometers as described in the next section.
Because of the different daily modulations of solar axion halo signals measured by different magnetometers, in order to prevent the signals from canceling each other out when combined in a global average, we time-shift the data from different magnetometers before taking the cross-correlation.

\begin{figure}[h!]
\center
\includegraphics[width=\columnwidth]{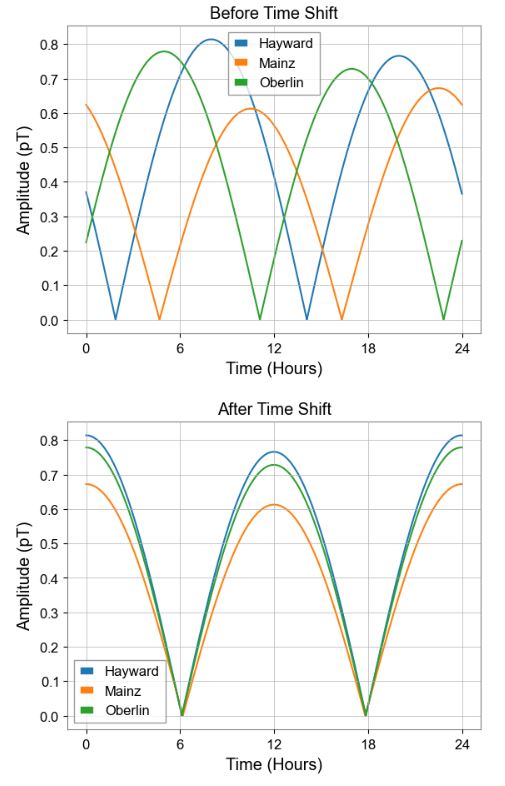}
\caption{The plot on the top shows the predicted signal amplitude from a solar axion halo throughout a day for the Hayward, Oberlin, and Mainz GNOME stations. The simulation is for the 34th day of Science Run 5 (mostly overlapping 26 September 2021 for the majority of GNOME stations) for a signal at $\nu=6$~Hz generated by the linear coupling [Eqs.~\eqref{eq:Bpl-rad} and \eqref{eq:Bpl-tan}, combined with Eq.\,\eqref{eq:pseudo-magnetic-field-global}]. The plot on the bottom shows the signal amplitudes after they have been time-shifted relative to each other to align the times of the largest signal amplitudes.}
\label{Fig:HaywardMainz}
\end{figure}

To determine both (1) the necessary time shift $\Delta t_j$ for magnetometer $j$ and (2) the fitting function for the expected response of GNOME to solar axion signals (discussed in Sec.~\ref{sec:methods:fit-to-daily-modulation}), we carry out a simulation.
With this simulation, we can then time-shift the data to appropriately align the peak signal strengths in each station over the same (time-shifted) periods, as shown in Fig.~\ref{Fig:HaywardMainz}.
For each day of data, the modulation is simulated for each available station using the location of the Sun, the location of the Earth, $\sigma_j/g_{F,j}$ for each magnetometer, and the sensitive direction $\hat{\bs{m}}_j$ of each station.
The simulation is used to generate the expected signals in each station, which are then run through the full analysis to calculate for each day the overall network-averaged signal that would result from a solar axion halo.
Due to the signal's dependence on frequency, the simulation is done over a range of frequencies for each day.

Note that we use a deterministic model incorporating solar ephemerides and GNOME station orientations/sensitivities to compute the time-shift offsets that align the predicted daily modulation across stations and to derive the analytic fitting template discussed in Sec.\,\ref{sec:methods:fit-to-daily-modulation}.
These simulations do not include noise realizations and we did not run end-to-end Monte-Carlo signal-injection/recovery through the full pipeline.
Detection thresholds and upper limits are instead calibrated from the data themselves using the empirical noise statistics together with the distribution of the test statistic described in Sec.\,\ref{sec:interpretation}.

\subsection{Data analysis summary}
\label{sec:methods:analysis-summary}

\begin{figure*}
\center
\includegraphics[width=0.9\textwidth]{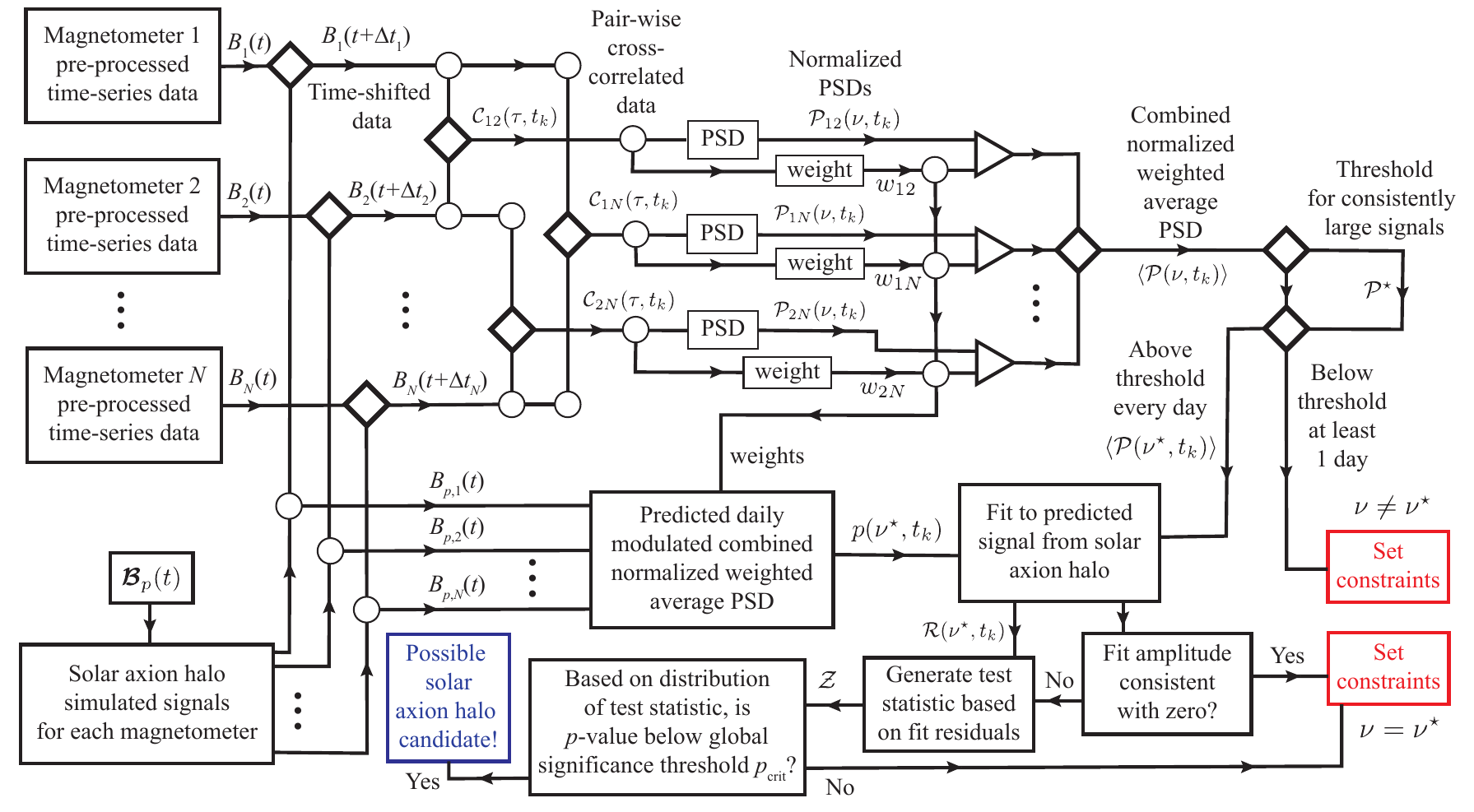}
\caption{Schematic flowchart diagram describing the data analysis procedure.}
\label{Fig:data-analysis-flowchart}
\end{figure*}

Our analysis pipeline, schematically diagrammed in Fig.\,\ref{Fig:data-analysis-flowchart}, is designed to identify a coherent, daily-modulated, single-frequency pseudo-magnetic field pattern consistent with the expected signal from a solar axion halo.
Following the simulation described in Sec.\,\ref{sec:methods:simulation}, which determines the expected temporal modulation of the signal amplitudes for each GNOME station, the pre-processed and time-shifted data are cross-correlated between all station pairs.
This approach enhances sensitivity to global correlated signals while averaging down local, uncorrelated noise.

For each 24-hour data segment, pairwise cross-correlations are computed in 20-minute windows, and the resulting cross-correlation signals are converted to power spectral densities (PSDs).
The PSDs are normalized and combined in a weighted average to generate a network-level signal power spectrum that is effectively ``whitened'' to minimize frequency-dependent systematics.
Candidate frequencies exhibiting consistently elevated PSD values across all days are identified and subjected to a final consistency check: their daily modulation pattern is compared to that predicted for a solar axion halo.
This final fit determines whether the observed modulation is statistically consistent with a true axion signal.
In the absence of a detection, upper limits are placed on the amplitude of the axion-induced pseudo-magnetic field, and corresponding constraints on the properties of the solar axion halo are derived.

\subsection{Pairwise cross-correlation}
\label{sec:methods:cross-correlation-time-shift}

In order to average away uncorrelated noise and thereby enhance detection of correlated signals appearing in all magnetometers, we calculate the cross-correlation between every pair of magnetometers $i,j$ as a function of relative time shift $\tau$.
The analysis is nominally done ``day-by-day,'' but as described in the previous section and the overview, the corresponding 24-hour segments of data from each magnetometer are taken from different time periods that are time-shifted with respect to one another by $\Delta t_{i,j}$ to phase-align the modulation of the expected solar axion halo signals.
The time shifts are calculated so that the solar axion halo signals for all stations have maxima at the same times $t\ts{max}$, respectively, as determined via modeling of simulated solar axion halo signals as discussed in Sec.~\ref{sec:methods:simulation} and shown in Fig.~\ref{Fig:HaywardMainz}.
The data are split into 20 minute intervals throughout each 24-hour segment.
For each interval, every station is cross-correlated with every other station available during that interval.
The cross-correlation centered at times $t_k$ over duration $T$ is given by:
\begin{align}
    \mc{C}_{ij}(\tau,t_k) = \sum_{t=t_k-T/2}^{t_k+T/2} B_i(t+\Delta t_i) B_j(t+\Delta t_j+\tau-T)~,
    \label{eq:cross-correlation}
\end{align}
where the data is appropriately zero-padded when the sum elements extend beyond the specified time range.

Note that in the cross-correlation $\mc{C}_{ij}(\tau,t_k)$ the contribution of a common mode signal scales approximately as the number of sampled points $N_t$ whereas random uncorrelated Gaussian noise contributes to $\mc{C}_{ij}(\tau,t_k)$ as $\sim \sqrt{N_t}$, and thus the signal-to-noise for a correlated signal in the data should improve as $\sim \sqrt{N_t}$ if the background noise is uncorrelated and Gaussian.

\begin{figure}[h!]
\center
\includegraphics[width=\columnwidth]{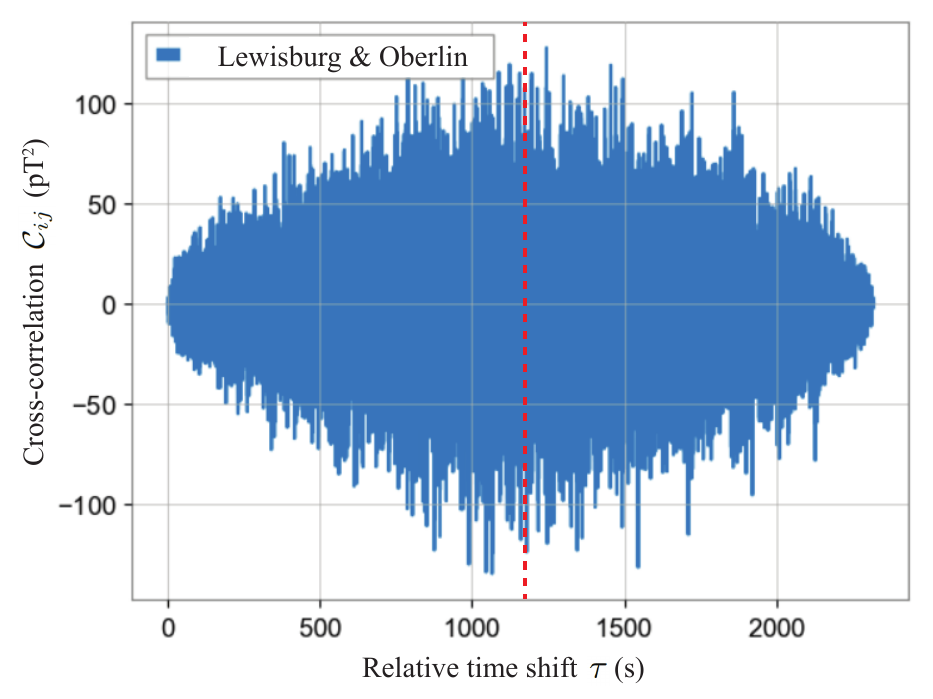}
\caption{Example of the cross-correlation [Eq.\,\eqref{eq:cross-correlation}] between two GNOME stations (Lewisburg and Oberlin) for a 20-minute data segment.
Note that the cross-correlation $\mc{C}_{ij}(\tau,t_k)$ involves the sum over all overlapping data, and as the magnitude of relative time shift $\tau$ increases the number of overlapping data points in the sum decreases, which gives the characteristic ``elliptical'' shape of $\mc{C}_{ij}$ as a function of $\tau$.
The red dashed line marks $\tau = T = 1200$\,s, where there is full overlap between the data segments and a maximum number of elements in the sum described by Eq.\,\eqref{eq:cross-correlation}.}
\label{Fig:CrossC}
\end{figure}
~\


~\

\subsection{Power spectral density}
\label{sec:methods:PSD}

\begin{figure}[h!]
\center
\includegraphics[width=\columnwidth]{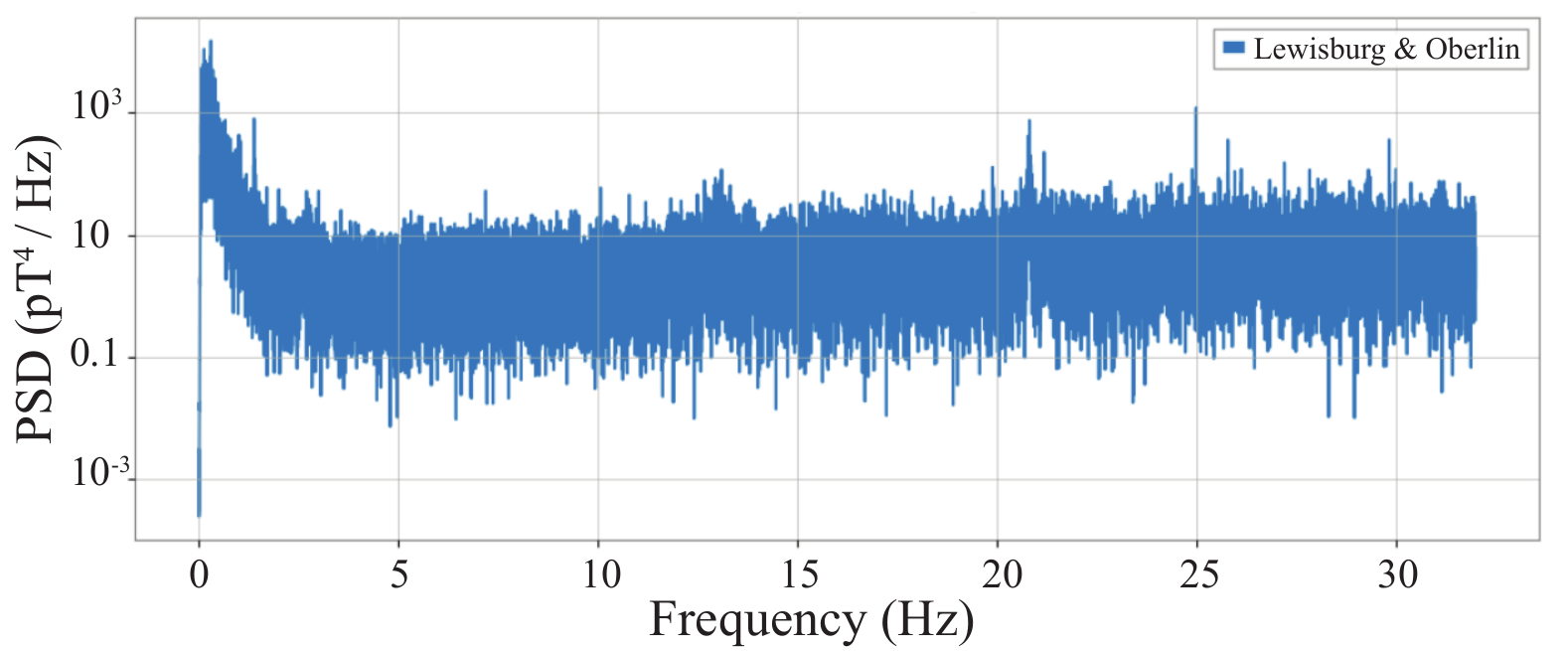}
\caption{Example of the power spectral density (PSD), $P_{ij}(\nu)_{T=20~\rm{min}}$, calculated from the cross-correlation signal shown in Fig.~\ref{Fig:CrossC} for a 20-minute period of data from the Lewisburg and Oberlin GNOME stations.}
\label{Fig:PSDCrossC}
\end{figure}

Next, the PSD $P_{ij}(\nu,t_k)$ of the cross-correlation $\mc{C}_{ij}(\tau,t_k)$ is calculated as a function of frequency $\nu$.
An example is shown in Fig.~\ref{Fig:PSDCrossC}.
The units of the PSD of the cross-correlation are magnetic field to the fourth power over frequency (${\rm{pT^4/Hz}}$), since $\mc{C}_{ij}(\tau,t_k)$ has units of magnetic field squared.
Note that there is a sharp decrease in the PSD values at low frequencies, this is due to the high-pass filter applied in pre-processing as described in Sec.~\ref{sec:methods:pre-process}.
The PSD is calculated for each $T = 20$~minute segment centered at time $t_k$ (so there are 72 discrete times $t_k$ sampled throughout 24 hours), and the 20-minute segments are averaged over the whole day to generate a daily average PSD, $P_{ij}(\nu)_{T=1~\rm{day}} = \sum_{t_k} P_{ij}(\nu,t_k)_{T=20~\rm{min}}/72$.
From these PSDs, the normalized PSD is calculated for each 20~minute segment:
\begin{align}
    \mc{P}_{ij}(\nu,t_k) = \frac{P_{ij}(\nu,t_k)_{T=20~\rm{min}}}{P_{ij}(\nu)_{T=1~\rm{day}}}~.
    \label{eq:normalized-PSD}
\end{align}

\begin{figure}[h!]
\center
\includegraphics[width=\columnwidth]{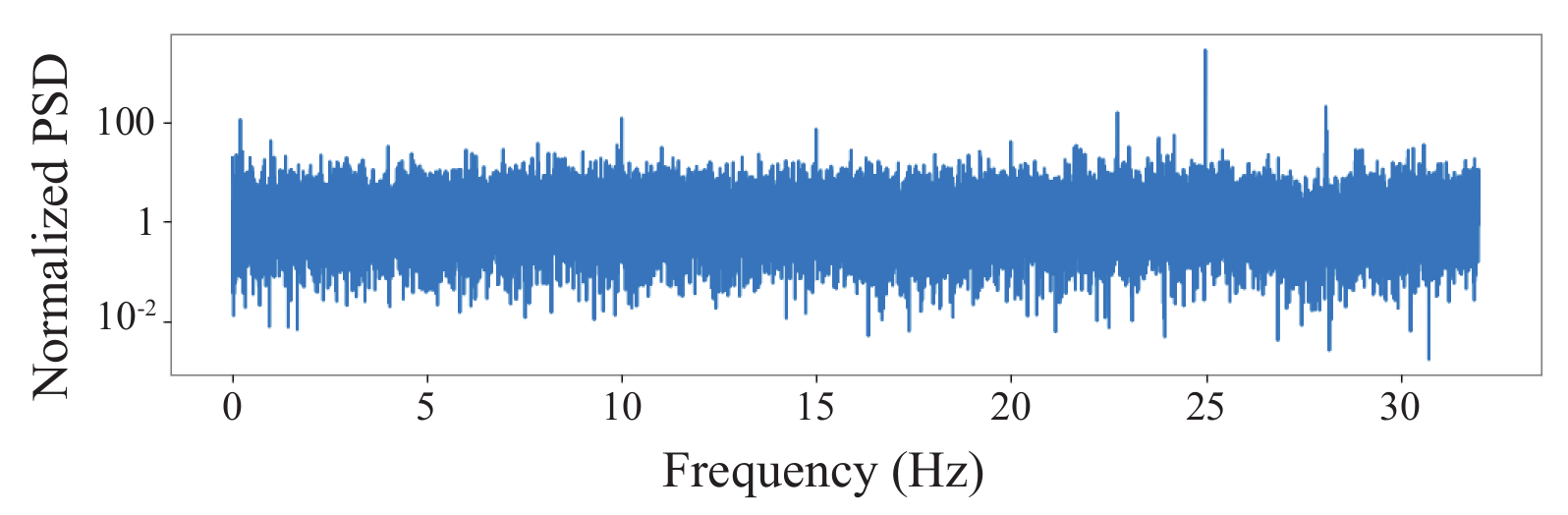}
\caption{Example of a normalized PSD $\mc{P}_{ij}(\nu,t_k)$, as described by Eq.\,\eqref{eq:normalized-PSD}, for a 20-minute data segment from the cross-correlation between the Lewisburg and Oberlin GNOME stations.}
\label{Fig:NormPSD}
\end{figure}

This procedure effectively ``whitens'' the frequency dependence of the normalized PSDs, as demonstrated in the example shown in Fig.~\ref{Fig:NormPSD}.
Because the solar axion halo signals are modulated throughout the day, if the maximum amplitudes of the axion signals at $t\ts{max}$ are sufficiently large, they can generate a $\mc{P}_{ij}(\nu,t_k)$ above the background noise at $t_k=t\ts{max}$ and possibly other times.
Note that this will occur for a particular frequency $\nu$ that matches the Compton frequency of the axion field.

\begin{figure}[h!]
\center
\includegraphics[width=\columnwidth]{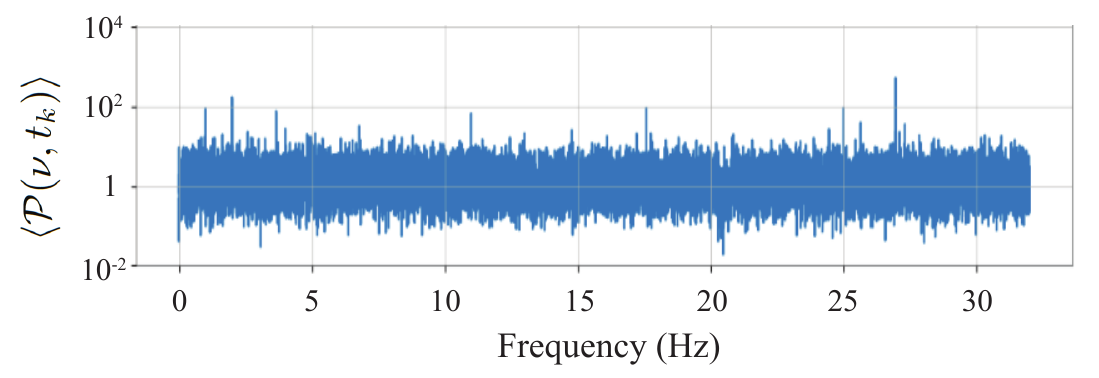}
\caption{Example of the combined normalized weighted PSD $\abrk{ \mc{P}(\nu,t_k) }$, as described by Eq.\,\eqref{eq:combined-weighted-normalized-PSD}, for a 20-minute data segment.}
\label{Fig:WeightedPSD}
\end{figure}

The normalized PSDs for all pairwise cross-correlations are combined in a weighted average:
\begin{align}
    \abrk{ \mc{P}(\nu,t_k) } = \frac{ \sum_{i,j>i} w_{ij} \mc{P}_{ij}(\nu,t_k) }{ \sum_{i,j>i} w_{ij} }~,
    \label{eq:combined-weighted-normalized-PSD}
\end{align}
where the weights $w_{ij}$ are given by the inverse of the square of the standard deviation of $\mc{C}_{ij}(\tau,t\ts{min})$ over $T=20$~minutes, corresponding to the time $t_k=t\ts{min}$ when the contribution from a solar axion halo signal is predicted to be the smallest.
The times $t_k=t\ts{min}$ used to calculate the weights are chosen so that the variance for each cross-correlation is guaranteed to be dominated mainly by noise rather than any possible solar axion halo signal.
Note that different weights $w_{ij}$ are calculated for each day.
An example of a combined normalized weighted average PSD for a 20-minute segment is shown in Fig.~\ref{Fig:WeightedPSD}.

The combined normalized weighted average PSD $\abrk{ \mc{P}(\nu,t_k) }$, incorporating pairwise cross-correlation from all GNOME magnetometers, can then be analyzed to find the frequencies each day where the largest values of $\abrk{ \mc{P}(\nu,t_k) }$ occur.
This is done by setting a threshold $\mc{P}^\star$ and selecting the frequencies $\nu^\star$ for which $\abrk{ \mc{P}(\nu^\star,t_k) } \geq \mc{P}^\star$ for at least one value of $t_k$ in all days analyzed.
The value of $\mc{P}^\star$ is determined so that the top 1\% of powers are passed to the next stage of the algorithm for further analysis.

\subsection{Fitting to daily modulation}
\label{sec:methods:fit-to-daily-modulation}

\begin{figure}[h!]
\center
\includegraphics[width=\columnwidth]{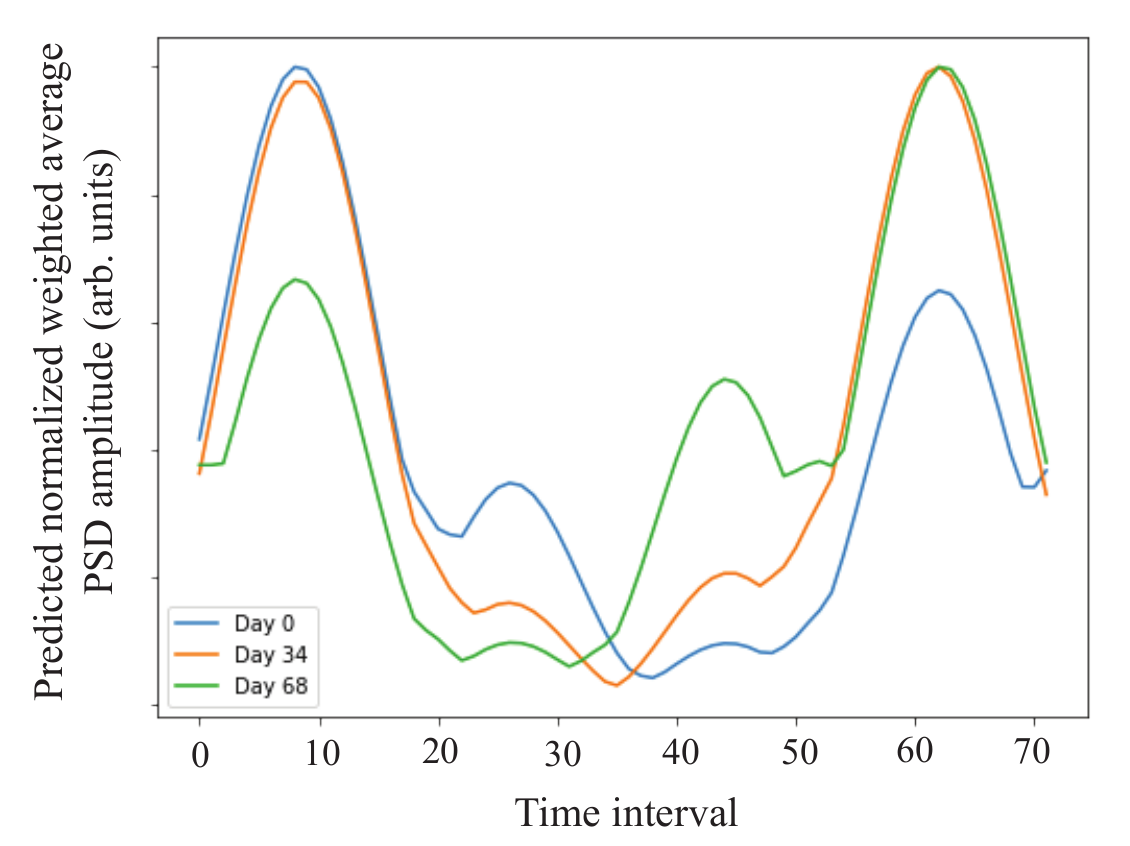}
\caption{Predicted daily modulation of the combined normalized weighted average PSD $p(4~{\rm Hz},t_k)$ from an axion halo signal from a linear gradient coupling [Eqs.~\eqref{eq:Bpl-rad} and \eqref{eq:Bpl-tan}]. Plot shows the results of the simulation of a solar axion signal appearing at frequency $\nu = 4~{\rm Hz}$ for three sample days over the 72 different 20~minute time intervals, illustrating the variation of the form of the daily modulation throughout Science Run 5. The irregular shape of the predicted $\abrk{ \mc{P}(4~{\rm Hz},t_k) }$ is due in part to magnetometers which for various reasons drop out of the average due to technical issues. There is also be significant day-to-day variation in the predicted pattern due to changing weights $w_{ij}$ as well as the changing orientation of stations with respect to the Sun due to the orbit of Earth and the tilt of its axis.}
\label{Fig:simulated-combined-power-daily-mod}
\end{figure}

In the last stage of the analysis, the modulation of $\abrk{ \mc{P}(\nu^\star,t_k) }$ throughout the day as a function of $t_k$ is compared to the predicted daily modulation $p(\nu,t_k)$ based on the modeling described in Sec.~\ref{sec:methods:simulation}.
The predicted axion solar halo signal in each GNOME magnetometer $j$ is assumed to be given by Eq.\,\eqref{eq:pseudo-magnetic-field-local} with the background magnetic field (noise) $B_{n,j}(t) = 0$.
Based on Eq.\,\eqref{eq:cross-correlation}, the predicted cross-correlation between the solar-axion-halo-induced magnetic fields measured by two GNOME magnetometers $i$ and $j$ is
\begin{align}
    \mc{C}^a_{ij}(\tau) = \sum_{t} \sbrk{ \frac{\sigma_i}{g_{F,i}} \mc{B}_p(t) \hat{\bs{\beta}} \cdot \hat{\bs{m}}_i }\sbrk{ \frac{\sigma_j}{g_{F,j}} \mc{B}_p(t+\tau) \hat{\bs{\beta}} \cdot \hat{\bs{m}}_j }~,
    \label{eq:cross-correlation-ALP}
\end{align}
We then take the PSD of $\mc{C}^a_{ij}(\tau)$:
\begin{align}
P^a_{ij}(\nu) = \prn{ \frac{\sigma_i\sigma_j}{g_{F,i}g_{F,j}}  }^2 \prn{\hat{\bs{\beta}} \cdot \hat{\bs{m}}_i}^2 \prn{\hat{\bs{\beta}} \cdot \hat{\bs{m}}_j}^2 \tilde{\mc{B}}_p^4(\nu)~,
\label{eq:PSD-of-ALP-CCs}
\end{align}
where $\tilde{\mc{B}}_p(\nu)$ is the Fourier transform of $\mc{B}_p(t)$ as given by Eqs.~\eqref{eq:Bpl-rad} -- \eqref{eq:Bpq-tan}, respectively.
In our analysis, we are considering 20-minute segments of data, and in this case the factors $\hat{\bs{\beta}} \cdot \hat{\bs{m}}_{i,j}$ vary slowly enough in time that they can be regarded as approximately constant for the purposes of calculating the PSD over each 20-minute time segment.
From Eq.\,\eqref{eq:PSD-of-ALP-CCs}, we can then obtain the associated normalized power according to Eq.\,\eqref{eq:normalized-PSD}:
\begin{align}
    \mc{P}^a_{ij}(\nu,t_k) &= \frac{P^a_{ij}(\nu,t_k)}{P_{ij}(\nu)_{T=1~\rm{day}}}~, \\
    &= \prn{ \frac{\sigma_i\sigma_j}{g_{F,i}g_{F,j}}  }^2 \frac{\tilde{\mc{B}}_p^4(\nu)}{P_{ij}(\nu)_{T=1~\rm{day}}} f_{ij}(t_k)~,
    \label{eq:normalized-PSD-ALP}
\end{align}
where the function $f_{ij}(t_k) = \prn{\hat{\bs{\beta}} \cdot \hat{\bs{m}}_i}^2 \prn{\hat{\bs{\beta}} \cdot \hat{\bs{m}}_j}^2$ describes the time dependence of $\mc{P}^a_{ij}(\nu,t_k)$.
Finally, we calculate the associated combined normalized weighted average PSD following Eq.\,\eqref{eq:combined-weighted-normalized-PSD}:
\begin{align}
    p(\nu,t_k) & = \frac{ \sum_{i,j>i} w_{ij} \mc{P}^a_{ij}(\nu,t_k) }{ \sum_{i,j>i} w_{ij} }~, \label{eq:combined-weighted-normalized-PSD-ALPs-1} \\
    & = \tilde{\mc{B}}_p^4(\nu) \frac{ \sum_{i,j>i} w_{ij} \prn{ \frac{\sigma_i\sigma_j}{g_{F,i}g_{F,j}}  }^2 \frac{f_{ij}(t_k)}{P_{ij}(\nu)_{T=1~\rm{day}}} }{ \sum_{i,j>i} w_{ij} }.
    \label{eq:combined-weighted-normalized-PSD-ALPs-2}
\end{align}

The agreement between the data and the model is evaluated by fitting $\abrk{ \mc{P}(\nu^\star,t_k) }$ to the model $p(\nu^\star,t_k)$ given by Eq.\,\eqref{eq:combined-weighted-normalized-PSD-ALPs-2}, with $\tilde{\mc{B}}_p(\nu)$ as a free parameter.
The quality of the resulting fit $p\ts{fit}(\nu^\star,t_k)$ is used to find whether the data are consistent with a signal generated by a solar axion halo.

\begin{figure}[h!]
\center
\includegraphics[width=\columnwidth]{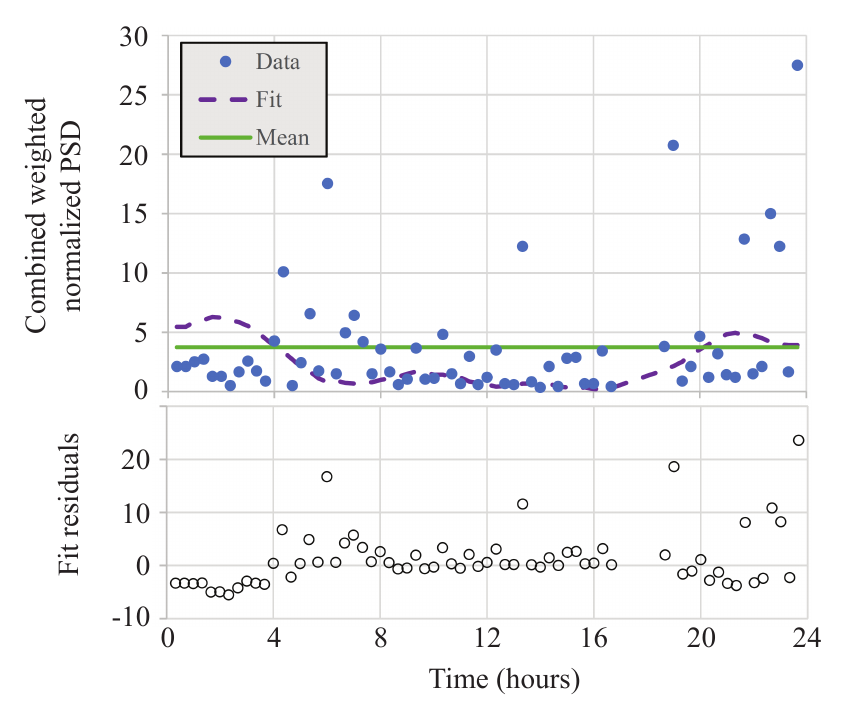}
\caption{Plot of the combined normalized weighted average PSD $\abrk{ \mc{P}(\nu^\star,t_k) }$ (blue dots) and the fitted predicted signal $p\ts{fit}(\nu^\star,t_k)$ (dashed purple line) at a representative frequency for one day of Science Run 5, along with the mean of the data (green line). Lower plot shows the fit residuals $\mc{R}(\nu^\star,t_k)$ (open black circles).}
\label{Fig:example-fit-to-predicted-signal}
\end{figure}

Over the course of the multiday measurement, the predicted pattern of the daily modulation of a solar axion signal changes significantly (Fig.~\ref{Fig:simulated-combined-power-daily-mod}).
Therefore, for simplicity, we evaluate the fit quality over the whole run by examining the fit residuals $\mc{R}(\nu^\star,t_k)$, namely:
\begin{align}
\mc{R}(\nu^\star,t_k) = \abrk{ \mc{P}(\nu^\star,t_k) } - p\ts{fit}(\nu^\star,t_k)~.
\label{eq:fit-residuals}
\end{align}
An example is shown in Fig.~\ref{Fig:example-fit-to-predicted-signal}.

\begin{figure}
\center
\includegraphics[width=\columnwidth]{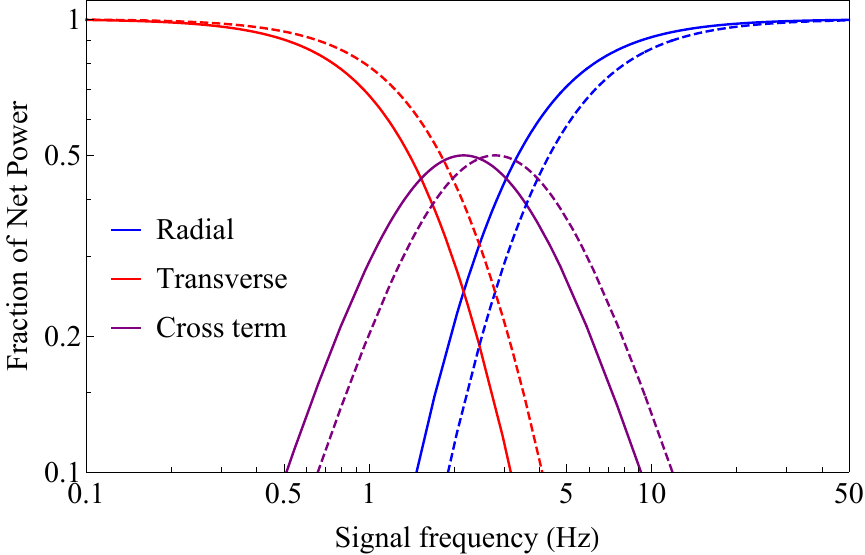}
\caption{The plot shows the fraction of the net power of the cross-correlation signal due to a solar axion halo purely in the form of the radial (blue lines) or transverse (red lines) components, as well as power associated with the cross-terms between the radial and transverse magnetic fields (purple lines). The solid lines represent the results for the linear coupling to proton spins ($\propto \mc{B}_{p,l}^4$) and the dashed lines represent the results for the quadratic coupling to proton spins ($\propto \mc{B}_{p,q}^4$). The fraction of power in the two components is independent of $\rho\ts{halo}$, so the results are independent of the assumed capture model. This illustrates that there are three regimes: (1) for sufficiently low frequencies, $\lesssim 1~{\rm Hz}$, the signal is dominated by the transverse component, (2) for sufficiently high frequencies, $\gtrsim 5~{\rm Hz}$ for the linear coupling and $\gtrsim 7~{\rm Hz}$ for the quadratic coupling, the signal is dominated by the radial component, and (3) for intermediate frequencies, both components contribute. }
\label{Fig:cross-correlation-power}
\end{figure}

For efficiency, the analysis is split into three parts corresponding to (a) a low-frequency range: 0.01--1~Hz; (b) a mid-frequency range: 1--7~Hz, and (c) a high-frequency range: 7--20~Hz.
The reason for adopting this strategy is illustrated in Fig.~\ref{Fig:cross-correlation-power}.
In the low-frequency range, the PSD describing a solar axion halo signal is dominated by the transverse component of the pseudo-magnetic field, so we can assume $\bs{\mc{B}}_p(t) \approx \mc{B}_p^{(\perp)}(t) \hat{\bs{k}}$ for both linear and quadratic interactions.
In the high-frequency range, the solar axion signal PSD is dominated by the radial component, so for both linear and quadratic interactions we can assume $\bs{\mc{B}}_p(t) \approx \mc{B}_p^{(r)}(t) \hat{\bs{r}}$.
However, over most (all) of the mid-frequency range for the linear (quadratic) interaction, the solar axion signal PSD has non-negligible contributions from both radial and transverse components, and also cross terms between these components.\footnote{The reason for cross terms in the measured power is that, although the radial and transverse fields are perpendicular to one another, when they are projected along a particular magnetometer sensitive axis pointing along $\hat{\bs{m}}_j$ that is, in general, at a non-zero angle to both field components, there are contributions from both $\mc{B}_p^{(r)}$ and $\mc{B}_p^{(\perp)}$. See Eq.\,\eqref{eq:pseudo-magnetic-field-local} and surrounding discussion.}
Furthermore, the relative contributions from radial and transverse components change with frequency.
Therefore, we individually model the solar axion signal for each $\nu^\star$ in the mid-frequency range.

It is straightforward to average the fit residuals $\mc{R}(\nu^\star,t_k)$ from day-to-day and determine the point-by-point statistical distribution of the residuals for every value of $t_k$.
Examining the results shows that for our data there is a non-Gaussian distribution of the fit residuals $\mc{R}(\nu^\star,t_k)$, so we turn to a likelihood analysis to see if there is relatively strong evidence of a signal matching what we would expect due to a solar axion halo at one particular frequency matching the axion Compton frequency $\omega_a$ [Eq.\,\eqref{eq:ComptonFrequency}].

We construct a test statistic $\mathcal{Z}$ based on comparing the reduced chi-squared value for the residuals $\mc{R}(\nu^\star,t_k)$ from the two-parameter (amplitude and offset) fit to our solar axion model, $\chi^2({\rm{model}})$, to the reduced chi-squared value for the residuals from a one-parameter fit to a constant value, $\chi^2({\rm{constant}})$:
\begin{align}
    \mathcal{Z} \equiv \frac{\chi^2({\rm{constant}})-1}{\chi^2({\rm{model}})-1}~.
\label{eq:test-statistic}
\end{align}
The idea is that if the fit to the solar axion halo model is significantly better than simply fitting to a constant value, then $\mathcal{Z}$ will be comparably larger (since the denominator will be closer to zero as compared to the numerator).
Even if the residuals do not follow a chi-squared distribution, $\mathcal{Z}$ will still be larger for signals more closely matching that expected from a solar axion halo.

Thus, a signal pattern indicating observation of a solar axion halo coupled to proton spins should possess two key characteristics: (1) the fitted amplitude of the signal pattern should be larger than the uncertainty of the fit and (2) the test statistic $\mathcal{Z}$ should be significantly larger than at other frequencies.
If the amplitude of the fitted signal pattern is smaller than the fit uncertainty, failing test (1), the result is interpreted as an upper limit.

For signal patterns satisfying condition (1), we empirically determine the distribution of the test statistic $\mathcal{Z}$ (for both the linear and quadratic couplings, it is found to match a student's t-distribution).
From the distribution, we determine the cumulative distribution function ${\rm{CDF}}$ and the corresponding local $p$-value at each frequency:
\begin{align}
p = 1 - {\rm{CDF}}\prn{\mathcal{Z}}~.
\end{align}
We consider there to be evidence for a solar axion halo coupled to proton spins if the local $p$-value is below the critical threshold $p\ts{crit}$ defined by
\begin{align}
    \prn{1 - p\ts{crit}}^{N_z} = 0.95~,
    \label{eq:global-p-value-threshold}
\end{align}
where $N_z$ is the number of $\mathcal{Z}$ values from fits satisfying criteria (1) from which the empirical distribution is determined.

\section{Results and Interpretation}
\label{sec:interpretation}

\subsection{Results}
\label{sec:interpretation:results}

We define a threshold $\mc{P}^\star$ for the combined normalized weighted average power.
Specifically, $\mc{P}^\star$ is chosen so that, for 99\% of frequencies, there exists at least one day in GNOME Science Run 5 for which the combined normalized weighted average PSD $\abrk{ \mc{P}(\nu,t_k) }$ [see Eq.\,\eqref{eq:combined-weighted-normalized-PSD} and Fig.~\ref{Fig:WeightedPSD}] never exceeds $\mc{P}^\star$ in any of the 20-minute segments.
Applying this criterion yields $\mc{P}^\star=3.35$.
The value of $\mc{P}^\star$ determines constraints on the properties a solar axion halo for all frequencies $\nu \neq \nu^\star$.

For the identified frequencies $\nu^\star$ where for every day of GNOME Science Run 5 $\abrk{ \mc{P}(\nu,t_k) } \geq \mc{P}^\star=3.35$ for at least one time segment $t_k$, none of the signal patterns passed the two tests identified in Sec.~\ref{sec:methods:fit-to-daily-modulation}: either (1) the fitted amplitude was smaller than the uncertainty of the fit or (2) the reduced $\chi^2$ value of the fit residuals $\mc{R}(\nu^\star,t_k)$ [Eq.\,\eqref{eq:fit-residuals}] indicated the solar axion halo model did not fit the data.
The constraints on the solar axion halo properties for $\nu = \nu^\star$ are then derived from the determined upper limits on the modulated signal amplitude.
This is done based on calculating the upper limit on $\tilde{\mc{B}}_p(\nu)$ using Eq.\,\eqref{eq:combined-weighted-normalized-PSD-ALPs-2}.

\subsection{Constraint on pseudo-magnetic field for frequencies $\nu \neq \nu^\star$}
\label{sec:interpretation:Bp-constraint-below-threshold}

The procedure described in Sec.~\ref{sec:methods:PSD} identifies the frequencies $\nu^\star$ where the normalized average PSD values, $\abrk{ \mc{P}(\nu,t_k) }$, exceed the threshold $\mc{P}^\star=3.35$ for at least one value of $t_k$ for all days analyzed.
This sets a bound on power below which the associated maximum normalized power from a solar axion halo, $\mc{P}_a(t\ts{max})$, would not be detectable.
This is because the signals at the frequencies $\nu \neq \nu^\star$ are not analyzed at the second stage of analysis (discussed in Sec.~\ref{sec:methods:fit-to-daily-modulation}) which checks the daily modulation.
Thus the constraint on the pseudo-magnetic field $\mc{B}_p$ at all frequencies $\nu \neq \nu^\star$ is determined by finding the associated normalized power $\mc{P}_a(t\ts{max})$ from a solar axion halo that would cause the measured signal to exceed $\mc{P}^\star=3.35$ for \emph{at least} one value of $t_k$ every day with 95\% probability.

\begin{figure}
\center
\includegraphics[width=\columnwidth]{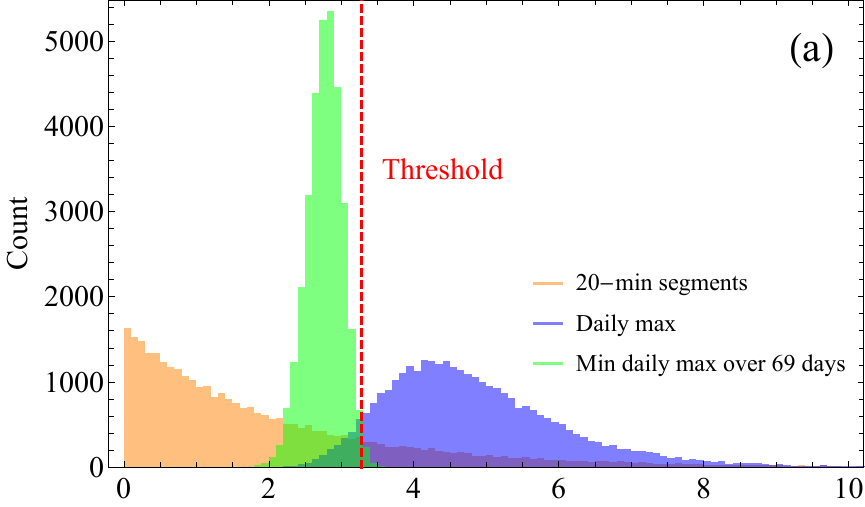}
\includegraphics[width=\columnwidth]{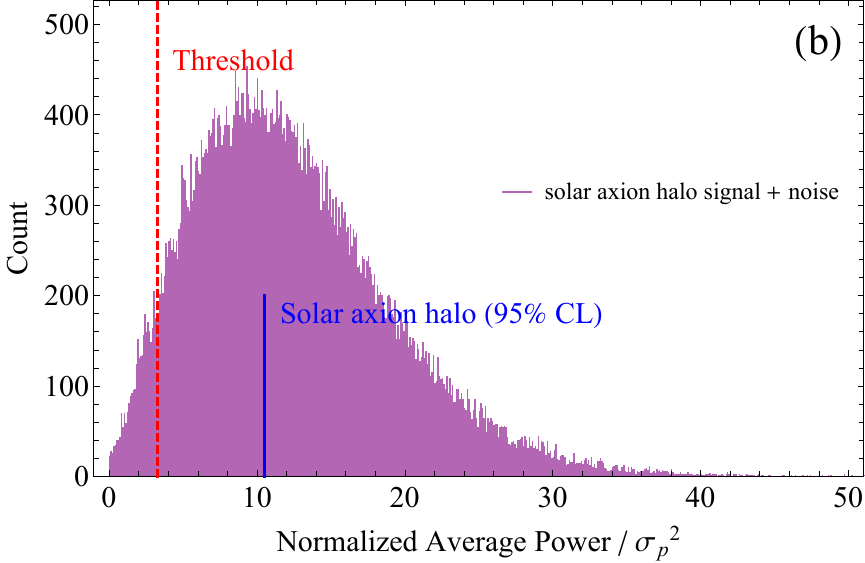}
\caption{Histograms of a Monte Carlo simulation describing the analyzed signals at various stages of analysis. The red dashed lines in both plots mark the power threshold $\mc{P}^\star$. In plot (a) the gold-shaded histogram shows the expected probability distribution of $\abrk{ \mc{P}(\nu,t_k) }$ values for a single 20-minute data block assuming noise-only data that is well-described by a central $\chi^2$ distribution with two degrees of freedom; the blue-shaded histogram shows the largest values of $\abrk{ \mc{P}(\nu,t_k) }$ at each frequency over the course of a single day; the green-shaded histogram shows the minimum largest daily values of $\abrk{ \mc{P}(\nu,t_k) }$ over 69 days. In plot (b) the purple-shaded histogram shows a non-central $\chi^2$ distribution with two degrees of freedom that estimates the probability distribution for solar axion halo signals plus noise such that 95\% of the powers exceed $\mc{P}^\star$.  This enables us to estimate the power in the solar axion halo signal (marked by the blue line) such that it would be identified and passed to the next stage of analysis for fitting of the predicted daily modulation at the 95\% confidence level.}
\label{Fig:ideal-histogram}
\end{figure}

We first illustrate the basic concepts of our analysis using a Monte Carlo simulation.
If there is a signal from a solar axion halo, it is assumed to appear at only one frequency.
Therefore, the distribution of values of $\abrk{ \mc{P}(\nu,t_k) }$ is dominated by noise even in the presence of a solar axion halo signal.
The averaging and normalization procedures discussed in Sec.~\ref{sec:methods:PSD} effectively whiten the PSD data such that the distribution of $\abrk{ \mc{P}(\nu,t_k) }$ values is reasonably well-described by a central $\chi^2$ distribution with two degrees of freedom.
The gold-shaded histogram in Fig.~\ref{Fig:ideal-histogram}(a) shows the results of a noise-only Monte Carlo simulation of the $\abrk{ \mc{P}(\nu,t_k) }$ values for a single 20-minute data block.
This approximately matches the observed distribution of real data; if it is assumed that a solar axion signal appears only at a single frequency its presence would not significantly affect the probability distribution.
The blue-shaded histogram shows the highest values of $\abrk{ \mc{P}(\nu,t_k) }$ at each frequency over the course of a single day, corresponding to 72 different data blocks and thus 72 different values of $t_k$ in Fig.~\ref{Fig:ideal-histogram}.
These highest values of $\abrk{ \mc{P}(\nu,t_k) }$ on a daily basis are checked against the threshold $\mc{P}^\star=3.35$ each day.
Ultimately, to be passed to the next stage of analysis, the largest daily value of $\abrk{ \mc{P}(\nu,t_k) }$ must be $>3.35$ for every day of the 69-day Science Run.

Assuming that the noise characteristics are the same from day-to-day, the green-shaded histogram in Fig.~\ref{Fig:ideal-histogram} shows the results of a Monte Carlo simulation for the {\emph{minimum}} largest daily values of $\abrk{ \mc{P}(\nu,t_k) }$ over the course of a 69-day Science Run.
The dashed red line indicates the value of $\mc{P}^\star$ for which 1\% of the frequencies are passed to the next stage of analysis discussed in Sec.~\ref{sec:methods:fit-to-daily-modulation}.
Note that $\mc{P}^\star$ is lower than the most probable largest value of $\abrk{ \mc{P}(\nu,t_k) }$ over the course of a single day [the blue-shaded histogram in Fig.~\ref{Fig:ideal-histogram}(a)], illustrating that multi-day analysis improves sensitivity.\footnote{Note that $\mc{P}^\star$ cannot be smaller than leftmost tail of the blue-shaded histogram in Fig.\,\ref{Fig:ideal-histogram}, which shows that the search algorithm is ultimately limited by the daily noise of $\abrk{ \mc{P}(\nu,t_k) }$.}

In practice, however, the assumption that the noise characteristics are the same from day-to-day does not hold for the real GNOME data.
In fact, day-to-day noise varies significantly and we find that the actual histogram of the minimum largest daily values of $\abrk{ \mc{P}(\nu,t_k) }$ for Science Run 5 most closely matches the histogram associated with the quietest single day.

We can use our Monte Carlo simulation to estimate the normalized average power in the solar axion halo signal $\mc{P}_a(t\ts{max})$ at the associated frequency $\nu_a$ such that, taking into account noise, $\abrk{ \mc{P}(\nu_a,t\ts{max}) } \geq \mc{P}^\star$ for the quietest day in 95\% of trials.
The distribution of the values of $\abrk{ \mc{P}(\nu_a,t\ts{max}) }$ over many trials is approximately described by a non-central $\chi^2$ distribution with two degrees of freedom, shown by the purple-shaded histogram in Fig.~\ref{Fig:ideal-histogram}(b).
The reason that we use a non-central $\chi^2$ distribution here is that we model the normalized power statistic at a particular frequency $\nu$ as the squared magnitude of an approximately Gaussian Fourier coefficient.
For the noise-only case this yields a central $\chi^2$ distribution with two degrees of freedom as noted above.
A coherent axion signal at frequency $\nu$ adds a deterministic mean to the coefficient, so the statistic follows a non‑central $\chi^2$ distribution where the non-centrality parameter determined by the axion signal.
We use this form to compute the probability of exceeding the selection threshold and, by inversion, the 95\% C.L. upper limits for frequencies that do not pass the first stage of the analysis.
This Monte Carlo simulation shows that if $\mc{P}_a(t\ts{max})$ exceeds $3.2\mc{P}^\star$ [as indicated by the blue line in Fig.~\ref{Fig:ideal-histogram}(b)], it will be passed to the second stage of analysis in 95\% of trials.
Thus we conclude that for all frequencies $\nu \neq \nu^\star$, there is a constraint at the 95\% confidence level that $\mc{P}_a(t\ts{max}) < 3.2 \mc{P}^\star$.

The next step is to translate the constraint on $\mc{P}_a(t\ts{max})$ into a frequency-dependent constraint on $\mc{B}_p(\nu \neq \nu^\star)$.
In order to calculate the constraint, we conservatively assume that at the frequency of interest $\nu$ the entire signal is due to a solar axion halo.
(Since if we assumed that there was an additional magnetic field signal at that frequency, it would imply that $\mc{B}_p$ contributed only part of the observed power and consequently the constraint on $\mc{B}_p$ would be more stringent.)
Based on the above reasoning, we can find the normalized average power due to a pseudo-magnetic field $\mc{B}_p$ from a solar axion halo by following the steps of our analysis described in Sec.~\ref{sec:methods}.

For at least one day, even at the time $t_k = t\ts{max}$ when the predicted solar axion halo signal is at its highest value\footnote{Note that the time-shifting of the data ensures that $f_{ij}(t_k) = \prn{\hat{\bs{\beta}} \cdot \hat{\bs{m}}_i}^2 \prn{\hat{\bs{\beta}} \cdot \hat{\bs{m}}_j}^2$ appearing in Eq.\,\eqref{eq:combined-weighted-normalized-PSD-ALPs-2} reaches its maximum at the same $t\ts{max}$ for all station pairs.}, $\abrk{ \mc{P}^a(\nu,t\ts{max}) } < 3.2 \mc{P}^\star$ with 95\% confidence.
Thus, based on Eq.\,\eqref{eq:combined-weighted-normalized-PSD-ALPs-2}, we obtain an upper limit on $\tilde{\mc{B}}_p^4(\nu)$ given by
\begin{align}
\tilde{\mc{B}}_p^4(\nu \neq \nu^\star) < \frac{ 3.2 \mc{P}^\star \sum_{i,j>i} w_{ij} }{ \sum_{i,j>i} w_{ij} \prn{ \frac{\sigma_i\sigma_j}{g_{F,i}g_{F,j}}  }^2 \frac{f_{ij}(t\ts{max})}{P_{ij}(\nu)_{T=1~\rm{day}}} }~.
\label{eq:limit-on-Bp-for-nu-not-star}
\end{align}
Note that in Eq.\,\eqref{eq:limit-on-Bp-for-nu-not-star}, the constraint on $\tilde{\mc{B}}_p^4(\nu)$ depends on the day since $P_{ij}(\nu)_{T=1~\rm{day}}$ varies from day-to-day.
Therefore, we calculate the constraint on $\tilde{\mc{B}}_p^4(\nu)$ for all days where the combined normalized average power never exceeds $\mc{P}^\star=3.35$ and choose the most stringent limit from among these days.

\begin{figure}[h!]
\center
\includegraphics[width=\columnwidth]{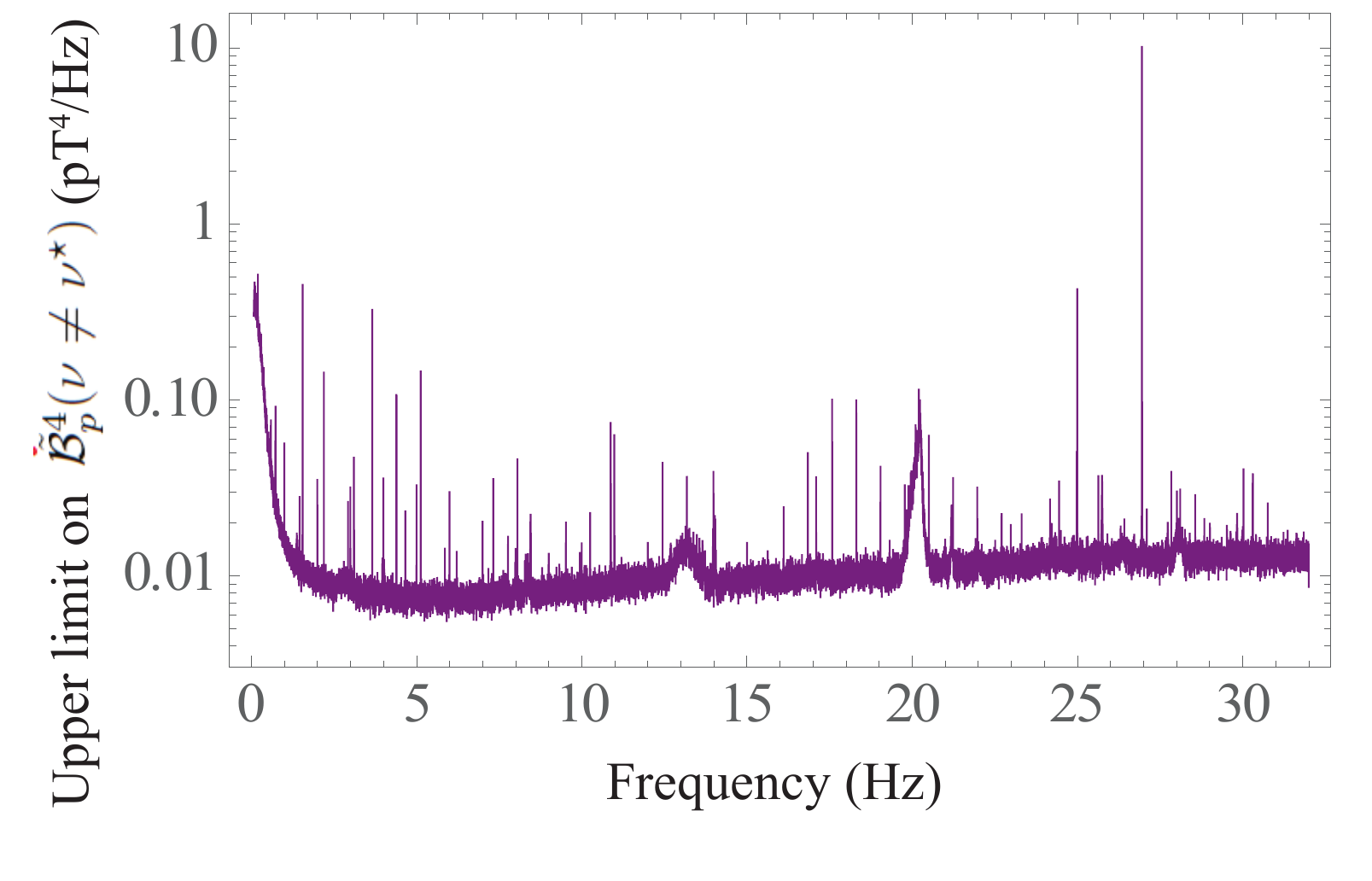}
\caption{Calculated upper limit on $\tilde{\mc{B}}_p^4(\nu)$ at the 95\% statistical confidence level as determined from the GNOME Science Run 5 data set. The data includes the calculated upper limits both for frequencies below and above the threshold $\mc{P}^\star$.}
\label{Fig:upper-limit-on-B4}
\end{figure}

The calculated upper limit on $\tilde{\mc{B}}_p^4(\nu \neq \nu^\star)$ at the 95\% confidence level is used to calculate the bounds on the properties of the solar axion halo for frequencies $\nu \neq \nu^\star$ (Fig.\,\ref{Fig:upper-limit-on-B4}). Due to the high-pass filtering of the data at the pre-processing level (Sec.~\ref{sec:methods:pre-process}), we found that magnetic field noise was abnormally suppressed at low frequencies, and so we conservatively cut off our analysis at a lowest frequency of 0.05~Hz, far above where the effects of the high-pass filter manifest.

\subsection{Constraint on pseudo-magnetic field for frequencies $\nu = \nu^\star$}
\label{sec:interpretation:Bp-constraint-above-threshold}

\begin{figure}[h!]
\center
\includegraphics[width=\columnwidth]{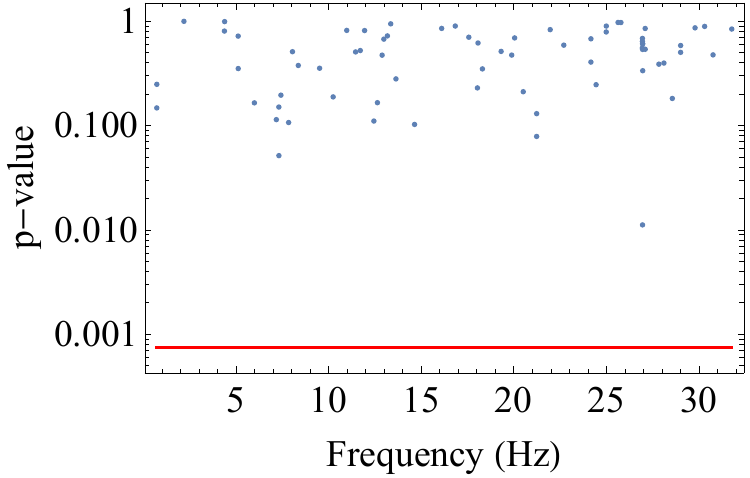} \\
\includegraphics[width=\columnwidth]{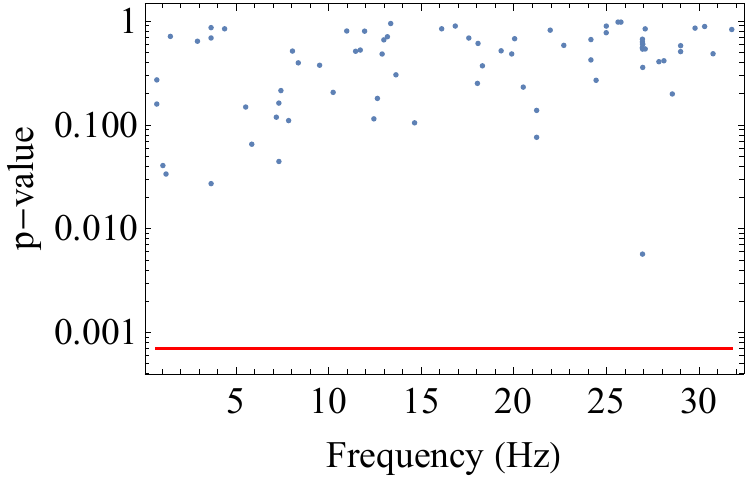}
\caption{The $p$-values for the test statistic $\mathcal{Z}$ for solar axion halo signals at frequencies $\nu^\star$ from the linear (upper plot) and quadratic (lower plot) gradient interactions. The red lines indicate the critical threshold $p\ts{crit}$ for 95\% global significance as defined by Eq.\,\eqref{eq:global-p-value-threshold}. $p$-values below $p\ts{crit}$ would indicate a test statistic unlikely to come from the distribution of test statistics at other frequencies, and would be regarded as solar axion halo signal candidates. However, no observed $p$-values are below $p\ts{crit}$ in either case, and therefore we report upper limits. The plots only show cases where the fitted amplitude of the solar axion halo signal exceeds the amplitude fit uncertainty, meaning that the value is inconsistent with zero amplitude.}
\label{Fig:global-significance}
\end{figure}

The procedure described in Sec.\,\ref{sec:methods:fit-to-daily-modulation} is employed to establish an upper limit on the presence of any signal originating from a solar axion halo for the selected 1\% of the analyzed frequencies $\nu^\star$.
Recall that this selection is based on the condition that the normalized average PSD values, $\abrk{ \mc{P}(\nu^\star,t_k) }$, surpass the threshold $\mc{P}^\star=3.35$ for at least one value of $t_k$ for all days under analysis.
It was determined from the fits to our model that none of the observed signals had amplitudes exceeding the fit uncertainty or, if they did, the reduced $\chi^2$ values of the fit indicated poor agreement with the model.
The conclusion that there is no evidence for a signal from a solar axion halo is based on the fact that, as shown in Fig.\,\ref{Fig:global-significance}, none of the $p$-values for the test statistics $\mathcal{Z}$ (discussed in Sec.\,\ref{sec:methods:fit-to-daily-modulation}) surpass the critical threshold $p\ts{crit}$ for 95\% global significance as defined by Eq.\,\eqref{eq:global-p-value-threshold}.
In the latter case, for frequencies that fail the global p-value test, a non-zero best-fit amplitude is most plausibly due to unmodeled technical noise rather than a true signal.
To reflect this, we conservatively inflate the uncertainties by rescaling all segment errors by a common factor until the fitted amplitude is statistically consistent with zero.
Then, based on the fitted amplitudes and their uncertainty, we assign an upper limit on the normalized average power of a possible solar axion signal at the 95\% confidence level, $\mc{P}\ts{max}(\nu^\star)$ (Fig.\,\ref{Fig:upper-limit-on-B4}).
This upper limit is then used to calculate the constraint on the pseudomagnetic field $\tilde{\mc{B}}_p(\nu^\star)$ using a nearly identical approach to that described in Sec.~\ref{sec:interpretation:Bp-constraint-below-threshold} and summarized in Eq.\,\eqref{eq:limit-on-Bp-for-nu-not-star}:
\begin{align}
\tilde{\mc{B}}_p^4(\nu^\star) < \frac{ \mc{P}\ts{max}(\nu^\star) \sum_{i,j>i} w_{ij} }{ \sum_{i,j>i} w_{ij} \prn{ \frac{\sigma_i\sigma_j}{g_{F,i}g_{F,j}}  }^2 \frac{f_{ij}(t\ts{max})}{P_{ij}(\nu)_{T=1~\rm{day}}} }~.
\label{eq:limit-on-Bp-for-nu-star}
\end{align}

\subsection{Constraints on solar axion halo properties}
\label{sec:interpretation:solar-ALP-halo-constraint}

\begin{figure*}
\center
\includegraphics[width=\columnwidth]{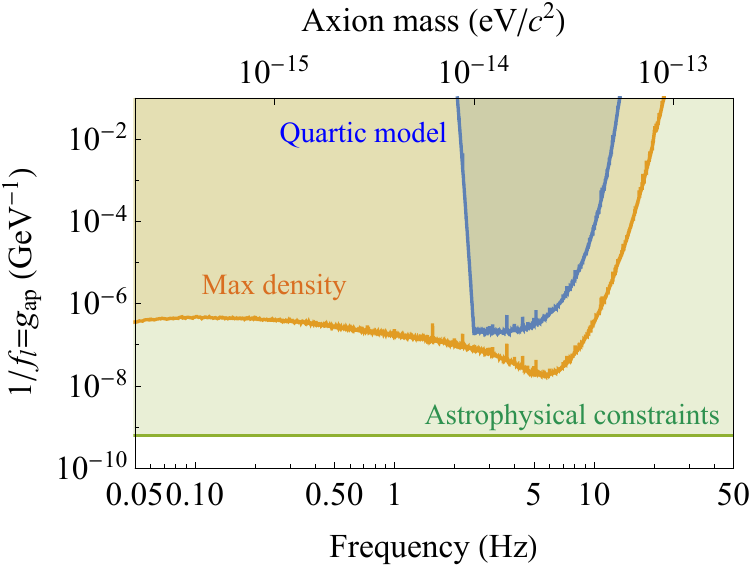}
\includegraphics[width=\columnwidth]{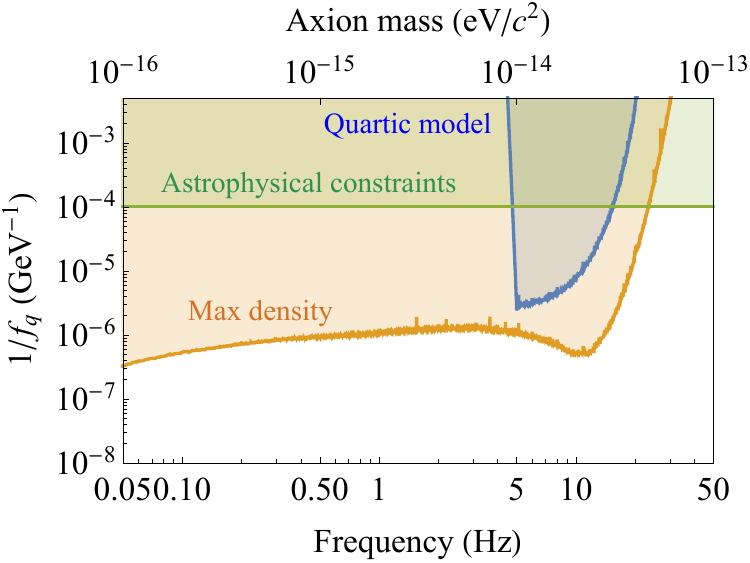}
\caption{Derived constraints on the linear, $1/f_l = g_{ap}$, (left plot) and quadratic, $1/f_q$, (right plot) couplings of axions to proton spins as described by Eqs.~\eqref{eq:linear-Hamiltonian} and \eqref{eq:quadratic-Hamiltonian}. The constraints shown in the plots are, conservatively, 60\% weaker than the 95\% statistical confidence level in order to account for systematic uncertainties in frequency-dependent magnetometer calibration and nuclear theory. The blue line and blue shaded region shows the constraints assuming a solar axion halo formed by the gravitational capture model described in Ref.\,\cite{budker2023generic}. The orange line and orange shaded region shows the constraints assuming the maximum possible solar axion halo overdensity based on limits from observed planetary ephemerides \cite{banerjee2020relaxion,banerjee2020searching}. The green shaded region shows astrophysical constraints derived from observation of supernova 1987a \cite{lella2024getting,Pos13}.}
\label{Fig:constraints}
\end{figure*}

The final step of the analysis is to use the calculated constraints on $\tilde{\mc{B}}_p(\nu)$ for all frequencies [Eqs.~\eqref{eq:limit-on-Bp-for-nu-not-star} and \eqref{eq:limit-on-Bp-for-nu-star}] with the estimated axion overdensities calculated in Refs.\,\cite{banerjee2020relaxion,banerjee2020searching,budker2023generic} (plotted in Fig.~\ref{Fig:max-solar-ALP-halo-density}) to deduce constraints on the axion-proton coupling constants $1/f_l = g_{ap}$ and $1/f_q$ ($g_{ap}$ is the axion-proton coupling constant often referred to in the literature, see, e.g., Refs.\,\cite{graham2015experimental,kimball2022search}).
We do this by using Eq.\,\eqref{eq:pseudo-magnetic-field-global} along with Eqs.~\eqref{eq:Bpl-rad}-\eqref{eq:Bpq-tan} to solve for the constraints on the coupling constants assuming the values of $\rho\ts{halo}(R\ts{ES})$ derived in Refs.\,\cite{banerjee2020relaxion,banerjee2020searching,budker2023generic} (Fig.~\ref{Fig:max-solar-ALP-halo-density}).

The upper limits on $\tilde{\mc{B}}_p(\nu)$ at the 95\% confidence level described in Secs.\,\ref{sec:interpretation:Bp-constraint-below-threshold} and \ref{sec:interpretation:Bp-constraint-above-threshold} are derived from the empirical distribution of the noise–only test statistic and are therefore purely statistical, conditional on our assumed signal model.
The main systematic effects enter as overall multiplicative factors when converting from magnetic field units to a pseudo‑magnetic field $\tilde{\mc{B}}_p(\nu)$, and from $\tilde{\mc{B}}_p(\nu)$ to the axion–proton couplings.\footnote{Note that all the selection and fitting in the analysis is done on normalized cross‑correlation PSDs [Eqs.\,\eqref{eq:normalized-PSD} -- \eqref{eq:combined-weighted-normalized-PSD-ALPs-2}], so a calibration error at a station cancels between numerator and denominator to first order, and so nominally does not affect the analysis until the translation into limits on the axion-proton interaction.}
First, re-scaling of data based on the hourly calibration pulses at 1~Hz, as discussed in Sec.\,\ref{sec:methods:overview}, reduces calibration drifts over all frequencies to $\lesssim 10\%$, so we estimate that the $\tilde{\mc{B}}_p(\nu)$ limits contain at most a $10\%$ correlated calibration uncertainty.
Second, the nuclear‑structure factors $\sigma_p/g$ that relate the pseudo‑magnetic field to the axion/proton‑spin coupling [Eq.\,\eqref{eq:pseudo-mag-field-in-each-sensor}] have theoretical uncertainties at the level of $\sim 20 - 50\%$ as discussed in Ref.\,\cite{Kim15} and summarized in Table\,\ref{table:SR5-mags}.
These nuclear uncertainties propagate directly into a comparable correlated scale uncertainty on the coupling limits $1/f_l$ and $1/f_q$, which should be understood to be in addition to the 95\% statistical confidence level.
Therefore we conservatively weaken the upper limits on $1/f_l$ and $1/f_q$ by 60\% as compared to those derived purely from the 95\% statistical confidence level, taking the worst case scenario where systematic errors in nuclear theory conspire with frequency-dependent magnetometer calibration errors.
The resulting constraints, assuming the existence of a solar axion halo, are shown in Fig.~\ref{Fig:constraints}.
By contrast, the uncertainty in the solar halo density $\rho\ts{halo}(R\ts{ES})$ is \emph{much} larger and is treated by presenting results for two benchmark halo models rather than assigning a single error bar.
Note that in the case of the quadratic coupling, the derived constraints surpass the astrophysical limits by more than two orders of magnitude over a wide range of frequencies.

Not shown in Fig.\,\ref{Fig:constraints} are constraints from the large number of dark matter haloscope experiments with spin-based sensors \cite{cong2025spin,jackson2023probing,jiang2024searches} that have been performed in recent years \cite{gavilan2025searching,jiang2021search,jiang2024long,abel2017search,aybas2021search,xu2024constraining,garcon2019constraints,walter2025search,abel2023search,bloch2023constraints,wei2025dark,bloch2022new,wu2019search,crescini2020axion}.\footnote{Although, it should be noted that most of these experiments probe couplings of axion dark matter to neutrons, which, depending on the axion or ALP model, can have vastly different coupling strengths as compared to protons as probed in the GNOME experiment.}
We expect that the data from many of these experiments could be reanalyzed to either set even more stringent constraints, or, perhaps, discover evidence pointing to the existence of a solar axion halo.

\section{Conclusion and outlook}

In this work, we have carried out the first dedicated search for a gravitationally bound solar axion Bose–Einstein condensate halo using the fifth science run of the Global Network of Optical Magnetometers for Exotic physics searches (GNOME).
We developed a detailed model for pseudo-magnetic fields generated by axion-field-gradient coupling to proton spins, both linear and quadratic, due to a solar halo, including the daily modulation pattern expected in a global magnetometer network.
The analysis pipeline combined time-shifted cross-correlations between geographically distributed stations with frequency-domain selection and modulation-pattern fitting to identify possible candidates.

Our analysis of 69 days of data from 12 GNOME stations revealed no statistically significant signatures consistent with a solar axion halo. We therefore set 95\% confidence-level upper limits on the amplitude of axion-induced pseudo-magnetic fields over the frequency range 0.05–20 Hz. These limits were translated into constraints on the axion–proton couplings for two benchmark halo densities: the maximum overdensity allowed by planetary-ephemeris constraints \cite{banerjee2020relaxion,banerjee2020searching} and the densities predicted by a quartic self-interaction gravitational-capture model \cite{budker2023generic}.
For quadratic couplings, the resulting limits exceed existing astrophysical bounds by more than two orders of magnitude across much of the accessible frequency range.

Looking forward, GNOME is entering a new phase with the deployment of alkali–noble-gas comagnetometers in the Advanced GNOME network \cite{afach2023WhatCanGNOMEdo}.
These self-compensating sensors offer enhanced sensitivity to exotic spin couplings while suppressing magnetic-field noise, and can probe axion interactions with both proton and neutron spins (and, in certain scenarios, electron spins) with sensitivities orders of magnitude beyond that achieved with the GNOME magnetometers used in the study described here.
The first dark-matter search with a two-station comagnetometer interferometer has already demonstrated this capability \cite{gavilan2025searching}, and a global array of such sensors will greatly extend the reach of future searches for solar axion halos, virialized dark-matter fields, and transient exotic-field events.

Beyond the solar-halo search presented here, GNOME continues to pursue a broad program of exotic-physics searches \cite{afach2023WhatCanGNOMEdo}, including tests for topological-defect dark matter such as axion domain walls \cite{Pos13,masia2020analysis,afach2021search,kim2022machine}, compact composite objects like axion stars and Q-balls \cite{kimball2018searching}, stochastic ultralight-boson fields \cite{masia2023intensity}, and bursts of exotic radiation from astrophysical events \cite{dailey2021quantum,khamis2024multi,eby2022probing,arakawa2025multimessenger}.
Another direction to explore is the theoretical possibility of solar halos formed by spin-1 bosonic dark matter such as hidden photons.

\acknowledgments

The authors are sincerely grateful to Yuzhe Zhang and Volodymyr Takhistov for useful discussions.

This work was supported by the U.S. National Science Foundation under grants PHYS-2510625, PHYS-2510627, PHYS-2110388, PHYS-2110370, and PHYS-1707803, the German Research Foundation (DFG) under grant no.\,439720477 (RioGNOME), the Cluster of Excellence ``Precision Physics, Fundamental Interactions, and Structure of Matter'' (PRISMA++ EXC 2118/2) funded by DFG within the German Excellence Strategy (Project ID 390831469), the COST Action within the project COSMIC WISPers (Grant No. CA21106), the Australian Research Council (ARC) DP240100534, the Israeli Science Foundation, Minerva, the European Research Council (ERC), and the National Natural Science Foundation of China (Grant No. 62301377).
Z. D. G. acknowledges institutional funding provided by the Institute of Physics Belgrade through a grant by the Ministry of Science, Technological Development and Innovations of the Republic of Serbia.
The work of S. P. is supported by the National Science Centre of Poland within the Opus program (Grant No. 2019/34/E/ST2/00440) and G. L. acknowledges the support of the Excellence Initiative - Research University of the Jagiellonian University in Krak\'ow.
D.F.J.K. also acknowledges support from a generous donation by Rosemary Rodd Spitzer.

The data that support the findings of this article are openly available \cite{GNOMEwebsite}.

\newpage

\appendix

\section{Translating GNOME magnetometer data to pseudo-magnetic field characteristics}
\label{appendix:GNOME-mag-characteristics}

GNOME magnetometers are calibrated to report data in magnetic field units (pT).
However, as noted in the discussion surrounding Eq.\,\eqref{eq:pseudo-mag-field-in-each-sensor}, the coupling of an atomic spin to an axion field is not proportional to the gyromagnetic ratio.
Thus to interpret the effect an axion field $a\prn{ \bs{r},t }$ on a magnetometer reading, the GNOME data must be recalibrated to account for the details of the setup and the atomic/nuclear structure of the probed atoms.

Because all GNOME magnetometers have multilayer passive shielding systems built of mu-metal or ferrite \cite{afach2018characterization}, their sensitivity to axion interactions with electron spins is highly suppressed as discussed in Ref.\,\cite{kimball2016magnetic}.
Thus, GNOME magnetometers are primarily sensitivity to axion interactions with nuclear spins.
For GNOME Science Runs 1-5, all GNOME magnetometers use alkali atoms.
According to the shell model, the nuclei of alkali atoms have valence protons and neutron spins are paired off, suppressing neutron spin polarization in alkali nuclei (see discussion in, e.g., Ref.\,\cite{Kim15}).
Therefore, GNOME magnetometers are predominantly sensitive to axion interactions with proton spins.

In the present search for correlated signals from a solar axion halo using GNOME, we search for either a linear coupling of the axion field gradient $\bs{\nabla}a(\bs{r},t)$ to spins $\bs{S}$ or a quadratic coupling of the gradient of the axion field intensity $\bs{\nabla}a^2(\bs{r},t)$ to spins as described by Eqs.~\eqref{eq:linear-Hamiltonian} and \eqref{eq:quadratic-Hamiltonian} [and also Eqs.\,\eqref{eq:pseudo-mag-field-linear} and \eqref{eq:pseudo-mag-field-quadratic}].
In either case, based on analogy with the Zeeman effect, the pattern of signal amplitudes due to an axion field measured by GNOME can be characterized by a pseudo-magnetic field $B_j$ measured by magnetometer $j$ as discussed in Ref.\,\cite{afach2021search} and presented in Eq.\,\eqref{eq:pseudo-mag-field-in-each-sensor}.
The ratio between the effective proton spin and the Land\'e g-factor ($\sigma_{j}/g_{F,j}$) calibrates for the proton-spin coupling in the respective magnetometer.
This ratio depends on the atomic and nuclear structure as well as details of the magnetometry scheme, as described in Appendix B of Ref.\,\cite{afach2021search}.
Estimates of $\sigma_p/g$ for all the GNOME magnetometers active during Science Run 2 are described in Ref.\,\cite{afach2021search}.

Between the time of Science Run 2 and Science Run 5, several new stations joined GNOME and several stations modified or upgraded their magnetometers as described in the next subsections.

\subsection{New and modified GNOME magnetometers in Science Run 4}
\label{Sec:new-mags-SR4}

In Science Run 4 several new magnetometers joined GNOME (Los Angeles, Moxa, and Oberlin) and one magnetometer (Hayward) changed its scheme and species.
Estimates for $\sigma_p/g$ for the new GNOME magnetometers are discussed below.

\subsubsection{Hayward}
\label{app:subsec:Hayward-mag-SR4}

Between Science Runs 3 and 4 Hayward switched to a $^{87}$Rb SERF magnetometer (QuSpin Zero-Field Magnetometer, QZFM \cite{osborne2018fully}).
The QZFM operates with a relatively high spin polarization $P \approx 0.5$, a regime discussed in Refs.\,\cite{savukov2005effects,appelt1998theory}.
The QZFM is similar to the Lewisburg magnetometer discussed in Ref.\,\cite{afach2021search}.
For a nucleus with $I=3/2$, the effective Land\'e $g$-factor is given by
\begin{align}
    \abrk{g}\ts{hf} = g_s \frac{1+P^2}{6+2P^2}~,
    \label{eq:g-HiPol}
\end{align}
and the effective proton spin polarization is given by
\begin{align}
    \abrk{\sigma_{p}}\ts{hf} = \sigma_p \frac{5+P^2}{6+2P^2}~.
    \label{eq:p-spin-HiPol}
\end{align}
Based on Eqs.~\eqref{eq:g-HiPol} and \eqref{eq:p-spin-HiPol}, we find that for the Hayward QZFM that $\abrk{\sigma_{p}}\ts{hf}/\abrk{g}\ts{hf} = 0.70^{+0.00}_{-0.15}$.

\subsubsection{Los Angeles}

Los Angeles operates an rf-driven $^{85}$Rb magnetometer probing the $F=2$ ground state hyperfine level.
A reasonable estimate for $\sigma_p/g$ in this case can be derived based on the single-particle Schmidt model for nuclear spin \cite{schmidt1937magnetischen} and the Russell-Saunders scheme for atomic states (see, for example, Ref.\,\cite{budker2008atomic}), and is given in Table~II of Ref.\,\cite{afach2021search}.

\subsubsection{Moxa}

Moxa operates an rf-driven $^{133}$Cs magnetometer probing the $F=4$ ground state hyperfine level (similar to Fribourg's station from Science Run 2).
Again, $\sigma_p/g$ is given in Table~II from Ref.\,\cite{afach2021search}.

\subsubsection{Beersheba and Oberlin}

The Beersheba and Oberlin stations operate SERF magnetometers with a mixture of alkalis in the low-spin-polarization mode.
The vapor cells contain $^{39}$K, $^{85}$Rb, and $^{87}$Rb atoms: with 90\% K and 10\% Rb. Spin-exchange collisions average over both ground-state hyperfine levels of all three species.
Taking into account the relative abundances of the different atomic species ($\approx 90\%$~$^{39}$K, $\approx 7.2\%$~$^{85}$Rb, $\approx 2.8\%$~$^{87}$Rb), we find that $\abrk{g}\ts{hf} \approx 0.309$, $\abrk{\sigma_{p}}\ts{hf} = -0.15^{+0.05}_{-0.00}$, and thus $\abrk{\sigma_{p}}\ts{hf}/\abrk{g}\ts{hf} = -0.49^{+0.18}_{-0.00}$ for the Oberlin magnetometer.

\subsection{New and modified GNOME magnetometers in Science Run 5}
\label{Sec:new-mags-SR5}

In Science Run 5 there are yet again new magnetometers (Belgrade and Canberra) and two magnetometers (Krakow and Mainz) changed their scheme and species.
Estimates for $\sigma_p/g$ for the new GNOME magnetometers are discussed below and the results are summarized in Table~\ref{table:SR5-mags}.

\subsubsection{Belgrade}

Belgrade operates an rf-driven $^{133}$Cs magnetometer probing the $F=4$ ground-state hyperfine level (similar to Fribourg's station from Science Run 2 and Moxa's station).

\subsubsection{Canberra}

Canberra operates a $^{87}$Rb SERF magnetometer in the high-polarization scheme. The magnetometer is similar to that operated by Lewisburg.

\subsubsection{Krakow and Mainz}

Between Science Runs 4 and 5, Krakow and Mainz switched to commerical $^{87}$Rb SERF magnetometers (QuSpin Zero-Field Magnetometer, QZFM \cite{osborne2018fully}), the same models as the one operated by Hayward and described in Sec.~\ref{app:subsec:Hayward-mag-SR4}.

\begin{table*}
\caption{Characteristics of the magnetometers active during Science Run~5. The station name, location in longitude and latitude, orientation of the sensitive axis, type of magnetometer (NMOR~\cite{budker2002nonlinear, kimball2009magnetometric, gawlik2006nonlinear}, rf-driven~\cite{afach2018characterization}, or SERF~\cite{allred2002high}), and probed transition are listed. The rightmost column lists the estimated ratio between the effective proton spin polarization and the Land\'e $g$-factor for the magnetometer, $\sigma_p/g$, which depends on the atomic species and the magnetometry scheme as described in Appendix~B of Ref.\,\cite{afach2021search}. The $\sigma_p/g$ value is used to relate the measured magnetic field to the signal expected from the interaction of an axion field with proton spins. The uncertainty indicated describes the range of values from different theoretical calculations~\cite{Kim15}. Stations annotated with superscript $\star$ symbols denote those with adequate data quality and operational time percentage during Science Run 5, rendering them suitable for the solar axion halo search.}
\begin{tabular}{ l D{.}{.}{8} D{.}{.}{7} D{.}{.}{0} D{.}{.}{0} l l r}
\hline
\hline
          &                  \multicolumn{2}{c}{Location}                 &           \multicolumn{2}{c}{Orientation}           & \multicolumn{1}{l}{~} & \multicolumn{1}{l}{~} \\
Station   &  \multicolumn{1}{l}{Longitude} &  \multicolumn{1}{l}{Latitude} &   \multicolumn{1}{l}{Az} & \multicolumn{1}{l}{Alt} & \multicolumn{1}{l}{Type} & \multicolumn{1}{l}{Probed transition} & \multicolumn{1}{l}{~~~~$\sigma_p/g$}\\
\hline
\rule{0ex}{3.6ex} Beersheba &  34.8043\textrm{\textdegree~E} & 31.2612\textrm{\textdegree~N} &    0\textrm{\textdegree} & 90\textrm{\textdegree} & \textrm{SERF} & \textrm{$^{39}$K~D1} & $-0.49^{+0.18}_{-0.00}$\\
\rule{0ex}{3.6ex} Beijing & 116.1868\textrm{\textdegree~E} & 40.2457\textrm{\textdegree~N} &    251\textrm{\textdegree} & 0\textrm{\textdegree} & \textrm{NMOR} & \textrm{$^{133}$Cs~D2~} F=4 & $-0.39^{+0.19}_{-0.00}$\\
\rule{0ex}{3.6ex} Belgrade &  20.3928\textrm{\textdegree~E} & 44.8546\textrm{\textdegree~N} &    300\textrm{\textdegree} & 0\textrm{\textdegree} & \textrm{rf-driven} & \textrm{$^{133}$Cs~D1~} F=4 & $-0.39^{+0.19}_{-0.00}$\\
\rule{0ex}{3.6ex} Berkeley 2 & 122.2570\textrm{\textdegree~W} & 37.8723\textrm{\textdegree~N} &    0\textrm{\textdegree} & +90\textrm{\textdegree} & \textrm{SERF} & \textrm{$^{87}$Rb~D1~} & $~0.70^{+0.00}_{-0.15}$\\
\rule{0ex}{3.6ex} Canberra & 149.1185\textrm{\textdegree~E} & 35.2745\textrm{\textdegree~S} &    0\textrm{\textdegree} & +90\textrm{\textdegree} & \textrm{SERF} & \textrm{$^{87}$Rb~D1~} & $~0.70^{+0.00}_{-0.15}$\\
\rule{0ex}{3.6ex} Daejeon   & 127.3987\textrm{\textdegree~E} & 36.3909\textrm{\textdegree~N} &    0\textrm{\textdegree} & +90\textrm{\textdegree} & \textrm{NMOR} & \textrm{$^{133}$Cs~D2~} F=4 & $-0.39^{+0.19}_{-0.00}$\\
\rule{0ex}{3.6ex} Hayward$^\star$   & 122.0539\textrm{\textdegree~W} & 37.6564\textrm{\textdegree~N} &    0\textrm{\textdegree} & +90\textrm{\textdegree} & \textrm{SERF} & \textrm{$^{87}$Rb~D1~} & $~0.70^{+0.00}_{-0.15}$\\
\rule{0ex}{3.6ex} Krakow$^\star$    &  19.9048\textrm{\textdegree~E} & 50.0289\textrm{\textdegree~N} &  0\textrm{\textdegree} & +90\textrm{\textdegree} & \textrm{SERF} & \textrm{$^{87}$Rb~D1~} & $~0.70^{+0.00}_{-0.15}$\\
\rule{0ex}{3.6ex} Lewisburg$^\star$ &  76.8825\textrm{\textdegree~W} & 40.9557\textrm{\textdegree~N} &    0\textrm{\textdegree} & +90\textrm{\textdegree} & \textrm{SERF} & \textrm{$^{87}$Rb~D2} & $~0.70^{+0.00}_{-0.15}$\\
\rule{0ex}{3.6ex} Los Angeles$^\star$ &  118.4407\textrm{\textdegree~W} & 34.0705\textrm{\textdegree~N} &    270\textrm{\textdegree} & 0\textrm{\textdegree} & \textrm{rf-driven} & \textrm{$^{85}$Rb~D2} F=2 & $~0.50^{+0.00}_{-0.07}$\\
\rule{0ex}{3.6ex} Mainz$^\star$     &   8.2354\textrm{\textdegree~E} & 49.9915\textrm{\textdegree~N} &    0\textrm{\textdegree} & -90\textrm{\textdegree} & \textrm{SERF} & \textrm{$^{87}$Rb~D1~} & $~0.70^{+0.00}_{-0.15}$\\
\rule{0ex}{3.6ex} Moxa$^\star$     &   11.6147\textrm{\textdegree~E} & 50.6450\textrm{\textdegree~N} &    270\textrm{\textdegree} & 0\textrm{\textdegree} & \textrm{rf-driven} & \textrm{$^{133}$Cs~D1~} F=4 & $-0.39^{+0.19}_{-0.00}$\\
\rule{0ex}{3.6ex} Oberlin$^\star$ &  81.7796\textrm{\textdegree~W} & 41.2950\textrm{\textdegree~N} &    276\textrm{\textdegree} & 0\textrm{\textdegree} & \textrm{SERF} & \textrm{$^{39}$K~D1} & $-0.49^{+0.18}_{-0.00}$\\
\hline
\hline
\end{tabular}
\label{table:SR5-mags}
\end{table*}


\begin{thebibliography}{77}%
\makeatletter
\providecommand \@ifxundefined [1]{%
 \@ifx{#1\undefined}
}%
\providecommand \@ifnum [1]{%
 \ifnum #1\expandafter \@firstoftwo
 \else \expandafter \@secondoftwo
 \fi
}%
\providecommand \@ifx [1]{%
 \ifx #1\expandafter \@firstoftwo
 \else \expandafter \@secondoftwo
 \fi
}%
\providecommand \natexlab [1]{#1}%
\providecommand \enquote  [1]{``#1''}%
\providecommand \bibnamefont  [1]{#1}%
\providecommand \bibfnamefont [1]{#1}%
\providecommand \citenamefont [1]{#1}%
\providecommand \href@noop [0]{\@secondoftwo}%
\providecommand \href [0]{\begingroup \@sanitize@url \@href}%
\providecommand \@href[1]{\@@startlink{#1}\@@href}%
\providecommand \@@href[1]{\endgroup#1\@@endlink}%
\providecommand \@sanitize@url [0]{\catcode `\\12\catcode `\$12\catcode
  `\&12\catcode `\#12\catcode `\^12\catcode `\_12\catcode `\%12\relax}%
\providecommand \@@startlink[1]{}%
\providecommand \@@endlink[0]{}%
\providecommand \url  [0]{\begingroup\@sanitize@url \@url }%
\providecommand \@url [1]{\endgroup\@href {#1}{\urlprefix }}%
\providecommand \urlprefix  [0]{URL }%
\providecommand \Eprint [0]{\href }%
\providecommand \doibase [0]{https://doi.org/}%
\providecommand \selectlanguage [0]{\@gobble}%
\providecommand \bibinfo  [0]{\@secondoftwo}%
\providecommand \bibfield  [0]{\@secondoftwo}%
\providecommand \translation [1]{[#1]}%
\providecommand \BibitemOpen [0]{}%
\providecommand \bibitemStop [0]{}%
\providecommand \bibitemNoStop [0]{.\EOS\space}%
\providecommand \EOS [0]{\spacefactor3000\relax}%
\providecommand \BibitemShut  [1]{\csname bibitem#1\endcsname}%
\let\auto@bib@innerbib\@empty
\bibitem [{\citenamefont {Graham}\ \emph {et~al.}(2015)\citenamefont {Graham},
  \citenamefont {Irastorza}, \citenamefont {Lamoreaux}, \citenamefont
  {Lindner},\ and\ \citenamefont {van Bibber}}]{graham2015experimental}%
  \BibitemOpen
  \bibfield  {author} {\bibinfo {author} {\bibfnamefont {P.~W.}\ \bibnamefont
  {Graham}}, \bibinfo {author} {\bibfnamefont {I.~G.}\ \bibnamefont
  {Irastorza}}, \bibinfo {author} {\bibfnamefont {S.~K.}\ \bibnamefont
  {Lamoreaux}}, \bibinfo {author} {\bibfnamefont {A.}~\bibnamefont {Lindner}},\
  and\ \bibinfo {author} {\bibfnamefont {K.~A.}\ \bibnamefont {van Bibber}},\
  }\bibfield  {title} {\bibinfo {title} {Experimental searches for the axion
  and axion-like particles},\ }\href@noop {} {\bibfield  {journal} {\bibinfo
  {journal} {Annu. Rev. Nucl. Part. Sci.}\ }\textbf {\bibinfo {volume} {65}},\
  \bibinfo {pages} {485} (\bibinfo {year} {2015})}\BibitemShut {NoStop}%
\bibitem [{\citenamefont {Jackson~Kimball}\ and\ \citenamefont {van
  Bibber}(2022)}]{kimball2022search}%
  \BibitemOpen
  \bibfield  {author} {\bibinfo {author} {\bibfnamefont {D.~F.}\ \bibnamefont
  {Jackson~Kimball}}\ and\ \bibinfo {author} {\bibfnamefont {K.}~\bibnamefont
  {van Bibber}},\ }\href@noop {} {\emph {\bibinfo {title} {{The Search for
  Ultralight Bosonic Dark Matter}}}}\ (\bibinfo  {publisher} {Springer},\
  \bibinfo {year} {2022})\BibitemShut {NoStop}%
\bibitem [{\citenamefont {Banerjee}\ \emph
  {et~al.}(2020{\natexlab{a}})\citenamefont {Banerjee}, \citenamefont {Budker},
  \citenamefont {Eby}, \citenamefont {Kim},\ and\ \citenamefont
  {Perez}}]{banerjee2020relaxion}%
  \BibitemOpen
  \bibfield  {author} {\bibinfo {author} {\bibfnamefont {A.}~\bibnamefont
  {Banerjee}}, \bibinfo {author} {\bibfnamefont {D.}~\bibnamefont {Budker}},
  \bibinfo {author} {\bibfnamefont {J.}~\bibnamefont {Eby}}, \bibinfo {author}
  {\bibfnamefont {H.}~\bibnamefont {Kim}},\ and\ \bibinfo {author}
  {\bibfnamefont {G.}~\bibnamefont {Perez}},\ }\bibfield  {title} {\bibinfo
  {title} {Relaxion stars and their detection via atomic physics},\ }\href@noop
  {} {\bibfield  {journal} {\bibinfo  {journal} {Commun. Phys.}\ }\textbf
  {\bibinfo {volume} {3}},\ \bibinfo {pages} {1} (\bibinfo {year}
  {2020}{\natexlab{a}})}\BibitemShut {NoStop}%
\bibitem [{\citenamefont {Banerjee}\ \emph
  {et~al.}(2020{\natexlab{b}})\citenamefont {Banerjee}, \citenamefont {Budker},
  \citenamefont {Eby}, \citenamefont {Flambaum}, \citenamefont {Kim},
  \citenamefont {Matsedonskyi},\ and\ \citenamefont
  {Perez}}]{banerjee2020searching}%
  \BibitemOpen
  \bibfield  {author} {\bibinfo {author} {\bibfnamefont {A.}~\bibnamefont
  {Banerjee}}, \bibinfo {author} {\bibfnamefont {D.}~\bibnamefont {Budker}},
  \bibinfo {author} {\bibfnamefont {J.}~\bibnamefont {Eby}}, \bibinfo {author}
  {\bibfnamefont {V.~V.}\ \bibnamefont {Flambaum}}, \bibinfo {author}
  {\bibfnamefont {H.}~\bibnamefont {Kim}}, \bibinfo {author} {\bibfnamefont
  {O.}~\bibnamefont {Matsedonskyi}},\ and\ \bibinfo {author} {\bibfnamefont
  {G.}~\bibnamefont {Perez}},\ }\bibfield  {title} {\bibinfo {title} {Searching
  for earth/solar axion halos},\ }\href@noop {} {\bibfield  {journal} {\bibinfo
   {journal} {J. High Energy Phys.}\ }\textbf {\bibinfo {volume}
  {2020}}\bibinfo  {number} { (9)},\ \bibinfo {pages} {4}}\BibitemShut
  {NoStop}%
\bibitem [{\citenamefont {Tretiak}\ \emph {et~al.}(2022)\citenamefont
  {Tretiak}, \citenamefont {Zhang}, \citenamefont {Figueroa}, \citenamefont
  {Antypas}, \citenamefont {Brogna}, \citenamefont {Banerjee}, \citenamefont
  {Perez},\ and\ \citenamefont {Budker}}]{tretiak2022improved}%
  \BibitemOpen
\bibfield  {number} {  }\bibfield  {author} {\bibinfo {author} {\bibfnamefont
  {O.}~\bibnamefont {Tretiak}}, \bibinfo {author} {\bibfnamefont
  {X.}~\bibnamefont {Zhang}}, \bibinfo {author} {\bibfnamefont {N.~L.}\
  \bibnamefont {Figueroa}}, \bibinfo {author} {\bibfnamefont {D.}~\bibnamefont
  {Antypas}}, \bibinfo {author} {\bibfnamefont {A.}~\bibnamefont {Brogna}},
  \bibinfo {author} {\bibfnamefont {A.}~\bibnamefont {Banerjee}}, \bibinfo
  {author} {\bibfnamefont {G.}~\bibnamefont {Perez}},\ and\ \bibinfo {author}
  {\bibfnamefont {D.}~\bibnamefont {Budker}},\ }\bibfield  {title} {\bibinfo
  {title} {Improved bounds on ultralight scalar dark matter in the
  radio-frequency range},\ }\href@noop {} {\bibfield  {journal} {\bibinfo
  {journal} {arXiv:2201.02042}\ } (\bibinfo {year} {2022})}\BibitemShut
  {NoStop}%
\bibitem [{\citenamefont {Xu}\ and\ \citenamefont {Siegel}(2008)}]{xu2008dark}%
  \BibitemOpen
  \bibfield  {author} {\bibinfo {author} {\bibfnamefont {X.}~\bibnamefont
  {Xu}}\ and\ \bibinfo {author} {\bibfnamefont {E.}~\bibnamefont {Siegel}},\
  }\bibfield  {title} {\bibinfo {title} {Dark matter in the solar system},\
  }\href@noop {} {\bibfield  {journal} {\bibinfo  {journal} {arXiv:0806.3767}\
  } (\bibinfo {year} {2008})}\BibitemShut {NoStop}%
\bibitem [{\citenamefont {Khriplovich}\ and\ \citenamefont
  {Shepelyansky}(2009)}]{khriplovich2009capture}%
  \BibitemOpen
  \bibfield  {author} {\bibinfo {author} {\bibfnamefont {I.}~\bibnamefont
  {Khriplovich}}\ and\ \bibinfo {author} {\bibfnamefont {D.}~\bibnamefont
  {Shepelyansky}},\ }\bibfield  {title} {\bibinfo {title} {Capture of dark
  matter by the solar system},\ }\href@noop {} {\bibfield  {journal} {\bibinfo
  {journal} {Int. J. Mod. Phys. D}\ }\textbf {\bibinfo {volume} {18}},\
  \bibinfo {pages} {1903} (\bibinfo {year} {2009})}\BibitemShut {NoStop}%
\bibitem [{\citenamefont
  {Khriplovich}(2011{\natexlab{a}})}]{khriplovich2011capture}%
  \BibitemOpen
  \bibfield  {author} {\bibinfo {author} {\bibfnamefont {I.}~\bibnamefont
  {Khriplovich}},\ }\bibfield  {title} {\bibinfo {title} {Capture of dark
  matter by the solar system: simple estimates},\ }\href@noop {} {\bibfield
  {journal} {\bibinfo  {journal} {Int. J. Mod. Phys. D}\ }\textbf {\bibinfo
  {volume} {20}},\ \bibinfo {pages} {17} (\bibinfo {year}
  {2011}{\natexlab{a}})}\BibitemShut {NoStop}%
\bibitem [{\citenamefont
  {Khriplovich}(2011{\natexlab{b}})}]{khriplovich2011capture2}%
  \BibitemOpen
  \bibfield  {author} {\bibinfo {author} {\bibfnamefont {I.}~\bibnamefont
  {Khriplovich}},\ }\bibfield  {title} {\bibinfo {title} {Capture of dark
  matter by the solar system. analytical estimates},\ }in\ \href@noop {} {\emph
  {\bibinfo {booktitle} {Gribov-80 Memorial Volume: Quantum Chromodynamics and
  Beyond}}}\ (\bibinfo  {publisher} {World Scientific},\ \bibinfo {year}
  {2011})\ pp.\ \bibinfo {pages} {471--478}\BibitemShut {NoStop}%
\bibitem [{\citenamefont {Brito}\ \emph {et~al.}(2015)\citenamefont {Brito},
  \citenamefont {Cardoso},\ and\ \citenamefont {Okawa}}]{brito2015accretion}%
  \BibitemOpen
  \bibfield  {author} {\bibinfo {author} {\bibfnamefont {R.}~\bibnamefont
  {Brito}}, \bibinfo {author} {\bibfnamefont {V.}~\bibnamefont {Cardoso}},\
  and\ \bibinfo {author} {\bibfnamefont {H.}~\bibnamefont {Okawa}},\ }\bibfield
   {title} {\bibinfo {title} {Accretion of dark matter by stars},\ }\href@noop
  {} {\bibfield  {journal} {\bibinfo  {journal} {Phys. Rev. Lett.}\ }\textbf
  {\bibinfo {volume} {115}},\ \bibinfo {pages} {111301} (\bibinfo {year}
  {2015})}\BibitemShut {NoStop}%
\bibitem [{\citenamefont {Brito}\ \emph {et~al.}(2016)\citenamefont {Brito},
  \citenamefont {Cardoso}, \citenamefont {Macedo}, \citenamefont {Okawa},\ and\
  \citenamefont {Palenzuela}}]{brito2016interaction}%
  \BibitemOpen
  \bibfield  {author} {\bibinfo {author} {\bibfnamefont {R.}~\bibnamefont
  {Brito}}, \bibinfo {author} {\bibfnamefont {V.}~\bibnamefont {Cardoso}},
  \bibinfo {author} {\bibfnamefont {C.~F.}\ \bibnamefont {Macedo}}, \bibinfo
  {author} {\bibfnamefont {H.}~\bibnamefont {Okawa}},\ and\ \bibinfo {author}
  {\bibfnamefont {C.}~\bibnamefont {Palenzuela}},\ }\bibfield  {title}
  {\bibinfo {title} {Interaction between bosonic dark matter and stars},\
  }\href@noop {} {\bibfield  {journal} {\bibinfo  {journal} {Phys. Rev. D}\
  }\textbf {\bibinfo {volume} {93}},\ \bibinfo {pages} {044045} (\bibinfo
  {year} {2016})}\BibitemShut {NoStop}%
\bibitem [{\citenamefont {Budker}\ \emph {et~al.}(2023)\citenamefont {Budker},
  \citenamefont {Eby}, \citenamefont {Gorghetto}, \citenamefont {Jiang},\ and\
  \citenamefont {Perez}}]{budker2023generic}%
  \BibitemOpen
  \bibfield  {author} {\bibinfo {author} {\bibfnamefont {D.}~\bibnamefont
  {Budker}}, \bibinfo {author} {\bibfnamefont {J.}~\bibnamefont {Eby}},
  \bibinfo {author} {\bibfnamefont {M.}~\bibnamefont {Gorghetto}}, \bibinfo
  {author} {\bibfnamefont {M.}~\bibnamefont {Jiang}},\ and\ \bibinfo {author}
  {\bibfnamefont {G.}~\bibnamefont {Perez}},\ }\bibfield  {title} {\bibinfo
  {title} {A generic formation mechanism of ultralight dark matter solar
  halos},\ }\href@noop {} {\bibfield  {journal} {\bibinfo  {journal} {J.
  Cosmol. Astropart. Phys.}\ }\textbf {\bibinfo {volume} {2023}}\bibinfo
  {number} { (12)},\ \bibinfo {pages} {021}}\BibitemShut {NoStop}%
\bibitem [{\citenamefont {Nielsen}\ \emph {et~al.}(2019)\citenamefont
  {Nielsen}, \citenamefont {Palessandro},\ and\ \citenamefont
  {Sloth}}]{nielsen2019gravitational}%
  \BibitemOpen
\bibfield  {number} {  }\bibfield  {author} {\bibinfo {author} {\bibfnamefont
  {N.~G.}\ \bibnamefont {Nielsen}}, \bibinfo {author} {\bibfnamefont
  {A.}~\bibnamefont {Palessandro}},\ and\ \bibinfo {author} {\bibfnamefont
  {M.~S.}\ \bibnamefont {Sloth}},\ }\bibfield  {title} {\bibinfo {title}
  {Gravitational atoms},\ }\href@noop {} {\bibfield  {journal} {\bibinfo
  {journal} {Phys. Rev. D}\ }\textbf {\bibinfo {volume} {99}},\ \bibinfo
  {pages} {123011} (\bibinfo {year} {2019})}\BibitemShut {NoStop}%
\bibitem [{\citenamefont {Arvanitaki}\ and\ \citenamefont
  {Dubovsky}(2011)}]{arvanitaki2011exploring}%
  \BibitemOpen
  \bibfield  {author} {\bibinfo {author} {\bibfnamefont {A.}~\bibnamefont
  {Arvanitaki}}\ and\ \bibinfo {author} {\bibfnamefont {S.}~\bibnamefont
  {Dubovsky}},\ }\bibfield  {title} {\bibinfo {title} {Exploring the string
  axiverse with precision black hole physics},\ }\href@noop {} {\bibfield
  {journal} {\bibinfo  {journal} {Phys. Rev. D}\ }\textbf {\bibinfo {volume}
  {83}},\ \bibinfo {pages} {044026} (\bibinfo {year} {2011})}\BibitemShut
  {NoStop}%
\bibitem [{\citenamefont {Baryakhtar}\ \emph {et~al.}(2021)\citenamefont
  {Baryakhtar}, \citenamefont {Galanis}, \citenamefont {Lasenby},\ and\
  \citenamefont {Simon}}]{baryakhtar2021black}%
  \BibitemOpen
  \bibfield  {author} {\bibinfo {author} {\bibfnamefont {M.}~\bibnamefont
  {Baryakhtar}}, \bibinfo {author} {\bibfnamefont {M.}~\bibnamefont {Galanis}},
  \bibinfo {author} {\bibfnamefont {R.}~\bibnamefont {Lasenby}},\ and\ \bibinfo
  {author} {\bibfnamefont {O.}~\bibnamefont {Simon}},\ }\bibfield  {title}
  {\bibinfo {title} {Black hole superradiance of self-interacting scalar
  fields},\ }\href@noop {} {\bibfield  {journal} {\bibinfo  {journal} {Phys.
  Rev. D}\ }\textbf {\bibinfo {volume} {103}},\ \bibinfo {pages} {095019}
  (\bibinfo {year} {2021})}\BibitemShut {NoStop}%
\bibitem [{\citenamefont {Kryemadhi}\ \emph {et~al.}(2023)\citenamefont
  {Kryemadhi}, \citenamefont {Maroudas}, \citenamefont {Mastronikolis},\ and\
  \citenamefont {Zioutas}}]{kryemadhi2023gravitational}%
  \BibitemOpen
  \bibfield  {author} {\bibinfo {author} {\bibfnamefont {A.}~\bibnamefont
  {Kryemadhi}}, \bibinfo {author} {\bibfnamefont {M.}~\bibnamefont {Maroudas}},
  \bibinfo {author} {\bibfnamefont {A.}~\bibnamefont {Mastronikolis}},\ and\
  \bibinfo {author} {\bibfnamefont {K.}~\bibnamefont {Zioutas}},\ }\bibfield
  {title} {\bibinfo {title} {Gravitational focusing effects on streaming dark
  matter as a new detection concept},\ }\href@noop {} {\bibfield  {journal}
  {\bibinfo  {journal} {Phys. Rev. D}\ }\textbf {\bibinfo {volume} {108}},\
  \bibinfo {pages} {123043} (\bibinfo {year} {2023})}\BibitemShut {NoStop}%
\bibitem [{\citenamefont {Kim}\ and\ \citenamefont
  {Lenoci}(2022)}]{kim2022gravitational}%
  \BibitemOpen
  \bibfield  {author} {\bibinfo {author} {\bibfnamefont {H.}~\bibnamefont
  {Kim}}\ and\ \bibinfo {author} {\bibfnamefont {A.}~\bibnamefont {Lenoci}},\
  }\bibfield  {title} {\bibinfo {title} {Gravitational focusing of wave dark
  matter},\ }\href@noop {} {\bibfield  {journal} {\bibinfo  {journal} {Phys.
  Rev. D}\ }\textbf {\bibinfo {volume} {105}},\ \bibinfo {pages} {063032}
  (\bibinfo {year} {2022})}\BibitemShut {NoStop}%
\bibitem [{\citenamefont {Patla}\ \emph {et~al.}(2013)\citenamefont {Patla},
  \citenamefont {Nemiroff}, \citenamefont {Hoffmann},\ and\ \citenamefont
  {Zioutas}}]{patla2013flux}%
  \BibitemOpen
  \bibfield  {author} {\bibinfo {author} {\bibfnamefont {B.~R.}\ \bibnamefont
  {Patla}}, \bibinfo {author} {\bibfnamefont {R.~J.}\ \bibnamefont {Nemiroff}},
  \bibinfo {author} {\bibfnamefont {D.~H.}\ \bibnamefont {Hoffmann}},\ and\
  \bibinfo {author} {\bibfnamefont {K.}~\bibnamefont {Zioutas}},\ }\bibfield
  {title} {\bibinfo {title} {Flux enhancement of slow-moving particles by sun
  or jupiter: can they be detected on earth?},\ }\href@noop {} {\bibfield
  {journal} {\bibinfo  {journal} {Astrophys. J.}\ }\textbf {\bibinfo {volume}
  {780}},\ \bibinfo {pages} {158} (\bibinfo {year} {2013})}\BibitemShut
  {NoStop}%
\bibitem [{\citenamefont {Foster}\ \emph {et~al.}(2018)\citenamefont {Foster},
  \citenamefont {Rodd},\ and\ \citenamefont {Safdi}}]{foster2018revealing}%
  \BibitemOpen
  \bibfield  {author} {\bibinfo {author} {\bibfnamefont {J.~W.}\ \bibnamefont
  {Foster}}, \bibinfo {author} {\bibfnamefont {N.~L.}\ \bibnamefont {Rodd}},\
  and\ \bibinfo {author} {\bibfnamefont {B.~R.}\ \bibnamefont {Safdi}},\
  }\bibfield  {title} {\bibinfo {title} {Revealing the dark matter halo with
  axion direct detection},\ }\href@noop {} {\bibfield  {journal} {\bibinfo
  {journal} {Phys. Rev. D}\ }\textbf {\bibinfo {volume} {97}},\ \bibinfo
  {pages} {123006} (\bibinfo {year} {2018})}\BibitemShut {NoStop}%
\bibitem [{\citenamefont {Cheong}\ \emph {et~al.}(2025)\citenamefont {Cheong},
  \citenamefont {Rodd},\ and\ \citenamefont {Wang}}]{cheong2025quantum}%
  \BibitemOpen
  \bibfield  {author} {\bibinfo {author} {\bibfnamefont {D.~Y.}\ \bibnamefont
  {Cheong}}, \bibinfo {author} {\bibfnamefont {N.~L.}\ \bibnamefont {Rodd}},\
  and\ \bibinfo {author} {\bibfnamefont {L.-T.}\ \bibnamefont {Wang}},\
  }\bibfield  {title} {\bibinfo {title} {Quantum description of wave dark
  matter},\ }\href@noop {} {\bibfield  {journal} {\bibinfo  {journal} {Phys.
  Rev. D}\ }\textbf {\bibinfo {volume} {111}},\ \bibinfo {pages} {015028}
  (\bibinfo {year} {2025})}\BibitemShut {NoStop}%
\bibitem [{\citenamefont {de~Salas}\ and\ \citenamefont
  {Widmark}(2021)}]{de2021dark}%
  \BibitemOpen
  \bibfield  {author} {\bibinfo {author} {\bibfnamefont {P.~F.}\ \bibnamefont
  {de~Salas}}\ and\ \bibinfo {author} {\bibfnamefont {A.}~\bibnamefont
  {Widmark}},\ }\bibfield  {title} {\bibinfo {title} {{Dark matter local
  density determination: recent observations and future prospects}},\
  }\href@noop {} {\bibfield  {journal} {\bibinfo  {journal} {Rep. Prog. Phys.}\
  }\textbf {\bibinfo {volume} {84}},\ \bibinfo {pages} {104901} (\bibinfo
  {year} {2021})}\BibitemShut {NoStop}%
\bibitem [{\citenamefont {Group}\ \emph {et~al.}(2022)\citenamefont {Group},
  \citenamefont {Workman}, \citenamefont {Burkert}, \citenamefont {Crede},
  \citenamefont {Klempt}, \citenamefont {Thoma}, \citenamefont {Tiator},
  \citenamefont {Agashe}, \citenamefont {Aielli}, \citenamefont {Allanach}
  \emph {et~al.}}]{particle2022review}%
  \BibitemOpen
  \bibfield  {author} {\bibinfo {author} {\bibfnamefont {P.~D.}\ \bibnamefont
  {Group}}, \bibinfo {author} {\bibfnamefont {R.}~\bibnamefont {Workman}},
  \bibinfo {author} {\bibfnamefont {V.}~\bibnamefont {Burkert}}, \bibinfo
  {author} {\bibfnamefont {V.}~\bibnamefont {Crede}}, \bibinfo {author}
  {\bibfnamefont {E.}~\bibnamefont {Klempt}}, \bibinfo {author} {\bibfnamefont
  {U.}~\bibnamefont {Thoma}}, \bibinfo {author} {\bibfnamefont
  {L.}~\bibnamefont {Tiator}}, \bibinfo {author} {\bibfnamefont
  {K.}~\bibnamefont {Agashe}}, \bibinfo {author} {\bibfnamefont
  {G.}~\bibnamefont {Aielli}}, \bibinfo {author} {\bibfnamefont
  {B.}~\bibnamefont {Allanach}}, \emph {et~al.},\ }\bibfield  {title} {\bibinfo
  {title} {{Review of particle physics}},\ }\href@noop {} {\bibfield  {journal}
  {\bibinfo  {journal} {Prog. Theor. Exp. Phys.}\ }\textbf {\bibinfo {volume}
  {2022}},\ \bibinfo {pages} {083C01} (\bibinfo {year} {2022})}\BibitemShut
  {NoStop}%
\bibitem [{\citenamefont {Pitjev}\ and\ \citenamefont
  {Pitjeva}(2013)}]{pitjev2013constraints}%
  \BibitemOpen
  \bibfield  {author} {\bibinfo {author} {\bibfnamefont {N.}~\bibnamefont
  {Pitjev}}\ and\ \bibinfo {author} {\bibfnamefont {E.}~\bibnamefont
  {Pitjeva}},\ }\bibfield  {title} {\bibinfo {title} {Constraints on dark
  matter in the solar system},\ }\href@noop {} {\bibfield  {journal} {\bibinfo
  {journal} {Astron. Lett.}\ }\textbf {\bibinfo {volume} {39}},\ \bibinfo
  {pages} {141} (\bibinfo {year} {2013})}\BibitemShut {NoStop}%
\bibitem [{\citenamefont {Graham}\ and\ \citenamefont
  {Rajendran}(2013)}]{graham2013new}%
  \BibitemOpen
  \bibfield  {author} {\bibinfo {author} {\bibfnamefont {P.~W.}\ \bibnamefont
  {Graham}}\ and\ \bibinfo {author} {\bibfnamefont {S.}~\bibnamefont
  {Rajendran}},\ }\bibfield  {title} {\bibinfo {title} {New observables for
  direct detection of axion dark matter},\ }\href@noop {} {\bibfield  {journal}
  {\bibinfo  {journal} {Phys. Rev. D}\ }\textbf {\bibinfo {volume} {88}},\
  \bibinfo {pages} {035023} (\bibinfo {year} {2013})}\BibitemShut {NoStop}%
\bibitem [{\citenamefont {Stadnik}\ and\ \citenamefont
  {Flambaum}(2014)}]{stadnik2014axion}%
  \BibitemOpen
  \bibfield  {author} {\bibinfo {author} {\bibfnamefont {Y.}~\bibnamefont
  {Stadnik}}\ and\ \bibinfo {author} {\bibfnamefont {V.}~\bibnamefont
  {Flambaum}},\ }\bibfield  {title} {\bibinfo {title} {Axion-induced effects in
  atoms, molecules, and nuclei: Parity nonconservation, anapole moments,
  electric dipole moments, and spin-gravity and spin-axion momentum
  couplings},\ }\href@noop {} {\bibfield  {journal} {\bibinfo  {journal} {Phys.
  Rev. D}\ }\textbf {\bibinfo {volume} {89}},\ \bibinfo {pages} {043522}
  (\bibinfo {year} {2014})}\BibitemShut {NoStop}%
\bibitem [{\citenamefont {Cong}\ \emph {et~al.}(2025)\citenamefont {Cong},
  \citenamefont {Ji}, \citenamefont {Fadeev}, \citenamefont {Ficek},
  \citenamefont {Jiang}, \citenamefont {Flambaum}, \citenamefont {Guan},
  \citenamefont {Jackson~Kimball}, \citenamefont {Kozlov}, \citenamefont
  {Stadnik} \emph {et~al.}}]{cong2025spin}%
  \BibitemOpen
  \bibfield  {author} {\bibinfo {author} {\bibfnamefont {L.}~\bibnamefont
  {Cong}}, \bibinfo {author} {\bibfnamefont {W.}~\bibnamefont {Ji}}, \bibinfo
  {author} {\bibfnamefont {P.}~\bibnamefont {Fadeev}}, \bibinfo {author}
  {\bibfnamefont {F.}~\bibnamefont {Ficek}}, \bibinfo {author} {\bibfnamefont
  {M.}~\bibnamefont {Jiang}}, \bibinfo {author} {\bibfnamefont {V.~V.}\
  \bibnamefont {Flambaum}}, \bibinfo {author} {\bibfnamefont {H.}~\bibnamefont
  {Guan}}, \bibinfo {author} {\bibfnamefont {D.~F.}\ \bibnamefont
  {Jackson~Kimball}}, \bibinfo {author} {\bibfnamefont {M.~G.}\ \bibnamefont
  {Kozlov}}, \bibinfo {author} {\bibfnamefont {Y.~V.}\ \bibnamefont {Stadnik}},
  \emph {et~al.},\ }\bibfield  {title} {\bibinfo {title} {Spin-dependent exotic
  interactions},\ }\href@noop {} {\bibfield  {journal} {\bibinfo  {journal}
  {Rev. Mod. Phys.}\ }\textbf {\bibinfo {volume} {97}},\ \bibinfo {pages}
  {025005} (\bibinfo {year} {2025})}\BibitemShut {NoStop}%
\bibitem [{\citenamefont {Pospelov}\ \emph {et~al.}(2013)\citenamefont
  {Pospelov}, \citenamefont {Pustelny}, \citenamefont {Ledbetter},
  \citenamefont {Jackson~Kimball}, \citenamefont {Gawlik},\ and\ \citenamefont
  {Budker}}]{Pos13}%
  \BibitemOpen
  \bibfield  {author} {\bibinfo {author} {\bibfnamefont {M.}~\bibnamefont
  {Pospelov}}, \bibinfo {author} {\bibfnamefont {S.}~\bibnamefont {Pustelny}},
  \bibinfo {author} {\bibfnamefont {M.~P.}\ \bibnamefont {Ledbetter}}, \bibinfo
  {author} {\bibfnamefont {D.~F.}\ \bibnamefont {Jackson~Kimball}}, \bibinfo
  {author} {\bibfnamefont {W.}~\bibnamefont {Gawlik}},\ and\ \bibinfo {author}
  {\bibfnamefont {D.}~\bibnamefont {Budker}},\ }\bibfield  {title} {\bibinfo
  {title} {Detecting domain walls of axionlike models using terrestrial
  experiments},\ }\href@noop {} {\bibfield  {journal} {\bibinfo  {journal}
  {Phys. Rev. Lett.}\ }\textbf {\bibinfo {volume} {110}},\ \bibinfo {pages}
  {021803} (\bibinfo {year} {2013})}\BibitemShut {NoStop}%
\bibitem [{\citenamefont {Safronova}\ \emph {et~al.}(2018)\citenamefont
  {Safronova}, \citenamefont {Budker}, \citenamefont {DeMille}, \citenamefont
  {Jackson~Kimball}, \citenamefont {Derevianko},\ and\ \citenamefont
  {Clark}}]{safronova2018search}%
  \BibitemOpen
  \bibfield  {author} {\bibinfo {author} {\bibfnamefont {M.}~\bibnamefont
  {Safronova}}, \bibinfo {author} {\bibfnamefont {D.}~\bibnamefont {Budker}},
  \bibinfo {author} {\bibfnamefont {D.}~\bibnamefont {DeMille}}, \bibinfo
  {author} {\bibfnamefont {D.~F.}\ \bibnamefont {Jackson~Kimball}}, \bibinfo
  {author} {\bibfnamefont {A.}~\bibnamefont {Derevianko}},\ and\ \bibinfo
  {author} {\bibfnamefont {C.~W.}\ \bibnamefont {Clark}},\ }\bibfield  {title}
  {\bibinfo {title} {Search for new physics with atoms and molecules},\
  }\href@noop {} {\bibfield  {journal} {\bibinfo  {journal} {Rev. Mod. Phys.}\
  }\textbf {\bibinfo {volume} {90}},\ \bibinfo {pages} {025008} (\bibinfo
  {year} {2018})}\BibitemShut {NoStop}%
\bibitem [{\citenamefont {Dailey}\ \emph {et~al.}(2021)\citenamefont {Dailey},
  \citenamefont {Bradley}, \citenamefont {Jackson~Kimball}, \citenamefont
  {Sulai}, \citenamefont {Pustelny}, \citenamefont {Wickenbrock},\ and\
  \citenamefont {Derevianko}}]{dailey2021quantum}%
  \BibitemOpen
  \bibfield  {author} {\bibinfo {author} {\bibfnamefont {C.}~\bibnamefont
  {Dailey}}, \bibinfo {author} {\bibfnamefont {C.}~\bibnamefont {Bradley}},
  \bibinfo {author} {\bibfnamefont {D.~F.}\ \bibnamefont {Jackson~Kimball}},
  \bibinfo {author} {\bibfnamefont {I.~A.}\ \bibnamefont {Sulai}}, \bibinfo
  {author} {\bibfnamefont {S.}~\bibnamefont {Pustelny}}, \bibinfo {author}
  {\bibfnamefont {A.}~\bibnamefont {Wickenbrock}},\ and\ \bibinfo {author}
  {\bibfnamefont {A.}~\bibnamefont {Derevianko}},\ }\bibfield  {title}
  {\bibinfo {title} {Quantum sensor networks as exotic field telescopes for
  multi-messenger astronomy},\ }\href@noop {} {\bibfield  {journal} {\bibinfo
  {journal} {Nature Astron.}\ }\textbf {\bibinfo {volume} {5}},\ \bibinfo
  {pages} {150} (\bibinfo {year} {2021})}\BibitemShut {NoStop}%
\bibitem [{\citenamefont {Afach}\ \emph {et~al.}(2023)\citenamefont {Afach},
  \citenamefont {Tumturk}, \citenamefont {Bekker}, \citenamefont {Buchler},
  \citenamefont {Budker}, \citenamefont {Cervantes}, \citenamefont
  {Derevianko}, \citenamefont {Eby}, \citenamefont {Figueroa}, \citenamefont
  {Folman} \emph {et~al.}}]{afach2023WhatCanGNOMEdo}%
  \BibitemOpen
  \bibfield  {author} {\bibinfo {author} {\bibfnamefont {S.}~\bibnamefont
  {Afach}}, \bibinfo {author} {\bibfnamefont {D.~A.}\ \bibnamefont {Tumturk}},
  \bibinfo {author} {\bibfnamefont {H.}~\bibnamefont {Bekker}}, \bibinfo
  {author} {\bibfnamefont {B.}~\bibnamefont {Buchler}}, \bibinfo {author}
  {\bibfnamefont {D.}~\bibnamefont {Budker}}, \bibinfo {author} {\bibfnamefont
  {K.}~\bibnamefont {Cervantes}}, \bibinfo {author} {\bibfnamefont
  {A.}~\bibnamefont {Derevianko}}, \bibinfo {author} {\bibfnamefont
  {J.}~\bibnamefont {Eby}}, \bibinfo {author} {\bibfnamefont {N.}~\bibnamefont
  {Figueroa}}, \bibinfo {author} {\bibfnamefont {R.}~\bibnamefont {Folman}},
  \emph {et~al.},\ }\bibfield  {title} {\bibinfo {title} {{What can a GNOME do?
  Search targets for the Global Network of Optical Magnetometers for Exotic
  physics searches}},\ }\href@noop {} {\bibfield  {journal} {\bibinfo
  {journal} {Ann. Phys. (Berl.)}\ }\textbf {\bibinfo {volume} {2023}},\
  \bibinfo {pages} {2300083} (\bibinfo {year} {2023})}\BibitemShut {NoStop}%
\bibitem [{\citenamefont {Jackson~Kimball}(2015)}]{Kim15}%
  \BibitemOpen
  \bibfield  {author} {\bibinfo {author} {\bibfnamefont {D.~F.}\ \bibnamefont
  {Jackson~Kimball}},\ }\bibfield  {title} {\bibinfo {title} {Nuclear spin
  content and constraints on exotic spin-dependent couplings},\ }\href@noop {}
  {\bibfield  {journal} {\bibinfo  {journal} {New J. Phys.}\ }\textbf {\bibinfo
  {volume} {17}},\ \bibinfo {pages} {073008} (\bibinfo {year}
  {2015})}\BibitemShut {NoStop}%
\bibitem [{\citenamefont {Jackson~Kimball}\ \emph {et~al.}(2016)\citenamefont
  {Jackson~Kimball}, \citenamefont {Dudley}, \citenamefont {Li}, \citenamefont
  {Thulasi}, \citenamefont {Pustelny}, \citenamefont {Budker},\ and\
  \citenamefont {Zolotorev}}]{kimball2016magnetic}%
  \BibitemOpen
  \bibfield  {author} {\bibinfo {author} {\bibfnamefont {D.}~\bibnamefont
  {Jackson~Kimball}}, \bibinfo {author} {\bibfnamefont {J.}~\bibnamefont
  {Dudley}}, \bibinfo {author} {\bibfnamefont {Y.}~\bibnamefont {Li}}, \bibinfo
  {author} {\bibfnamefont {S.}~\bibnamefont {Thulasi}}, \bibinfo {author}
  {\bibfnamefont {S.}~\bibnamefont {Pustelny}}, \bibinfo {author}
  {\bibfnamefont {D.}~\bibnamefont {Budker}},\ and\ \bibinfo {author}
  {\bibfnamefont {M.}~\bibnamefont {Zolotorev}},\ }\bibfield  {title} {\bibinfo
  {title} {Magnetic shielding and exotic spin-dependent interactions},\
  }\href@noop {} {\bibfield  {journal} {\bibinfo  {journal} {Phys. Rev. D}\
  }\textbf {\bibinfo {volume} {94}},\ \bibinfo {pages} {082005} (\bibinfo
  {year} {2016})}\BibitemShut {NoStop}%
\bibitem [{\citenamefont {Pustelny}\ \emph {et~al.}(2013)\citenamefont
  {Pustelny}, \citenamefont {Jackson~Kimball}, \citenamefont {Pankow},
  \citenamefont {Ledbetter}, \citenamefont {Wlodarczyk}, \citenamefont
  {Wcislo}, \citenamefont {Pospelov}, \citenamefont {Smith}, \citenamefont
  {Read}, \citenamefont {Gawlik} \emph {et~al.}}]{pustelny2013global}%
  \BibitemOpen
  \bibfield  {author} {\bibinfo {author} {\bibfnamefont {S.}~\bibnamefont
  {Pustelny}}, \bibinfo {author} {\bibfnamefont {D.~F.}\ \bibnamefont
  {Jackson~Kimball}}, \bibinfo {author} {\bibfnamefont {C.}~\bibnamefont
  {Pankow}}, \bibinfo {author} {\bibfnamefont {M.~P.}\ \bibnamefont
  {Ledbetter}}, \bibinfo {author} {\bibfnamefont {P.}~\bibnamefont
  {Wlodarczyk}}, \bibinfo {author} {\bibfnamefont {P.}~\bibnamefont {Wcislo}},
  \bibinfo {author} {\bibfnamefont {M.}~\bibnamefont {Pospelov}}, \bibinfo
  {author} {\bibfnamefont {J.~R.}\ \bibnamefont {Smith}}, \bibinfo {author}
  {\bibfnamefont {J.}~\bibnamefont {Read}}, \bibinfo {author} {\bibfnamefont
  {W.}~\bibnamefont {Gawlik}}, \emph {et~al.},\ }\bibfield  {title} {\bibinfo
  {title} {The global network of optical magnetometers for exotic physics
  (gnome): A novel scheme to search for physics beyond the standard model},\
  }\href@noop {} {\bibfield  {journal} {\bibinfo  {journal} {Annalen der
  Physik}\ }\textbf {\bibinfo {volume} {525}},\ \bibinfo {pages} {659}
  (\bibinfo {year} {2013})}\BibitemShut {NoStop}%
\bibitem [{\citenamefont {Jackson~Kimball}\ \emph {et~al.}(2018)\citenamefont
  {Jackson~Kimball}, \citenamefont {Budker}, \citenamefont {Eby}, \citenamefont
  {Pospelov}, \citenamefont {Pustelny}, \citenamefont {Scholtes}, \citenamefont
  {Stadnik}, \citenamefont {Weis},\ and\ \citenamefont
  {Wickenbrock}}]{kimball2018searching}%
  \BibitemOpen
  \bibfield  {author} {\bibinfo {author} {\bibfnamefont {D.}~\bibnamefont
  {Jackson~Kimball}}, \bibinfo {author} {\bibfnamefont {D.}~\bibnamefont
  {Budker}}, \bibinfo {author} {\bibfnamefont {J.}~\bibnamefont {Eby}},
  \bibinfo {author} {\bibfnamefont {M.}~\bibnamefont {Pospelov}}, \bibinfo
  {author} {\bibfnamefont {S.}~\bibnamefont {Pustelny}}, \bibinfo {author}
  {\bibfnamefont {T.}~\bibnamefont {Scholtes}}, \bibinfo {author}
  {\bibfnamefont {Y.}~\bibnamefont {Stadnik}}, \bibinfo {author} {\bibfnamefont
  {A.}~\bibnamefont {Weis}},\ and\ \bibinfo {author} {\bibfnamefont
  {A.}~\bibnamefont {Wickenbrock}},\ }\bibfield  {title} {\bibinfo {title}
  {Searching for axion stars and q-balls with a terrestrial magnetometer
  network},\ }\href@noop {} {\bibfield  {journal} {\bibinfo  {journal} {Phys.
  Rev. D}\ }\textbf {\bibinfo {volume} {97}},\ \bibinfo {pages} {043002}
  (\bibinfo {year} {2018})}\BibitemShut {NoStop}%
\bibitem [{\citenamefont {Afach}\ \emph {et~al.}(2018)\citenamefont {Afach},
  \citenamefont {Budker}, \citenamefont {DeCamp}, \citenamefont {Dumont},
  \citenamefont {Gruji{\'c}}, \citenamefont {Guo}, \citenamefont
  {Jackson~Kimball}, \citenamefont {Kornack}, \citenamefont {Lebedev},
  \citenamefont {Li} \emph {et~al.}}]{afach2018characterization}%
  \BibitemOpen
  \bibfield  {author} {\bibinfo {author} {\bibfnamefont {S.}~\bibnamefont
  {Afach}}, \bibinfo {author} {\bibfnamefont {D.}~\bibnamefont {Budker}},
  \bibinfo {author} {\bibfnamefont {G.}~\bibnamefont {DeCamp}}, \bibinfo
  {author} {\bibfnamefont {V.}~\bibnamefont {Dumont}}, \bibinfo {author}
  {\bibfnamefont {Z.~D.}\ \bibnamefont {Gruji{\'c}}}, \bibinfo {author}
  {\bibfnamefont {H.}~\bibnamefont {Guo}}, \bibinfo {author} {\bibfnamefont
  {D.}~\bibnamefont {Jackson~Kimball}}, \bibinfo {author} {\bibfnamefont
  {T.}~\bibnamefont {Kornack}}, \bibinfo {author} {\bibfnamefont
  {V.}~\bibnamefont {Lebedev}}, \bibinfo {author} {\bibfnamefont
  {W.}~\bibnamefont {Li}}, \emph {et~al.},\ }\bibfield  {title} {\bibinfo
  {title} {Characterization of the global network of optical magnetometers to
  search for exotic physics (gnome)},\ }\href@noop {} {\bibfield  {journal}
  {\bibinfo  {journal} {Phys. Dark Universe}\ }\textbf {\bibinfo {volume}
  {22}},\ \bibinfo {pages} {162} (\bibinfo {year} {2018})}\BibitemShut
  {NoStop}%
\bibitem [{\citenamefont {Masia-Roig}\ \emph {et~al.}(2020)\citenamefont
  {Masia-Roig}, \citenamefont {Smiga}, \citenamefont {Budker}, \citenamefont
  {Dumont}, \citenamefont {Grujic}, \citenamefont {Kim}, \citenamefont
  {Jackson~Kimball}, \citenamefont {Lebedev}, \citenamefont {Monroy},
  \citenamefont {Pustelny} \emph {et~al.}}]{masia2020analysis}%
  \BibitemOpen
  \bibfield  {author} {\bibinfo {author} {\bibfnamefont {H.}~\bibnamefont
  {Masia-Roig}}, \bibinfo {author} {\bibfnamefont {J.~A.}\ \bibnamefont
  {Smiga}}, \bibinfo {author} {\bibfnamefont {D.}~\bibnamefont {Budker}},
  \bibinfo {author} {\bibfnamefont {V.}~\bibnamefont {Dumont}}, \bibinfo
  {author} {\bibfnamefont {Z.}~\bibnamefont {Grujic}}, \bibinfo {author}
  {\bibfnamefont {D.}~\bibnamefont {Kim}}, \bibinfo {author} {\bibfnamefont
  {D.~F.}\ \bibnamefont {Jackson~Kimball}}, \bibinfo {author} {\bibfnamefont
  {V.}~\bibnamefont {Lebedev}}, \bibinfo {author} {\bibfnamefont
  {M.}~\bibnamefont {Monroy}}, \bibinfo {author} {\bibfnamefont
  {S.}~\bibnamefont {Pustelny}}, \emph {et~al.},\ }\bibfield  {title} {\bibinfo
  {title} {Analysis method for detecting topological defect dark matter with a
  global magnetometer network},\ }\href@noop {} {\bibfield  {journal} {\bibinfo
   {journal} {Phys. Dark Universe}\ }\textbf {\bibinfo {volume} {28}},\
  \bibinfo {pages} {100494} (\bibinfo {year} {2020})}\BibitemShut {NoStop}%
\bibitem [{\citenamefont {Afach}\ \emph {et~al.}(2021)\citenamefont {Afach},
  \citenamefont {Buchler}, \citenamefont {Budker}, \citenamefont {Dailey},
  \citenamefont {Derevianko}, \citenamefont {Dumont}, \citenamefont {Figueroa},
  \citenamefont {Gerhardt}, \citenamefont {Gruji{\'c}}, \citenamefont {Guo}
  \emph {et~al.}}]{afach2021search}%
  \BibitemOpen
  \bibfield  {author} {\bibinfo {author} {\bibfnamefont {S.}~\bibnamefont
  {Afach}}, \bibinfo {author} {\bibfnamefont {B.~C.}\ \bibnamefont {Buchler}},
  \bibinfo {author} {\bibfnamefont {D.}~\bibnamefont {Budker}}, \bibinfo
  {author} {\bibfnamefont {C.}~\bibnamefont {Dailey}}, \bibinfo {author}
  {\bibfnamefont {A.}~\bibnamefont {Derevianko}}, \bibinfo {author}
  {\bibfnamefont {V.}~\bibnamefont {Dumont}}, \bibinfo {author} {\bibfnamefont
  {N.~L.}\ \bibnamefont {Figueroa}}, \bibinfo {author} {\bibfnamefont
  {I.}~\bibnamefont {Gerhardt}}, \bibinfo {author} {\bibfnamefont {Z.~D.}\
  \bibnamefont {Gruji{\'c}}}, \bibinfo {author} {\bibfnamefont
  {H.}~\bibnamefont {Guo}}, \emph {et~al.},\ }\bibfield  {title} {\bibinfo
  {title} {Search for topological defect dark matter with a global network of
  optical magnetometers},\ }\href@noop {} {\bibfield  {journal} {\bibinfo
  {journal} {Nature Physics}\ }\textbf {\bibinfo {volume} {17}},\ \bibinfo
  {pages} {1396} (\bibinfo {year} {2021})}\BibitemShut {NoStop}%
\bibitem [{\citenamefont {Budker}\ and\ \citenamefont
  {Romalis}(2007)}]{budker2007optical}%
  \BibitemOpen
  \bibfield  {author} {\bibinfo {author} {\bibfnamefont {D.}~\bibnamefont
  {Budker}}\ and\ \bibinfo {author} {\bibfnamefont {M.}~\bibnamefont
  {Romalis}},\ }\bibfield  {title} {\bibinfo {title} {Optical magnetometry},\
  }\href@noop {} {\bibfield  {journal} {\bibinfo  {journal} {Nature Phys.}\
  }\textbf {\bibinfo {volume} {3}},\ \bibinfo {pages} {227} (\bibinfo {year}
  {2007})}\BibitemShut {NoStop}%
\bibitem [{\citenamefont {Budker}\ and\ \citenamefont
  {Jackson~Kimball}(2013)}]{Bud13}%
  \BibitemOpen
  \bibfield  {author} {\bibinfo {author} {\bibfnamefont {D.}~\bibnamefont
  {Budker}}\ and\ \bibinfo {author} {\bibfnamefont {D.~F.}\ \bibnamefont
  {Jackson~Kimball}},\ }\href@noop {} {\emph {\bibinfo {title} {Optical
  Magnetometry}}}\ (\bibinfo  {publisher} {Cambridge University Press},\
  \bibinfo {year} {2013})\BibitemShut {NoStop}%
\bibitem [{\citenamefont {Budker}\ \emph
  {et~al.}(2002{\natexlab{a}})\citenamefont {Budker}, \citenamefont {Gawlik},
  \citenamefont {Kimball}, \citenamefont {Rochester}, \citenamefont
  {Yashchuk},\ and\ \citenamefont {Weis}}]{budker2002resonant}%
  \BibitemOpen
  \bibfield  {author} {\bibinfo {author} {\bibfnamefont {D.}~\bibnamefont
  {Budker}}, \bibinfo {author} {\bibfnamefont {W.}~\bibnamefont {Gawlik}},
  \bibinfo {author} {\bibfnamefont {D.}~\bibnamefont {Kimball}}, \bibinfo
  {author} {\bibfnamefont {S.}~\bibnamefont {Rochester}}, \bibinfo {author}
  {\bibfnamefont {V.}~\bibnamefont {Yashchuk}},\ and\ \bibinfo {author}
  {\bibfnamefont {A.}~\bibnamefont {Weis}},\ }\bibfield  {title} {\bibinfo
  {title} {Resonant nonlinear magneto-optical effects in atoms},\ }\href@noop
  {} {\bibfield  {journal} {\bibinfo  {journal} {Rev. Mod. Phys.}\ }\textbf
  {\bibinfo {volume} {74}},\ \bibinfo {pages} {1153} (\bibinfo {year}
  {2002}{\natexlab{a}})}\BibitemShut {NoStop}%
\bibitem [{\citenamefont {Graham}\ \emph {et~al.}(2018)\citenamefont {Graham},
  \citenamefont {Kaplan}, \citenamefont {Mardon}, \citenamefont {Rajendran},
  \citenamefont {Terrano}, \citenamefont {Trahms},\ and\ \citenamefont
  {Wilkason}}]{graham2018spin}%
  \BibitemOpen
  \bibfield  {author} {\bibinfo {author} {\bibfnamefont {P.~W.}\ \bibnamefont
  {Graham}}, \bibinfo {author} {\bibfnamefont {D.~E.}\ \bibnamefont {Kaplan}},
  \bibinfo {author} {\bibfnamefont {J.}~\bibnamefont {Mardon}}, \bibinfo
  {author} {\bibfnamefont {S.}~\bibnamefont {Rajendran}}, \bibinfo {author}
  {\bibfnamefont {W.~A.}\ \bibnamefont {Terrano}}, \bibinfo {author}
  {\bibfnamefont {L.}~\bibnamefont {Trahms}},\ and\ \bibinfo {author}
  {\bibfnamefont {T.}~\bibnamefont {Wilkason}},\ }\bibfield  {title} {\bibinfo
  {title} {Spin precession experiments for light axionic dark matter},\
  }\href@noop {} {\bibfield  {journal} {\bibinfo  {journal} {Phys. Rev. D}\
  }\textbf {\bibinfo {volume} {97}},\ \bibinfo {pages} {055006} (\bibinfo
  {year} {2018})}\BibitemShut {NoStop}%
\bibitem [{\citenamefont {del Castillo}\ \emph {et~al.}(2025)\citenamefont {del
  Castillo}, \citenamefont {Hammett},\ and\ \citenamefont
  {Jaeckel}}]{del2025enhanced}%
  \BibitemOpen
  \bibfield  {author} {\bibinfo {author} {\bibfnamefont {Y.~G.}\ \bibnamefont
  {del Castillo}}, \bibinfo {author} {\bibfnamefont {B.}~\bibnamefont
  {Hammett}},\ and\ \bibinfo {author} {\bibfnamefont {J.}~\bibnamefont
  {Jaeckel}},\ }\bibfield  {title} {\bibinfo {title} {{Enhanced Axion-wind near
  Earth's Surface}},\ }\href@noop {} {\bibfield  {journal} {\bibinfo  {journal}
  {arXiv:2502.04456}\ } (\bibinfo {year} {2025})}\BibitemShut {NoStop}%
\bibitem [{\citenamefont {Banerjee}\ \emph {et~al.}(2025)\citenamefont
  {Banerjee}, \citenamefont {Bloch}, \citenamefont {Bonnefoy}, \citenamefont
  {Ellis}, \citenamefont {Perez}, \citenamefont {Savoray}, \citenamefont
  {Springmann},\ and\ \citenamefont {Stadnik}}]{banerjee2025momentum}%
  \BibitemOpen
  \bibfield  {author} {\bibinfo {author} {\bibfnamefont {A.}~\bibnamefont
  {Banerjee}}, \bibinfo {author} {\bibfnamefont {I.~M.}\ \bibnamefont {Bloch}},
  \bibinfo {author} {\bibfnamefont {Q.}~\bibnamefont {Bonnefoy}}, \bibinfo
  {author} {\bibfnamefont {S.~A.}\ \bibnamefont {Ellis}}, \bibinfo {author}
  {\bibfnamefont {G.}~\bibnamefont {Perez}}, \bibinfo {author} {\bibfnamefont
  {I.}~\bibnamefont {Savoray}}, \bibinfo {author} {\bibfnamefont
  {K.}~\bibnamefont {Springmann}},\ and\ \bibinfo {author} {\bibfnamefont
  {Y.~V.}\ \bibnamefont {Stadnik}},\ }\bibfield  {title} {\bibinfo {title}
  {Momentum and matter matter for axion dark matter matters on earth},\
  }\href@noop {} {\bibfield  {journal} {\bibinfo  {journal} {arXiv:2502.04455}\
  } (\bibinfo {year} {2025})}\BibitemShut {NoStop}%
\bibitem [{\citenamefont {Hees}\ \emph {et~al.}(2018)\citenamefont {Hees},
  \citenamefont {Minazzoli}, \citenamefont {Savalle}, \citenamefont {Stadnik},\
  and\ \citenamefont {Wolf}}]{hees2018violation}%
  \BibitemOpen
  \bibfield  {author} {\bibinfo {author} {\bibfnamefont {A.}~\bibnamefont
  {Hees}}, \bibinfo {author} {\bibfnamefont {O.}~\bibnamefont {Minazzoli}},
  \bibinfo {author} {\bibfnamefont {E.}~\bibnamefont {Savalle}}, \bibinfo
  {author} {\bibfnamefont {Y.~V.}\ \bibnamefont {Stadnik}},\ and\ \bibinfo
  {author} {\bibfnamefont {P.}~\bibnamefont {Wolf}},\ }\bibfield  {title}
  {\bibinfo {title} {Violation of the equivalence principle from light scalar
  dark matter},\ }\href@noop {} {\bibfield  {journal} {\bibinfo  {journal}
  {Phys. Rev. D}\ }\textbf {\bibinfo {volume} {98}},\ \bibinfo {pages} {064051}
  (\bibinfo {year} {2018})}\BibitemShut {NoStop}%
\bibitem [{\citenamefont {Banerjee}\ \emph {et~al.}(2023)\citenamefont
  {Banerjee}, \citenamefont {Perez}, \citenamefont {Safronova}, \citenamefont
  {Savoray},\ and\ \citenamefont {Shalit}}]{banerjee2023phenomenology}%
  \BibitemOpen
  \bibfield  {author} {\bibinfo {author} {\bibfnamefont {A.}~\bibnamefont
  {Banerjee}}, \bibinfo {author} {\bibfnamefont {G.}~\bibnamefont {Perez}},
  \bibinfo {author} {\bibfnamefont {M.}~\bibnamefont {Safronova}}, \bibinfo
  {author} {\bibfnamefont {I.}~\bibnamefont {Savoray}},\ and\ \bibinfo {author}
  {\bibfnamefont {A.}~\bibnamefont {Shalit}},\ }\bibfield  {title} {\bibinfo
  {title} {The phenomenology of quadratically coupled ultra light dark
  matter},\ }\href@noop {} {\bibfield  {journal} {\bibinfo  {journal} {J. High
  Energy Phys.}\ }\textbf {\bibinfo {volume} {2023}}\bibinfo  {number} {
  (10)},\ \bibinfo {pages} {42}}\BibitemShut {NoStop}%
\bibitem [{\citenamefont {Khamis}\ \emph {et~al.}(2025)\citenamefont {Khamis},
  \citenamefont {Sulai}, \citenamefont {Hamilton}, \citenamefont {Afach},
  \citenamefont {Buchler}, \citenamefont {Budker}, \citenamefont {Figueroa},
  \citenamefont {Folman}, \citenamefont {Gavil{\'a}n-Mart{\'\i}n},
  \citenamefont {Givon} \emph {et~al.}}]{khamis2024multi}%
  \BibitemOpen
\bibfield  {number} {  }\bibfield  {author} {\bibinfo {author} {\bibfnamefont
  {S.~S.}\ \bibnamefont {Khamis}}, \bibinfo {author} {\bibfnamefont {I.~A.}\
  \bibnamefont {Sulai}}, \bibinfo {author} {\bibfnamefont {P.}~\bibnamefont
  {Hamilton}}, \bibinfo {author} {\bibfnamefont {S.}~\bibnamefont {Afach}},
  \bibinfo {author} {\bibfnamefont {B.}~\bibnamefont {Buchler}}, \bibinfo
  {author} {\bibfnamefont {D.}~\bibnamefont {Budker}}, \bibinfo {author}
  {\bibfnamefont {N.}~\bibnamefont {Figueroa}}, \bibinfo {author}
  {\bibfnamefont {R.}~\bibnamefont {Folman}}, \bibinfo {author} {\bibfnamefont
  {D.}~\bibnamefont {Gavil{\'a}n-Mart{\'\i}n}}, \bibinfo {author}
  {\bibfnamefont {M.}~\bibnamefont {Givon}}, \emph {et~al.},\ }\bibfield
  {title} {\bibinfo {title} {{Multimessenger Search for Exotic Field Emission
  with a Global Magnetometer Network}},\ }\href@noop {} {\bibfield  {journal}
  {\bibinfo  {journal} {Phys. Rev. X}\ }\textbf {\bibinfo {volume} {15}},\
  \bibinfo {pages} {031048} (\bibinfo {year} {2025})}\BibitemShut {NoStop}%
\bibitem [{\citenamefont {Lella}\ \emph {et~al.}(2024)\citenamefont {Lella},
  \citenamefont {Carenza}, \citenamefont {Co’}, \citenamefont {Lucente},
  \citenamefont {Giannotti}, \citenamefont {Mirizzi},\ and\ \citenamefont
  {Rauscher}}]{lella2024getting}%
  \BibitemOpen
  \bibfield  {author} {\bibinfo {author} {\bibfnamefont {A.}~\bibnamefont
  {Lella}}, \bibinfo {author} {\bibfnamefont {P.}~\bibnamefont {Carenza}},
  \bibinfo {author} {\bibfnamefont {G.}~\bibnamefont {Co’}}, \bibinfo
  {author} {\bibfnamefont {G.}~\bibnamefont {Lucente}}, \bibinfo {author}
  {\bibfnamefont {M.}~\bibnamefont {Giannotti}}, \bibinfo {author}
  {\bibfnamefont {A.}~\bibnamefont {Mirizzi}},\ and\ \bibinfo {author}
  {\bibfnamefont {T.}~\bibnamefont {Rauscher}},\ }\bibfield  {title} {\bibinfo
  {title} {Getting the most on supernova axions},\ }\href@noop {} {\bibfield
  {journal} {\bibinfo  {journal} {Phys. Rev. D}\ }\textbf {\bibinfo {volume}
  {109}},\ \bibinfo {pages} {023001} (\bibinfo {year} {2024})}\BibitemShut
  {NoStop}%
\bibitem [{\citenamefont {Jackson~Kimball}\ \emph {et~al.}(2023)\citenamefont
  {Jackson~Kimball}, \citenamefont {Budker}, \citenamefont {Chupp},
  \citenamefont {Geraci}, \citenamefont {Kolkowitz}, \citenamefont {Singh},\
  and\ \citenamefont {Sushkov}}]{jackson2023probing}%
  \BibitemOpen
  \bibfield  {author} {\bibinfo {author} {\bibfnamefont {D.~F.}\ \bibnamefont
  {Jackson~Kimball}}, \bibinfo {author} {\bibfnamefont {D.}~\bibnamefont
  {Budker}}, \bibinfo {author} {\bibfnamefont {T.~E.}\ \bibnamefont {Chupp}},
  \bibinfo {author} {\bibfnamefont {A.~A.}\ \bibnamefont {Geraci}}, \bibinfo
  {author} {\bibfnamefont {S.}~\bibnamefont {Kolkowitz}}, \bibinfo {author}
  {\bibfnamefont {J.~T.}\ \bibnamefont {Singh}},\ and\ \bibinfo {author}
  {\bibfnamefont {A.~O.}\ \bibnamefont {Sushkov}},\ }\bibfield  {title}
  {\bibinfo {title} {Probing fundamental physics with spin-based quantum
  sensors},\ }\href@noop {} {\bibfield  {journal} {\bibinfo  {journal} {Phys.
  Rev. A}\ ,\ \bibinfo {pages} {010101}} (\bibinfo {year} {2023})}\BibitemShut
  {NoStop}%
\bibitem [{\citenamefont {Jiang}\ \emph
  {et~al.}(2024{\natexlab{a}})\citenamefont {Jiang}, \citenamefont {Su},
  \citenamefont {Chen}, \citenamefont {Jiao}, \citenamefont {Huang},
  \citenamefont {Wang}, \citenamefont {Rong}, \citenamefont {Peng},\ and\
  \citenamefont {Du}}]{jiang2024searches}%
  \BibitemOpen
  \bibfield  {author} {\bibinfo {author} {\bibfnamefont {M.}~\bibnamefont
  {Jiang}}, \bibinfo {author} {\bibfnamefont {H.}~\bibnamefont {Su}}, \bibinfo
  {author} {\bibfnamefont {Y.}~\bibnamefont {Chen}}, \bibinfo {author}
  {\bibfnamefont {M.}~\bibnamefont {Jiao}}, \bibinfo {author} {\bibfnamefont
  {Y.}~\bibnamefont {Huang}}, \bibinfo {author} {\bibfnamefont
  {Y.}~\bibnamefont {Wang}}, \bibinfo {author} {\bibfnamefont {X.}~\bibnamefont
  {Rong}}, \bibinfo {author} {\bibfnamefont {X.}~\bibnamefont {Peng}},\ and\
  \bibinfo {author} {\bibfnamefont {J.}~\bibnamefont {Du}},\ }\bibfield
  {title} {\bibinfo {title} {Searches for exotic spin-dependent interactions
  with spin sensors},\ }\href@noop {} {\bibfield  {journal} {\bibinfo
  {journal} {Rep. Prog. Phys.}\ }\textbf {\bibinfo {volume} {88}},\ \bibinfo
  {pages} {016401} (\bibinfo {year} {2024}{\natexlab{a}})}\BibitemShut
  {NoStop}%
\bibitem [{\citenamefont {Gavilan-Martin}\ \emph {et~al.}(2025)\citenamefont
  {Gavilan-Martin}, \citenamefont {{\L}ukasiewicz}, \citenamefont {Padniuk},
  \citenamefont {Klinger}, \citenamefont {Smolis}, \citenamefont {Figueroa},
  \citenamefont {Jackson~Kimball}, \citenamefont {Sushkov}, \citenamefont
  {Pustelny}, \citenamefont {Budker} \emph {et~al.}}]{gavilan2025searching}%
  \BibitemOpen
  \bibfield  {author} {\bibinfo {author} {\bibfnamefont {D.}~\bibnamefont
  {Gavilan-Martin}}, \bibinfo {author} {\bibfnamefont {G.}~\bibnamefont
  {{\L}ukasiewicz}}, \bibinfo {author} {\bibfnamefont {M.}~\bibnamefont
  {Padniuk}}, \bibinfo {author} {\bibfnamefont {E.}~\bibnamefont {Klinger}},
  \bibinfo {author} {\bibfnamefont {M.}~\bibnamefont {Smolis}}, \bibinfo
  {author} {\bibfnamefont {N.~L.}\ \bibnamefont {Figueroa}}, \bibinfo {author}
  {\bibfnamefont {D.~F.}\ \bibnamefont {Jackson~Kimball}}, \bibinfo {author}
  {\bibfnamefont {A.~O.}\ \bibnamefont {Sushkov}}, \bibinfo {author}
  {\bibfnamefont {S.}~\bibnamefont {Pustelny}}, \bibinfo {author}
  {\bibfnamefont {D.}~\bibnamefont {Budker}}, \emph {et~al.},\ }\bibfield
  {title} {\bibinfo {title} {Searching for dark matter with a spin-based
  interferometer},\ }\href@noop {} {\bibfield  {journal} {\bibinfo  {journal}
  {Nature Communications}\ }\textbf {\bibinfo {volume} {16}},\ \bibinfo {pages}
  {4953} (\bibinfo {year} {2025})}\BibitemShut {NoStop}%
\bibitem [{\citenamefont {Jiang}\ \emph {et~al.}(2021)\citenamefont {Jiang},
  \citenamefont {Su}, \citenamefont {Garcon}, \citenamefont {Peng},\ and\
  \citenamefont {Budker}}]{jiang2021search}%
  \BibitemOpen
  \bibfield  {author} {\bibinfo {author} {\bibfnamefont {M.}~\bibnamefont
  {Jiang}}, \bibinfo {author} {\bibfnamefont {H.}~\bibnamefont {Su}}, \bibinfo
  {author} {\bibfnamefont {A.}~\bibnamefont {Garcon}}, \bibinfo {author}
  {\bibfnamefont {X.}~\bibnamefont {Peng}},\ and\ \bibinfo {author}
  {\bibfnamefont {D.}~\bibnamefont {Budker}},\ }\bibfield  {title} {\bibinfo
  {title} {Search for axion-like dark matter with spin-based amplifiers},\
  }\href@noop {} {\bibfield  {journal} {\bibinfo  {journal} {Nature Phys.}\
  }\textbf {\bibinfo {volume} {17}},\ \bibinfo {pages} {1402} (\bibinfo {year}
  {2021})}\BibitemShut {NoStop}%
\bibitem [{\citenamefont {Jiang}\ \emph
  {et~al.}(2024{\natexlab{b}})\citenamefont {Jiang}, \citenamefont {Hong},
  \citenamefont {Hu}, \citenamefont {Chen}, \citenamefont {Yang}, \citenamefont
  {Hu}, \citenamefont {Yang}, \citenamefont {Shu}, \citenamefont {Zhao},
  \citenamefont {Peng} \emph {et~al.}}]{jiang2024long}%
  \BibitemOpen
  \bibfield  {author} {\bibinfo {author} {\bibfnamefont {M.}~\bibnamefont
  {Jiang}}, \bibinfo {author} {\bibfnamefont {T.}~\bibnamefont {Hong}},
  \bibinfo {author} {\bibfnamefont {D.}~\bibnamefont {Hu}}, \bibinfo {author}
  {\bibfnamefont {Y.}~\bibnamefont {Chen}}, \bibinfo {author} {\bibfnamefont
  {F.}~\bibnamefont {Yang}}, \bibinfo {author} {\bibfnamefont {T.}~\bibnamefont
  {Hu}}, \bibinfo {author} {\bibfnamefont {X.}~\bibnamefont {Yang}}, \bibinfo
  {author} {\bibfnamefont {J.}~\bibnamefont {Shu}}, \bibinfo {author}
  {\bibfnamefont {Y.}~\bibnamefont {Zhao}}, \bibinfo {author} {\bibfnamefont
  {X.}~\bibnamefont {Peng}}, \emph {et~al.},\ }\bibfield  {title} {\bibinfo
  {title} {Long-baseline quantum sensor network as dark matter haloscope},\
  }\href@noop {} {\bibfield  {journal} {\bibinfo  {journal} {Nature Commun.}\
  }\textbf {\bibinfo {volume} {15}},\ \bibinfo {pages} {3331} (\bibinfo {year}
  {2024}{\natexlab{b}})}\BibitemShut {NoStop}%
\bibitem [{\citenamefont {Abel}\ \emph {et~al.}(2017)\citenamefont {Abel},
  \citenamefont {Ayres}, \citenamefont {Ban}, \citenamefont {Bison},
  \citenamefont {Bodek}, \citenamefont {Bondar}, \citenamefont {Daum},
  \citenamefont {Fairbairn}, \citenamefont {Flambaum}, \citenamefont
  {Geltenbort} \emph {et~al.}}]{abel2017search}%
  \BibitemOpen
  \bibfield  {author} {\bibinfo {author} {\bibfnamefont {C.}~\bibnamefont
  {Abel}}, \bibinfo {author} {\bibfnamefont {N.~J.}\ \bibnamefont {Ayres}},
  \bibinfo {author} {\bibfnamefont {G.}~\bibnamefont {Ban}}, \bibinfo {author}
  {\bibfnamefont {G.}~\bibnamefont {Bison}}, \bibinfo {author} {\bibfnamefont
  {K.}~\bibnamefont {Bodek}}, \bibinfo {author} {\bibfnamefont
  {V.}~\bibnamefont {Bondar}}, \bibinfo {author} {\bibfnamefont
  {M.}~\bibnamefont {Daum}}, \bibinfo {author} {\bibfnamefont {M.}~\bibnamefont
  {Fairbairn}}, \bibinfo {author} {\bibfnamefont {V.~V.}\ \bibnamefont
  {Flambaum}}, \bibinfo {author} {\bibfnamefont {P.}~\bibnamefont
  {Geltenbort}}, \emph {et~al.},\ }\bibfield  {title} {\bibinfo {title} {Search
  for axionlike dark matter through nuclear spin precession in electric and
  magnetic fields},\ }\href@noop {} {\bibfield  {journal} {\bibinfo  {journal}
  {Phys. Rev. X}\ }\textbf {\bibinfo {volume} {7}},\ \bibinfo {pages} {041034}
  (\bibinfo {year} {2017})}\BibitemShut {NoStop}%
\bibitem [{\citenamefont {Aybas}\ \emph {et~al.}(2021)\citenamefont {Aybas},
  \citenamefont {Adam}, \citenamefont {Blumenthal}, \citenamefont {Gramolin},
  \citenamefont {Johnson}, \citenamefont {Kleyheeg}, \citenamefont {Afach},
  \citenamefont {Blanchard}, \citenamefont {Centers}, \citenamefont {Garcon}
  \emph {et~al.}}]{aybas2021search}%
  \BibitemOpen
  \bibfield  {author} {\bibinfo {author} {\bibfnamefont {D.}~\bibnamefont
  {Aybas}}, \bibinfo {author} {\bibfnamefont {J.}~\bibnamefont {Adam}},
  \bibinfo {author} {\bibfnamefont {E.}~\bibnamefont {Blumenthal}}, \bibinfo
  {author} {\bibfnamefont {A.~V.}\ \bibnamefont {Gramolin}}, \bibinfo {author}
  {\bibfnamefont {D.}~\bibnamefont {Johnson}}, \bibinfo {author} {\bibfnamefont
  {A.}~\bibnamefont {Kleyheeg}}, \bibinfo {author} {\bibfnamefont
  {S.}~\bibnamefont {Afach}}, \bibinfo {author} {\bibfnamefont {J.~W.}\
  \bibnamefont {Blanchard}}, \bibinfo {author} {\bibfnamefont {G.~P.}\
  \bibnamefont {Centers}}, \bibinfo {author} {\bibfnamefont {A.}~\bibnamefont
  {Garcon}}, \emph {et~al.},\ }\bibfield  {title} {\bibinfo {title} {Search for
  axionlike dark matter using solid-state nuclear magnetic resonance},\
  }\href@noop {} {\bibfield  {journal} {\bibinfo  {journal} {Phys. Rev. Lett.}\
  }\textbf {\bibinfo {volume} {126}},\ \bibinfo {pages} {141802} (\bibinfo
  {year} {2021})}\BibitemShut {NoStop}%
\bibitem [{\citenamefont {Xu}\ \emph {et~al.}(2024)\citenamefont {Xu},
  \citenamefont {Ma}, \citenamefont {Wei}, \citenamefont {He}, \citenamefont
  {Heng}, \citenamefont {Huang}, \citenamefont {Ai}, \citenamefont {Liao},
  \citenamefont {Ji}, \citenamefont {Liu} \emph {et~al.}}]{xu2024constraining}%
  \BibitemOpen
  \bibfield  {author} {\bibinfo {author} {\bibfnamefont {Z.}~\bibnamefont
  {Xu}}, \bibinfo {author} {\bibfnamefont {X.}~\bibnamefont {Ma}}, \bibinfo
  {author} {\bibfnamefont {K.}~\bibnamefont {Wei}}, \bibinfo {author}
  {\bibfnamefont {Y.}~\bibnamefont {He}}, \bibinfo {author} {\bibfnamefont
  {X.}~\bibnamefont {Heng}}, \bibinfo {author} {\bibfnamefont {X.}~\bibnamefont
  {Huang}}, \bibinfo {author} {\bibfnamefont {T.}~\bibnamefont {Ai}}, \bibinfo
  {author} {\bibfnamefont {J.}~\bibnamefont {Liao}}, \bibinfo {author}
  {\bibfnamefont {W.}~\bibnamefont {Ji}}, \bibinfo {author} {\bibfnamefont
  {J.}~\bibnamefont {Liu}}, \emph {et~al.},\ }\bibfield  {title} {\bibinfo
  {title} {Constraining ultralight dark matter through an accelerated resonant
  search},\ }\href@noop {} {\bibfield  {journal} {\bibinfo  {journal} {Commun.
  Phys.}\ }\textbf {\bibinfo {volume} {7}},\ \bibinfo {pages} {226} (\bibinfo
  {year} {2024})}\BibitemShut {NoStop}%
\bibitem [{\citenamefont {Garcon}\ \emph {et~al.}(2019)\citenamefont {Garcon},
  \citenamefont {Blanchard}, \citenamefont {Centers}, \citenamefont {Figueroa},
  \citenamefont {Graham}, \citenamefont {Jackson~Kimball}, \citenamefont
  {Rajendran}, \citenamefont {Sushkov}, \citenamefont {Stadnik}, \citenamefont
  {Wickenbrock} \emph {et~al.}}]{garcon2019constraints}%
  \BibitemOpen
  \bibfield  {author} {\bibinfo {author} {\bibfnamefont {A.}~\bibnamefont
  {Garcon}}, \bibinfo {author} {\bibfnamefont {J.~W.}\ \bibnamefont
  {Blanchard}}, \bibinfo {author} {\bibfnamefont {G.~P.}\ \bibnamefont
  {Centers}}, \bibinfo {author} {\bibfnamefont {N.~L.}\ \bibnamefont
  {Figueroa}}, \bibinfo {author} {\bibfnamefont {P.~W.}\ \bibnamefont
  {Graham}}, \bibinfo {author} {\bibfnamefont {D.~F.}\ \bibnamefont
  {Jackson~Kimball}}, \bibinfo {author} {\bibfnamefont {S.}~\bibnamefont
  {Rajendran}}, \bibinfo {author} {\bibfnamefont {A.~O.}\ \bibnamefont
  {Sushkov}}, \bibinfo {author} {\bibfnamefont {Y.~V.}\ \bibnamefont
  {Stadnik}}, \bibinfo {author} {\bibfnamefont {A.}~\bibnamefont
  {Wickenbrock}}, \emph {et~al.},\ }\bibfield  {title} {\bibinfo {title}
  {Constraints on bosonic dark matter from ultralow-field nuclear magnetic
  resonance},\ }\href@noop {} {\bibfield  {journal} {\bibinfo  {journal} {Sci.
  Adv.}\ ,\ \bibinfo {pages} {eaax4539}} (\bibinfo {year} {2019})}\BibitemShut
  {NoStop}%
\bibitem [{\citenamefont {Walter}\ \emph {et~al.}(2025)\citenamefont {Walter},
  \citenamefont {Maliaka}, \citenamefont {Zhang}, \citenamefont {Blanchard},
  \citenamefont {Centers}, \citenamefont {Dogan}, \citenamefont {Engler},
  \citenamefont {Figueroa}, \citenamefont {Kim}, \citenamefont {Kimball} \emph
  {et~al.}}]{walter2025search}%
  \BibitemOpen
  \bibfield  {author} {\bibinfo {author} {\bibfnamefont {J.}~\bibnamefont
  {Walter}}, \bibinfo {author} {\bibfnamefont {O.}~\bibnamefont {Maliaka}},
  \bibinfo {author} {\bibfnamefont {Y.}~\bibnamefont {Zhang}}, \bibinfo
  {author} {\bibfnamefont {J.~W.}\ \bibnamefont {Blanchard}}, \bibinfo {author}
  {\bibfnamefont {G.}~\bibnamefont {Centers}}, \bibinfo {author} {\bibfnamefont
  {A.}~\bibnamefont {Dogan}}, \bibinfo {author} {\bibfnamefont
  {M.}~\bibnamefont {Engler}}, \bibinfo {author} {\bibfnamefont {N.~L.}\
  \bibnamefont {Figueroa}}, \bibinfo {author} {\bibfnamefont {Y.}~\bibnamefont
  {Kim}}, \bibinfo {author} {\bibfnamefont {D.~F.~J.}\ \bibnamefont {Kimball}},
  \emph {et~al.},\ }\bibfield  {title} {\bibinfo {title} {Search for axionlike
  dark matter using liquid-state nuclear magnetic resonance},\ }\href@noop {}
  {\bibfield  {journal} {\bibinfo  {journal} {Phys. Rev. D}\ }\textbf {\bibinfo
  {volume} {112}},\ \bibinfo {pages} {052008} (\bibinfo {year}
  {2025})}\BibitemShut {NoStop}%
\bibitem [{\citenamefont {Abel}\ \emph {et~al.}(2023)\citenamefont {Abel},
  \citenamefont {Ayres}, \citenamefont {Ban}, \citenamefont {Bison},
  \citenamefont {Bodek}, \citenamefont {Bondar}, \citenamefont {Chanel},
  \citenamefont {Crawford}, \citenamefont {Daum}, \citenamefont {Dechenaux}
  \emph {et~al.}}]{abel2023search}%
  \BibitemOpen
  \bibfield  {author} {\bibinfo {author} {\bibfnamefont {C.}~\bibnamefont
  {Abel}}, \bibinfo {author} {\bibfnamefont {N.~J.}\ \bibnamefont {Ayres}},
  \bibinfo {author} {\bibfnamefont {G.}~\bibnamefont {Ban}}, \bibinfo {author}
  {\bibfnamefont {G.}~\bibnamefont {Bison}}, \bibinfo {author} {\bibfnamefont
  {K.}~\bibnamefont {Bodek}}, \bibinfo {author} {\bibfnamefont
  {V.}~\bibnamefont {Bondar}}, \bibinfo {author} {\bibfnamefont
  {E.}~\bibnamefont {Chanel}}, \bibinfo {author} {\bibfnamefont
  {C.}~\bibnamefont {Crawford}}, \bibinfo {author} {\bibfnamefont
  {M.}~\bibnamefont {Daum}}, \bibinfo {author} {\bibfnamefont {B.}~\bibnamefont
  {Dechenaux}}, \emph {et~al.},\ }\bibfield  {title} {\bibinfo {title} {{Search
  for ultralight axion dark matter in a side-band analysis of a Hg-199
  free-spin precession signal}},\ }\href@noop {} {\bibfield  {journal}
  {\bibinfo  {journal} {SciPost Phys.}\ }\textbf {\bibinfo {volume} {15}},\
  \bibinfo {pages} {058} (\bibinfo {year} {2023})}\BibitemShut {NoStop}%
\bibitem [{\citenamefont {Bloch}\ \emph {et~al.}(2023)\citenamefont {Bloch},
  \citenamefont {Shaham}, \citenamefont {Hochberg}, \citenamefont {Kuflik},
  \citenamefont {Volansky},\ and\ \citenamefont {Katz}}]{bloch2023constraints}%
  \BibitemOpen
  \bibfield  {author} {\bibinfo {author} {\bibfnamefont {I.~M.}\ \bibnamefont
  {Bloch}}, \bibinfo {author} {\bibfnamefont {R.}~\bibnamefont {Shaham}},
  \bibinfo {author} {\bibfnamefont {Y.}~\bibnamefont {Hochberg}}, \bibinfo
  {author} {\bibfnamefont {E.}~\bibnamefont {Kuflik}}, \bibinfo {author}
  {\bibfnamefont {T.}~\bibnamefont {Volansky}},\ and\ \bibinfo {author}
  {\bibfnamefont {O.}~\bibnamefont {Katz}},\ }\bibfield  {title} {\bibinfo
  {title} {Constraints on axion-like dark matter from a serf comagnetometer},\
  }\href@noop {} {\bibfield  {journal} {\bibinfo  {journal} {Nature Commun.}\
  }\textbf {\bibinfo {volume} {14}},\ \bibinfo {pages} {5784} (\bibinfo {year}
  {2023})}\BibitemShut {NoStop}%
\bibitem [{\citenamefont {Wei}\ \emph {et~al.}(2025)\citenamefont {Wei},
  \citenamefont {Xu}, \citenamefont {He}, \citenamefont {Ma}, \citenamefont
  {Heng}, \citenamefont {Huang}, \citenamefont {Quan}, \citenamefont {Ji},
  \citenamefont {Liu}, \citenamefont {Wang} \emph {et~al.}}]{wei2025dark}%
  \BibitemOpen
  \bibfield  {author} {\bibinfo {author} {\bibfnamefont {K.}~\bibnamefont
  {Wei}}, \bibinfo {author} {\bibfnamefont {Z.}~\bibnamefont {Xu}}, \bibinfo
  {author} {\bibfnamefont {Y.}~\bibnamefont {He}}, \bibinfo {author}
  {\bibfnamefont {X.}~\bibnamefont {Ma}}, \bibinfo {author} {\bibfnamefont
  {X.}~\bibnamefont {Heng}}, \bibinfo {author} {\bibfnamefont {X.}~\bibnamefont
  {Huang}}, \bibinfo {author} {\bibfnamefont {W.}~\bibnamefont {Quan}},
  \bibinfo {author} {\bibfnamefont {W.}~\bibnamefont {Ji}}, \bibinfo {author}
  {\bibfnamefont {J.}~\bibnamefont {Liu}}, \bibinfo {author} {\bibfnamefont
  {X.-P.}\ \bibnamefont {Wang}}, \emph {et~al.},\ }\bibfield  {title} {\bibinfo
  {title} {Dark matter search with a resonantly-coupled hybrid spin system},\
  }\href@noop {} {\bibfield  {journal} {\bibinfo  {journal} {Rep. Prog. Phys.}\
  }\textbf {\bibinfo {volume} {88}},\ \bibinfo {pages} {057801} (\bibinfo
  {year} {2025})}\BibitemShut {NoStop}%
\bibitem [{\citenamefont {Bloch}\ \emph {et~al.}(2022)\citenamefont {Bloch},
  \citenamefont {Ronen}, \citenamefont {Shaham}, \citenamefont {Katz},
  \citenamefont {Volansky},\ and\ \citenamefont {Katz}}]{bloch2022new}%
  \BibitemOpen
  \bibfield  {author} {\bibinfo {author} {\bibfnamefont {I.~M.}\ \bibnamefont
  {Bloch}}, \bibinfo {author} {\bibfnamefont {G.}~\bibnamefont {Ronen}},
  \bibinfo {author} {\bibfnamefont {R.}~\bibnamefont {Shaham}}, \bibinfo
  {author} {\bibfnamefont {O.}~\bibnamefont {Katz}}, \bibinfo {author}
  {\bibfnamefont {T.}~\bibnamefont {Volansky}},\ and\ \bibinfo {author}
  {\bibfnamefont {O.}~\bibnamefont {Katz}},\ }\bibfield  {title} {\bibinfo
  {title} {{New constraints on axion-like dark matter using a Floquet quantum
  detector}},\ }\href@noop {} {\bibfield  {journal} {\bibinfo  {journal} {Sci.
  Adv.}\ }\textbf {\bibinfo {volume} {8}},\ \bibinfo {pages} {eabl8919}
  (\bibinfo {year} {2022})}\BibitemShut {NoStop}%
\bibitem [{\citenamefont {Wu}\ \emph {et~al.}(2019)\citenamefont {Wu},
  \citenamefont {Blanchard}, \citenamefont {Centers}, \citenamefont {Figueroa},
  \citenamefont {Garcon}, \citenamefont {Graham}, \citenamefont {Kimball},
  \citenamefont {Rajendran}, \citenamefont {Stadnik}, \citenamefont {Sushkov}
  \emph {et~al.}}]{wu2019search}%
  \BibitemOpen
  \bibfield  {author} {\bibinfo {author} {\bibfnamefont {T.}~\bibnamefont
  {Wu}}, \bibinfo {author} {\bibfnamefont {J.~W.}\ \bibnamefont {Blanchard}},
  \bibinfo {author} {\bibfnamefont {G.~P.}\ \bibnamefont {Centers}}, \bibinfo
  {author} {\bibfnamefont {N.~L.}\ \bibnamefont {Figueroa}}, \bibinfo {author}
  {\bibfnamefont {A.}~\bibnamefont {Garcon}}, \bibinfo {author} {\bibfnamefont
  {P.~W.}\ \bibnamefont {Graham}}, \bibinfo {author} {\bibfnamefont {D.~F.~J.}\
  \bibnamefont {Kimball}}, \bibinfo {author} {\bibfnamefont {S.}~\bibnamefont
  {Rajendran}}, \bibinfo {author} {\bibfnamefont {Y.~V.}\ \bibnamefont
  {Stadnik}}, \bibinfo {author} {\bibfnamefont {A.~O.}\ \bibnamefont
  {Sushkov}}, \emph {et~al.},\ }\bibfield  {title} {\bibinfo {title} {Search
  for axionlike dark matter with a liquid-state nuclear spin comagnetometer},\
  }\href@noop {} {\bibfield  {journal} {\bibinfo  {journal} {Phys. Rev. Lett.}\
  }\textbf {\bibinfo {volume} {122}},\ \bibinfo {pages} {191302} (\bibinfo
  {year} {2019})}\BibitemShut {NoStop}%
\bibitem [{\citenamefont {Crescini}\ \emph {et~al.}(2020)\citenamefont
  {Crescini}, \citenamefont {Alesini}, \citenamefont {Braggio}, \citenamefont
  {Carugno}, \citenamefont {D’Agostino}, \citenamefont {Di~Gioacchino},
  \citenamefont {Falferi}, \citenamefont {Gambardella}, \citenamefont {Gatti},
  \citenamefont {Iannone} \emph {et~al.}}]{crescini2020axion}%
  \BibitemOpen
  \bibfield  {author} {\bibinfo {author} {\bibfnamefont {N.}~\bibnamefont
  {Crescini}}, \bibinfo {author} {\bibfnamefont {D.}~\bibnamefont {Alesini}},
  \bibinfo {author} {\bibfnamefont {C.}~\bibnamefont {Braggio}}, \bibinfo
  {author} {\bibfnamefont {G.}~\bibnamefont {Carugno}}, \bibinfo {author}
  {\bibfnamefont {D.}~\bibnamefont {D’Agostino}}, \bibinfo {author}
  {\bibfnamefont {D.}~\bibnamefont {Di~Gioacchino}}, \bibinfo {author}
  {\bibfnamefont {P.}~\bibnamefont {Falferi}}, \bibinfo {author} {\bibfnamefont
  {U.}~\bibnamefont {Gambardella}}, \bibinfo {author} {\bibfnamefont
  {C.}~\bibnamefont {Gatti}}, \bibinfo {author} {\bibfnamefont
  {G.}~\bibnamefont {Iannone}}, \emph {et~al.},\ }\bibfield  {title} {\bibinfo
  {title} {Axion search with a quantum-limited ferromagnetic haloscope},\
  }\href@noop {} {\bibfield  {journal} {\bibinfo  {journal} {Phys. Rev. Lett.}\
  }\textbf {\bibinfo {volume} {124}},\ \bibinfo {pages} {171801} (\bibinfo
  {year} {2020})}\BibitemShut {NoStop}%
\bibitem [{\citenamefont {Kim}\ \emph {et~al.}(2022)\citenamefont {Kim},
  \citenamefont {Kimball}, \citenamefont {Masia-Roig}, \citenamefont {Smiga},
  \citenamefont {Wickenbrock}, \citenamefont {Budker}, \citenamefont {Kim},
  \citenamefont {Shin},\ and\ \citenamefont {Semertzidis}}]{kim2022machine}%
  \BibitemOpen
  \bibfield  {author} {\bibinfo {author} {\bibfnamefont {D.}~\bibnamefont
  {Kim}}, \bibinfo {author} {\bibfnamefont {D.~F.~J.}\ \bibnamefont {Kimball}},
  \bibinfo {author} {\bibfnamefont {H.}~\bibnamefont {Masia-Roig}}, \bibinfo
  {author} {\bibfnamefont {J.~A.}\ \bibnamefont {Smiga}}, \bibinfo {author}
  {\bibfnamefont {A.}~\bibnamefont {Wickenbrock}}, \bibinfo {author}
  {\bibfnamefont {D.}~\bibnamefont {Budker}}, \bibinfo {author} {\bibfnamefont
  {Y.}~\bibnamefont {Kim}}, \bibinfo {author} {\bibfnamefont {Y.~C.}\
  \bibnamefont {Shin}},\ and\ \bibinfo {author} {\bibfnamefont {Y.~K.}\
  \bibnamefont {Semertzidis}},\ }\bibfield  {title} {\bibinfo {title} {A
  machine learning algorithm for direct detection of axion-like particle domain
  walls},\ }\href@noop {} {\bibfield  {journal} {\bibinfo  {journal} {Physics
  of the Dark Universe}\ }\textbf {\bibinfo {volume} {37}},\ \bibinfo {pages}
  {101118} (\bibinfo {year} {2022})}\BibitemShut {NoStop}%
\bibitem [{\citenamefont {Masia-Roig}\ \emph {et~al.}(2023)\citenamefont
  {Masia-Roig}, \citenamefont {Figueroa}, \citenamefont {Bordon}, \citenamefont
  {Smiga}, \citenamefont {Stadnik}, \citenamefont {Budker}, \citenamefont
  {Centers}, \citenamefont {Gramolin}, \citenamefont {Hamilton}, \citenamefont
  {Khamis} \emph {et~al.}}]{masia2023intensity}%
  \BibitemOpen
  \bibfield  {author} {\bibinfo {author} {\bibfnamefont {H.}~\bibnamefont
  {Masia-Roig}}, \bibinfo {author} {\bibfnamefont {N.~L.}\ \bibnamefont
  {Figueroa}}, \bibinfo {author} {\bibfnamefont {A.}~\bibnamefont {Bordon}},
  \bibinfo {author} {\bibfnamefont {J.~A.}\ \bibnamefont {Smiga}}, \bibinfo
  {author} {\bibfnamefont {Y.~V.}\ \bibnamefont {Stadnik}}, \bibinfo {author}
  {\bibfnamefont {D.}~\bibnamefont {Budker}}, \bibinfo {author} {\bibfnamefont
  {G.~P.}\ \bibnamefont {Centers}}, \bibinfo {author} {\bibfnamefont {A.~V.}\
  \bibnamefont {Gramolin}}, \bibinfo {author} {\bibfnamefont {P.~S.}\
  \bibnamefont {Hamilton}}, \bibinfo {author} {\bibfnamefont {S.}~\bibnamefont
  {Khamis}}, \emph {et~al.},\ }\bibfield  {title} {\bibinfo {title} {Intensity
  interferometry for ultralight bosonic dark matter detection},\ }\href@noop {}
  {\bibfield  {journal} {\bibinfo  {journal} {Phys. Rev. D}\ }\textbf {\bibinfo
  {volume} {108}},\ \bibinfo {pages} {015003} (\bibinfo {year}
  {2023})}\BibitemShut {NoStop}%
\bibitem [{\citenamefont {Eby}\ \emph {et~al.}(2022)\citenamefont {Eby},
  \citenamefont {Shirai}, \citenamefont {Stadnik},\ and\ \citenamefont
  {Takhistov}}]{eby2022probing}%
  \BibitemOpen
  \bibfield  {author} {\bibinfo {author} {\bibfnamefont {J.}~\bibnamefont
  {Eby}}, \bibinfo {author} {\bibfnamefont {S.}~\bibnamefont {Shirai}},
  \bibinfo {author} {\bibfnamefont {Y.~V.}\ \bibnamefont {Stadnik}},\ and\
  \bibinfo {author} {\bibfnamefont {V.}~\bibnamefont {Takhistov}},\ }\bibfield
  {title} {\bibinfo {title} {Probing relativistic axions from transient
  astrophysical sources},\ }\href@noop {} {\bibfield  {journal} {\bibinfo
  {journal} {Phys. Lett. B}\ }\textbf {\bibinfo {volume} {825}},\ \bibinfo
  {pages} {136858} (\bibinfo {year} {2022})}\BibitemShut {NoStop}%
\bibitem [{\citenamefont {Arakawa}\ \emph {et~al.}(2025)\citenamefont
  {Arakawa}, \citenamefont {Zaheer}, \citenamefont {Takhistov}, \citenamefont
  {Safronova}, \citenamefont {Eby},\ and\ \citenamefont
  {Cheung}}]{arakawa2025multimessenger}%
  \BibitemOpen
  \bibfield  {author} {\bibinfo {author} {\bibfnamefont {J.}~\bibnamefont
  {Arakawa}}, \bibinfo {author} {\bibfnamefont {M.~H.}\ \bibnamefont {Zaheer}},
  \bibinfo {author} {\bibfnamefont {V.}~\bibnamefont {Takhistov}}, \bibinfo
  {author} {\bibfnamefont {M.~S.}\ \bibnamefont {Safronova}}, \bibinfo {author}
  {\bibfnamefont {J.}~\bibnamefont {Eby}},\ and\ \bibinfo {author}
  {\bibfnamefont {C.}~\bibnamefont {Cheung}},\ }\bibfield  {title} {\bibinfo
  {title} {Multimessenger astronomy beyond the standard model: New window from
  quantum sensors},\ }\href@noop {} {\bibfield  {journal} {\bibinfo  {journal}
  {arXiv:2502.08716}\ } (\bibinfo {year} {2025})}\BibitemShut {NoStop}%
\bibitem [{GNO(2026)}]{GNOMEwebsite}%
  \BibitemOpen
  \href@noop {} {\bibinfo {title} {{{GNOME collaboration website}}}},\ \bibinfo
  {howpublished} {\url{https://budker.uni-mainz.de/gnome/}} (\bibinfo {year}
  {2026})\BibitemShut {NoStop}%
\bibitem [{\citenamefont {Osborne}\ \emph {et~al.}(2018)\citenamefont
  {Osborne}, \citenamefont {Orton}, \citenamefont {Alem},\ and\ \citenamefont
  {Shah}}]{osborne2018fully}%
  \BibitemOpen
  \bibfield  {author} {\bibinfo {author} {\bibfnamefont {J.}~\bibnamefont
  {Osborne}}, \bibinfo {author} {\bibfnamefont {J.}~\bibnamefont {Orton}},
  \bibinfo {author} {\bibfnamefont {O.}~\bibnamefont {Alem}},\ and\ \bibinfo
  {author} {\bibfnamefont {V.}~\bibnamefont {Shah}},\ }\bibfield  {title}
  {\bibinfo {title} {Fully integrated standalone zero field optically pumped
  magnetometer for biomagnetism},\ }in\ \href@noop {} {\emph {\bibinfo
  {booktitle} {Steep Dispersion Engineering and Opto-Atomic Precision Metrology
  XI}}},\ Vol.\ \bibinfo {volume} {10548}\ (\bibinfo {organization}
  {International Society for Optics and Photonics},\ \bibinfo {year} {2018})\
  p.\ \bibinfo {pages} {105481G}\BibitemShut {NoStop}%
\bibitem [{\citenamefont {Savukov}\ and\ \citenamefont
  {Romalis}(2005)}]{savukov2005effects}%
  \BibitemOpen
  \bibfield  {author} {\bibinfo {author} {\bibfnamefont {I.}~\bibnamefont
  {Savukov}}\ and\ \bibinfo {author} {\bibfnamefont {M.}~\bibnamefont
  {Romalis}},\ }\bibfield  {title} {\bibinfo {title} {Effects of spin-exchange
  collisions in a high-density alkali-metal vapor in low magnetic fields},\
  }\href@noop {} {\bibfield  {journal} {\bibinfo  {journal} {Phys. Rev. A}\
  }\textbf {\bibinfo {volume} {71}},\ \bibinfo {pages} {023405} (\bibinfo
  {year} {2005})}\BibitemShut {NoStop}%
\bibitem [{\citenamefont {Appelt}\ \emph {et~al.}(1998)\citenamefont {Appelt},
  \citenamefont {Baranga}, \citenamefont {Erickson}, \citenamefont {Romalis},
  \citenamefont {Young},\ and\ \citenamefont {Happer}}]{appelt1998theory}%
  \BibitemOpen
  \bibfield  {author} {\bibinfo {author} {\bibfnamefont {S.}~\bibnamefont
  {Appelt}}, \bibinfo {author} {\bibfnamefont {A.~B.-A.}\ \bibnamefont
  {Baranga}}, \bibinfo {author} {\bibfnamefont {C.}~\bibnamefont {Erickson}},
  \bibinfo {author} {\bibfnamefont {M.}~\bibnamefont {Romalis}}, \bibinfo
  {author} {\bibfnamefont {A.}~\bibnamefont {Young}},\ and\ \bibinfo {author}
  {\bibfnamefont {W.}~\bibnamefont {Happer}},\ }\bibfield  {title} {\bibinfo
  {title} {Theory of spin-exchange optical pumping of 3 he and 129 xe},\
  }\href@noop {} {\bibfield  {journal} {\bibinfo  {journal} {Phys. Rev. A}\
  }\textbf {\bibinfo {volume} {58}},\ \bibinfo {pages} {1412} (\bibinfo {year}
  {1998})}\BibitemShut {NoStop}%
\bibitem [{\citenamefont {Schmidt}(1937)}]{schmidt1937magnetischen}%
  \BibitemOpen
  \bibfield  {author} {\bibinfo {author} {\bibfnamefont {T.}~\bibnamefont
  {Schmidt}},\ }\bibfield  {title} {\bibinfo {title} {{\"U}ber die magnetischen
  momente der atomkerne},\ }\href@noop {} {\bibfield  {journal} {\bibinfo
  {journal} {Z. Phys.}\ }\textbf {\bibinfo {volume} {106}},\ \bibinfo {pages}
  {358} (\bibinfo {year} {1937})}\BibitemShut {NoStop}%
\bibitem [{\citenamefont {Budker}\ \emph {et~al.}(2008)\citenamefont {Budker},
  \citenamefont {Kimball},\ and\ \citenamefont {DeMille}}]{budker2008atomic}%
  \BibitemOpen
  \bibfield  {author} {\bibinfo {author} {\bibfnamefont {D.}~\bibnamefont
  {Budker}}, \bibinfo {author} {\bibfnamefont {D.~F.}\ \bibnamefont
  {Kimball}},\ and\ \bibinfo {author} {\bibfnamefont {D.~P.}\ \bibnamefont
  {DeMille}},\ }\href@noop {} {\emph {\bibinfo {title} {Atomic physics: an
  exploration through problems and solutions}}}\ (\bibinfo  {publisher} {Oxford
  University Press, USA},\ \bibinfo {year} {2008})\BibitemShut {NoStop}%
\bibitem [{\citenamefont {Budker}\ \emph
  {et~al.}(2002{\natexlab{b}})\citenamefont {Budker}, \citenamefont {Kimball},
  \citenamefont {Yashchuk},\ and\ \citenamefont
  {Zolotorev}}]{budker2002nonlinear}%
  \BibitemOpen
  \bibfield  {author} {\bibinfo {author} {\bibfnamefont {D.}~\bibnamefont
  {Budker}}, \bibinfo {author} {\bibfnamefont {D.}~\bibnamefont {Kimball}},
  \bibinfo {author} {\bibfnamefont {V.}~\bibnamefont {Yashchuk}},\ and\
  \bibinfo {author} {\bibfnamefont {M.}~\bibnamefont {Zolotorev}},\ }\bibfield
  {title} {\bibinfo {title} {Nonlinear magneto-optical rotation with
  frequency-modulated light},\ }\href@noop {} {\bibfield  {journal} {\bibinfo
  {journal} {Phys. Rev. A}\ }\textbf {\bibinfo {volume} {65}},\ \bibinfo
  {pages} {055403} (\bibinfo {year} {2002}{\natexlab{b}})}\BibitemShut
  {NoStop}%
\bibitem [{\citenamefont {Jackson~Kimball}\ \emph {et~al.}(2009)\citenamefont
  {Jackson~Kimball}, \citenamefont {Jacome}, \citenamefont {Guttikonda},
  \citenamefont {Bahr},\ and\ \citenamefont {Chan}}]{kimball2009magnetometric}%
  \BibitemOpen
  \bibfield  {author} {\bibinfo {author} {\bibfnamefont {D.~F.}\ \bibnamefont
  {Jackson~Kimball}}, \bibinfo {author} {\bibfnamefont {L.~R.}\ \bibnamefont
  {Jacome}}, \bibinfo {author} {\bibfnamefont {S.}~\bibnamefont {Guttikonda}},
  \bibinfo {author} {\bibfnamefont {E.~J.}\ \bibnamefont {Bahr}},\ and\
  \bibinfo {author} {\bibfnamefont {L.~F.}\ \bibnamefont {Chan}},\ }\bibfield
  {title} {\bibinfo {title} {Magnetometric sensitivity optimization for
  nonlinear optical rotation with frequency-modulated light: Rubidium d2
  line},\ }\href@noop {} {\bibfield  {journal} {\bibinfo  {journal} {J. Appl.
  Phys.}\ }\textbf {\bibinfo {volume} {106}},\ \bibinfo {pages} {063113}
  (\bibinfo {year} {2009})}\BibitemShut {NoStop}%
\bibitem [{\citenamefont {Gawlik}\ \emph {et~al.}(2006)\citenamefont {Gawlik},
  \citenamefont {Krzemie{\'n}}, \citenamefont {Pustelny}, \citenamefont
  {Sangla}, \citenamefont {Zachorowski}, \citenamefont {Graf}, \citenamefont
  {Sushkov},\ and\ \citenamefont {Budker}}]{gawlik2006nonlinear}%
  \BibitemOpen
  \bibfield  {author} {\bibinfo {author} {\bibfnamefont {W.}~\bibnamefont
  {Gawlik}}, \bibinfo {author} {\bibfnamefont {L.}~\bibnamefont
  {Krzemie{\'n}}}, \bibinfo {author} {\bibfnamefont {S.}~\bibnamefont
  {Pustelny}}, \bibinfo {author} {\bibfnamefont {D.}~\bibnamefont {Sangla}},
  \bibinfo {author} {\bibfnamefont {J.}~\bibnamefont {Zachorowski}}, \bibinfo
  {author} {\bibfnamefont {M.}~\bibnamefont {Graf}}, \bibinfo {author}
  {\bibfnamefont {A.}~\bibnamefont {Sushkov}},\ and\ \bibinfo {author}
  {\bibfnamefont {D.}~\bibnamefont {Budker}},\ }\bibfield  {title} {\bibinfo
  {title} {Nonlinear magneto-optical rotation with amplitude modulated light},\
  }\href@noop {} {\bibfield  {journal} {\bibinfo  {journal} {Appl. Phys.
  Lett.}\ }\textbf {\bibinfo {volume} {88}},\ \bibinfo {pages} {131108}
  (\bibinfo {year} {2006})}\BibitemShut {NoStop}%
\bibitem [{\citenamefont {Allred}\ \emph {et~al.}(2002)\citenamefont {Allred},
  \citenamefont {Lyman}, \citenamefont {Kornack},\ and\ \citenamefont
  {Romalis}}]{allred2002high}%
  \BibitemOpen
  \bibfield  {author} {\bibinfo {author} {\bibfnamefont {J.}~\bibnamefont
  {Allred}}, \bibinfo {author} {\bibfnamefont {R.}~\bibnamefont {Lyman}},
  \bibinfo {author} {\bibfnamefont {T.}~\bibnamefont {Kornack}},\ and\ \bibinfo
  {author} {\bibfnamefont {M.~V.}\ \bibnamefont {Romalis}},\ }\bibfield
  {title} {\bibinfo {title} {High-sensitivity atomic magnetometer unaffected by
  spin-exchange relaxation},\ }\href@noop {} {\bibfield  {journal} {\bibinfo
  {journal} {Phys. Rev. Lett.}\ }\textbf {\bibinfo {volume} {89}},\ \bibinfo
  {pages} {130801} (\bibinfo {year} {2002})}\BibitemShut {NoStop}%
\end{thebibliography}
%


\end{document}